\documentclass[twocolumn,showpacs, amsmath,amssymb,superscriptaddress]{revtex4-1}

\usepackage{microtype}
\usepackage{graphicx}

\usepackage{dcolumn}
\usepackage{subfigure}

\newcommand{\kb}[0]{k_\text{B}}
\DeclareMathOperator{\const}{const}
\DeclareMathOperator{\Tr}{Tr}

\newcommand{\D}[0]{\text{d}}

\begin{document}


\title{Logarithmic Finite-Size Effects on Interfacial Free Energies: Phenomenological Theory and Monte Carlo Studies}
\author{Fabian Schmitz}
\author{Peter Virnau}
\author{Kurt Binder}
\affiliation{Institute of Physics, Johannes Gutenberg University Mainz,\\
Staudingerweg 7, D-55128 Mainz, Germany}


\begin{abstract}
The computation of interfacial free energies between coexisting phases (e.g.~saturated vapor and liquid) by computer simulation methods is still a challenging problem due to the difficulty of an atomistic identification of an interface, and due to interfacial fluctuations on all length scales. The approach to estimate the interfacial tension from the free energy excess of a system with interfaces relative to corresponding single-phase systems does not suffer from the first problem but still suffers from the latter. Considering $d$-dimensional systems with interfacial area $L^{d-1}$ and linear dimension $L_z$ in the direction perpendicular to the interface, it is argued that the interfacial fluctuations cause logarithmic finite-size effects of order  $\ln (L) / L^{d-1}$ and order $\ln (L_z)/L ^{d-1}$, in addition to regular corrections (with leading order $\const /L^{d-1}$). A phenomenological theory predicts that the prefactors of the logarithmic terms are universal (but depend on the applied boundary conditions and the considered statistical ensemble). The physical origin of these corrections are the translational entropy of the interface as a whole, ``domain breathing'' (coupling of interfacial fluctuations to the bulk order parameter fluctuations of the coexisting domains), and capillary waves. Using a new variant of the ensemble switch method, interfacial tensions are found from Monte Carlo simulations of $d=2$ and $d=3$ Ising models and a Lennard Jones fluid. The simulation results are fully consistent with the theoretical predictions.
  \end{abstract}

\maketitle

\section{Introduction}
\label{sec: Introduction}

Interfacial phenomena are ubiquitous in the physics of condensed matter and materials science: nucleation of droplets \cite{1,2,3,4,5,6,7,8,9} in a supersaturated vapor (or nucleation of bubbles in an undersaturated liquid) is controlled by a competition between the free energy cost of forming an interface and gain in free energy (resulting from the fact that the stable phase has a lower free energy than the metastable one). Of course, related phenomena occur in more complex systems (crystal nucleation from the melt, formation of nematic or smectic droplets in fluids which can form liquid crystal phases etc.) and in various solid phases (nucleation of ferroelectric or ferromagnetic domains driven by appropriate fields, etc.). In complex fluids and biosystems heterogeneous structures (such as mesophases of strongly segregated block copolymers \cite{10}) are often maintained in thermal equilibrium, due to an interplay of various free energy contributions, one of them being an interfacial tension. Stable heterogeneous structures can also be stabilized in fluids due to the effect of confining walls, e.g. wetting layers \cite{11,12,13} and nanosystems \cite{14}.

Thus, the prediction of the excess free energy due to an interface between coexisting phases is a basic task of statistical mechanics \cite{15,16,17,18}. Although this has been recognized since a long time \cite{19}, and mean-field type approaches have been developed and are widely used e.g.~\cite{20,21,22,23}, such theories are not based on a firm ground: the existence of a well-defined ``intrinsic interfacial profile'' is doubtful \cite{15,16,17,24,25,26}; an inevitable input is the free energy density of homogeneous states \cite{27} throughout the two-phase  coexistence region: this is again a concept valid for systems with long range forces \cite{5,28,29,30}, but ill-defined in the short-range case \cite{5,9,30,31}. While the bulk phase behavior often can be accounted for rather well by mean-field type theories (apart from the neighborhood of critical points, of course, but there the neglected long wave length fluctuations and the effects caused by them can be well accounted for by renormalization group theory \cite{32}), this is not the case for interfacial phenomena. Interfaces (between fluid phases) have fluctuations on all length scales, and although their long wavelength part (capillary waves \cite{33,34,35,36,37,38}) is well understood, the interplay of short wavelengths with fluctuations in the bulk is not yet well understood \cite{26,36,37,38}. Thus one cannot improve the mean-field results by fluctuation corrections systematically.

In view of this dilemma, the prediction of interfacial free energies by computer simulation methods \cite{18,39,40,41,42,43,44,45,46,47,48,49,50,51,52,53,54,55,56,57,58,59,60,61,62,63,64,65,66,67,68,69,70,71,72,73,74,75,76,77,78} is very important. For many model systems of statistical mechanics, computer simulation methods can very accurately predict the equation of state, and thermodynamic properties derivable from it \cite{79,80,81}. Of course, computer simulations deal with systems of finite size, and hence finite-size effects need to be carefully considered \cite{82,83,84,85}, in particular near critical points or if dealing with phase coexistence. However, finite-size scaling concepts for such problems have been established since a long time \cite{81,82,83,84,85} and are very successful \cite{79,81}.

Unfortunately, with respect to finite-size effects on interfacial phenomena the situation is less satisfactory, although the problem has also been considered since a long time \cite{84,86,87,88,89,90,121}. Therefore, the present paper takes up this task again, reconsidering the finite-size effects on interfacial tensions for archetypical model systems, such as the Ising model in $d=2$ and $d=3$ dimensions, and the Lennard-Jones fluid. Our work is based on several ingredients:
	\begin{itemize}
	\item[(i)] By further adaptation of the recently developed ensemble switch method \cite{91,92,93}, a computationally very efficient alternative to existing approaches has become available.
	\item[(ii)] The computational power of recently available computer hardware exceeds the power that was available 20 to 30 years ago, when most previous studies of this problem were done but led to less conclusive results, by many orders of magnitude.
	\item[(iii)] Unlike most previous work, we vary both the linear dimension $L$ parallel to the interface and the linear dimension $L_z$ perpendicular to it systematically. We find that this aspect is crucial to unambiguously identify the sources of the various effects.
	\item[(iv)] We compare systematically the results obtained choosing different boundary conditions (e.g., periodic versus antiperiodic in the Ising model) and different ensembles  (conserved or nonconserved density when we interpret the Ising system as lattice gas).
	\end{itemize}
Due to these ingredients (i)-(iv), we have been able to discover a new mechanism of interfacial fluctuations (``domain breathing''), which has not been mentioned in the previous literature. Apart from the domain breathing mechanism, known effects like the translational entropy of the interface and capillary wave effects play a major role for our study.

As a disclaimer, we emphasize that some important aspects will not be studied in this work: we will not address the interesting crossover \cite{90} of these finite-size effects towards those associated with the critical point, where the interfacial tension vanishes; we also ignore the anisotropy of the interfacial tension (which is present also in the Ising model \cite{44,63,115}, and very important when approaching (in $d=3$) the roughening transition \cite{94} (or zero temperature, in $d=2$). Of course, this anisotropy must not be ignored when one considers crystal-fluid interfaces \cite{65,66,67,68,69,70,71,72,73,74,75,76,77,78}. We plan to study the latter in future work.

The outline of this paper is as follows: in Sec.~\ref{sec: PhenomenologicalTheory}, we describe in detail (a brief summary was already presented in a Letter \cite{95}) the phenomenological theory of the logarithmic finite-size effects on interfacial tensions. In Sec.~\ref{sec: ModelsAndSimulationMethods}, we briefly characterize the models that are studied, and describe the ensemble switch method that is used in the Monte Carlo simulations. Sec.~\ref{sec: NumericalResults} describes our numerical results for the $d=2$ and $d=3$ Ising model and a $d=3$ Lennard-Jones fluid, by which our theoretical predictions are tested. Sec.~\ref{sec: Conclusion} gives a summary and an outlook on open problems.

\section{Phenomenological Theory of Finite-Size Effects on Interfacial Free Energies}
\label{sec: PhenomenologicalTheory}

\subsection{System Geometry and Boundary Conditions}
\label{sec: SystemGeometryAndBoundaryConditions}

For simplicity, in most of our discussions we shall focus on the ferromagnetic Ising system with nearest neighbor interactions of strength $J$, i.e.~described by the Hamiltonian on a square or simple cubic lattice,
\begin{equation} \label{eq: IsingHamiltonian} 
\mathcal{H}=-J \sum\limits_{\langle i,j \rangle} \, S_i S_j - H\sum\limits_{i}  S_i , \quad S_i =\pm 1,
\end{equation}
where $\langle i, j\rangle $ denotes the sum over all nearest neighbor pairs, and $H$ is the magnetic field, which is set to zero throughout this work. We focus on coexisting phases, described for temperatures $T$ less than the critical temperature $T_c$ by states with positive or negative spontaneous magnetization, $\pm m_\text{coex}$. Motivated by the interpretation of the Ising magnet as a lattice gas model (where $S_i=-1$ means that the lattice site $i$ is empty, while $S_i=+1$ means that the lattice site $i$ is occupied by a particle), we denote the $(T,H)$ ensemble as ``grandcanonical'' (gc) and the $(T,m)$ ensemble as ``canonical'' (c). Here $m$ is defined as the magnetization per spin,
\begin{equation} \label{eq2}
m=\frac{1}{L_z L^{d-1}} \sum_i S_i \; ,
\end{equation}
where we have already anticipated that we take a lattice of linear dimension $L_z$ in the $z$-direction, while the linear dimension in the other direction(s) is taken to be~$L$. Remember that in the lattice gas version of the model, the density $\rho=(1 + m)/2$, and $H$ is related to the chemical potential difference relative to the chemical potential $\mu_\text{coex}$ where phase coexistence occurs, $H=(\mu-\mu_\text{coex})/2$.

Next we discuss the boundary conditions that are used to stabilize one or two interfaces between coexisting phases in the system. A very natural choice is the use of free surfaces with neighboring fixed spins in the $z$-direction: using the lattice spacing $a$ as unit of length, all spins in the plane (or row in $d=2$) $n=1$ are fixed at $S_i=+1$ and the spins in the plane $n=L_z$ are fixed at $S_i=-1$ (Fig.~\ref{fig: BoundaryConditionsFreeSurfaces}). Alternatively, we may use boundary magnetic fields $H_1 > 0$ in the plane (row) $n=1$ and $H_{L_z} =-H_1$ in the plane $n=L_z$, and spins in the planes $n=0$, $L_z +1$ are missing. In the remaining direction(s), periodic boundary conditions are used. This choice of boundary conditions is straightforwardly generalized to off-lattice systems which lack the special symmetry against spin reversal of the Ising model. E.g., for a Lennard-Jones fluid (or a polymer solution where the solvent is treated implicitly only \cite{55}), instead of the free surfaces with fixed spins, one uses two hard walls, where one wall is purely repulsive, favoring the vapor (or solvent-rich phase, in the case of the polymer solution) while the other wall has an attractive potential. Similar choices also apply when one studies systems containing a single solid-liquid interface \cite{96}.

It is clear that the properties of the system near these free surfaces or walls differ from the bulk properties over some range, and so $L_z$ has to be chosen large enough so that the effect of an effective potential that the wall exerts on the interface becomes negligible. The effect of this potential becomes appreciable under conditions where the system in the thermodynamic limit would undergo a wetting transition, while for $L_z$ finite but $L \rightarrow \infty$ interface localization/delocalization transitions can occur \cite{97,98,99}. One must then make sure to work under conditions deep inside the phase where the interface is preferentially in the center of the system, near $z=L_z/2$, and never close to the walls.

This problem can be avoided for the Ising model (and other symmetric systems, e.g. a symmetric binary Lennard Jones mixture \cite{64}) by using the antiperiodic boundary condition (APBC), Fig.~\ref{fig: BoundaryConditionsAPBC}, which is equivalent to the choice that spins in the planes $n=1$ and $n=L_z$ interact antiferromagnetically. Then the system retains its translational invariance in the $z$-direction.

However, perhaps the most frequently used choice is to use periodic boundary conditions in all directions, and focus on states of the system where both coexisting phases are present in the system, separated by two domain walls (Fig.~\ref{fig: BoundaryConditionsPBC}).

Note that we normally use $L_z$ larger than $L$ (sometimes it is advantageous to use $L_z \gg L$) but one has to be careful in not using a too large value of $L_z$: We wish to have a situation where in the grandcanonical ensemble systems with APBC (or with fixed spin boundary conditions) are dominated by states with two domains separated by a single interface (as anticipated in Fig.~\ref{fig: BoundaryConditions}) rather than by a larger even number of domains and hence a larger odd number of interfaces. Likewise, in the PBC case (Fig.~\ref{fig: BoundaryConditionsPBC}) the system in the grandcanonical ensemble will in fact be dominated by the pure phases $(m_+, m_-)$ without any interfaces, and the shown state with two interfaces (Fig.~\ref{fig: BoundaryConditionsPBC}) occurs as a rare fluctuation, but states with 4, 6 or more interfaces are comparatively negligible. In fact, for $L_z \rightarrow \infty$ at fixed $L$, the resulting quasi-one dimensional system splits into a sequence of infinitely many domains, the typical distance between domain walls (which is the correlation length of spin correlations in $z$-direction) is given by \cite{86}
\begin{equation} \label{eq3}
\xi_\parallel \propto w_L \exp (\gamma_\infty L^{d-1}) \quad ,
\end{equation}
with
\begin{equation} \label{eq4}
w_L \propto  \left\{
\begin{array}{ll}
\gamma^{-1/2}_\infty L ^{(3-d)/2} & d<3 \\
\gamma^{-1/2}_\infty \sqrt{\ln L} & d=3 
\end{array}
\right.
\end{equation}
where the length $w_L$ is the width of an interface with lateral dimension(s) $L$, and $\gamma_\infty$ is the interfacial tension in the limit $L \rightarrow \infty$. Here and in the following, the interfacial tension is always normalized by the thermal energy $\kb T$, $\kb$ being Boltzmann's constant, and is therefore given in units of inverse ($d-1$)-dimensional area. In Eq.~\eqref{eq3} the results from capillary wave broadening of the interface (see e.g.~\cite{100}) have been anticipated. Strictly speaking, the prefactor in Eq.~\eqref{eq4} for lattice systems is not $\gamma^{-1/2}_\infty$ but rather $\Gamma^{-1/2}$, where $\Gamma$ is the ``interfacial stiffness'' \cite{100}, but this difference is not of interest here. We shall discuss Eqs.~\eqref{eq3},~\eqref{eq4} in later subsections; here we only emphasize that the simulations need to be carried out in the regime $L_z \ll \xi_\parallel$ in order to ensure that only states with one interface (Figs.~\ref{fig: BoundaryConditionsFreeSurfaces} and~\ref{fig: BoundaryConditionsAPBC}) or at most two interfaces (Fig.~\ref{fig: BoundaryConditionsPBC}) are sampled. Apart from the critical region (remember that $\gamma_\infty \rightarrow 0$ as $T \rightarrow T_c$ \cite{15,17}), the exponential variation of $\xi_\parallel$ with the interfacial area $L ^{d-1}$ ensures that for reasonably large $L$ the length $\xi_\parallel$ is extremely large, and so the condition $L_z \ll \xi_\parallel$ is easily fulfilled. When one approaches the critical region, it is necessary to choose $L \gg \xi_b$, $\xi_b$ being the correlation length of order parameter fluctuations in the bulk. We also observe that sampling the order parameter distribution $P_{L, L_z}(m)$ in the grandcanonical ensemble using PBC (Fig.~\ref{fig: BoundaryConditionsPBC}) can also serve as a check that one works in the proper regime of $L$ and $L_z$ (Fig.~\ref{fig: ProbDistributions}). For studies of the interfacial tension, the distribution must have two sharp peaks at $m=\pm m_\text{coex}$ and a flat (essentially horizontal) minimum near $m=0$, with $P_{L,L_z} (m\approx 0)$ many orders of magnitude smaller than $P_{L, L_z}(\pm m_\text{coex})$; note the logarithmic scale of the ordinate in Fig.~\ref{fig: ProbDistributions}: If the minimum is shallow and rounded, we can conclude that $L$ is not large enough; if instead of a minimum we observe a broad maximum near $m=0$, we can conclude that for the chosen value of $L$ the perpendicular linear dimension $L_z$ is too large, and states with more than two domain walls contribute \cite{101,102}. In Fig.~\ref{fig: ProbDistributionsLargeLz}, where we have deliberately chosen a small value of $L$ ($L=6$), one can recognize that already for $L_z=48$, there is a flat local maximum at $\rho=0.5$, rather than a minimum, due to the fact that the sampling is ``contaminated" by states with 4 (rather than only 2) interfaces; for $L_z=96$ and $192$, this effect is so pronounced, that the method based on the analysis of $P_{L,L_z}(\rho=0.5)$ is inapplicable. For $L_z=384$, we have multi-domain states. As will be discussed below, the actual dependence of $P_{L, L_z}(m \approx 0)$ on $L$ and $L_z$ contains the desired information on the interfacial tension \cite{18,39,45,46,49,51,52,53,54,56,57,58,59,62,64}, but only if states with more than two domains make negligible contributions.

\subsection{Translational entropy of the whole interface}

When we consider an Ising chain at low temperatures, the correlation length is very large, $\xi \approx \exp (2J/\kb T)/2$ \cite{103}, and the associated free energy per spin is $F \approx -J-\kb T \exp (- 2 J/\kb T)$. The state of the system can be characterized by a sequence of large domains of parallel spins, with an average size \cite{104} $2 \xi$, separated by ``interfaces'' where the spin orientation changes. Thus, the system can be viewed as a dilute gas of randomly distributed interfaces. The cost of energy to create such an interface is $2J$, and the gain in (translational) entropy is $\kb  \exp (-2 J/\kb T)$.

As is well known, and can be shown explicitly by transfer matrix methods \cite{83}, this picture carries over to two-dimensional Ising strips of width $L$ (with PBC in the direction across the strip), where one finds
\begin{equation}\label{eq5}
\xi_\parallel \propto L^{1/2} \exp (L \gamma^{(d=2)}_\infty)
\end{equation}
with $\gamma^{(d=2)} _\infty$ being the exactly known \cite{105} interface tension of the two-dimensional Ising model, normalized by $\kb T$ (and hence having the dimension of inverse length, the unit of length being the lattice spacing $a$)
\begin{equation} \label{eq6}
\gamma_\infty^{(d=2)} =\frac{2J}{\kb T} - \ln\left( \frac{1+\exp (-2J/\kb T)}{1-\exp(-2J/\kb T)} \right) \; .
\end{equation}
Eq.~\eqref{eq5} coincides with the field-theoretic result Eq.~\eqref{eq3} in the case of $d=2$, as it should be. While the free energy cost of an interface in the Ising chain is $2J$, in the Ising strip it is
\begin{equation} \label{eq7}
F^\text{eff}_\text{int} = \kb T \gamma^{(d=2)}_\infty L + \frac{\kb T}{2} \ln\left( \frac{L}{\const} \right)
\end{equation}
The logarithmic correction in this expression was interpreted by Fisher \cite{106} as a result of an effective repulsive interaction between interfaces due to their capillary wave excitations.

If we again view the Ising strip at low temperatures as a dilute gas of domain walls separating large domains of opposite order parameter, it is natural to ask what the free energy difference $\Delta F$ between a system with one domain wall on a length $L_z$ and a system in a monodomain configuration on the same length scale is. Taking the entropy gain of putting the interface anywhere on this scale $L_z$ into account, we conclude \cite{107,108}
\begin{equation} \label{eq8}
\Delta F =F_\text{int} - \kb T \ln \left( \frac{L_z}{l_\text{int}} \right), \quad F_\text{int}=\kb T \gamma ^{(d=2)}_\infty L
\end{equation}
where we have normalized $L_z$ with some intrinsic length $l_\text{int}$ of the system, such that the ratio $L_z/l_\text{int}$ ``counts'' the number of distinct configurations containing one (coarse-grained) interface on the scale $L_z$. In the one-dimensional Ising chain, where no internal degrees of freedom are associated with the ``kink'' separating a domain of up spins from a domain of down spins, and the kink can appear between any two neighboring lattice sites, the length $l_\text{int}$ simply is the lattice spacing $a(=1)$. However, all the configurational degrees of freedom associated with an interface in higher dimensions are already included in $F_\text{int}$, and must not be included again in the translational entropy term in Eq.~\eqref{eq8}, to avoid double counting; thus we expect that $l_\text{int}$ will be much larger than the lattice spacing, and a plausible assumption is to identify $l_\text{int}$ with the interfacial width $w_L$, as written in Eqs.~\eqref{eq3},~\eqref{eq4}, see also Fig.~\ref{fig: TranslationalEntropy}. From Eq.~\eqref{eq8} we conclude that $\Delta F=0$ for $L_z=L_{z,0}$, with
\begin{multline} \label{eq9}
L_{z,0}=l_\text{int} \exp (F_\text{int}/\kb T)= l_\text{int} \exp (\gamma^{(d=2)}_\infty L) \\
=\exp (F^\text{eff}_\text{int}/\kb T)
\end{multline}
Thus, when we have a single interface in the system, an interpretation of correction terms as being due to repulsive interactions between interfaces lacks plausibility. If we rather use the interpretation used in Fig.~\ref{fig: TranslationalEntropySketch}, that we can work with non-interacting interfaces where an interface needs a space of extent $l_\text{int}=w_L$ in $z$-direction, any such problems are avoided, and Eq.~\eqref{eq3} is interpreted via Eq.~\eqref{eq8} as resulting from the translational entropy of the interface. We also note that Eq.~\eqref{eq8} is valid for is readily generalized to arbitrary dimension, by stating that the translational entropy gain of an interface in a $L^{d-1} \times L_z$ geometry causes a correction term to the interfacial tension $\gamma$ ($\gamma=\Delta F/L^{d-1}$), namely
\begin{equation} \label{eq10}
\Delta \gamma = - \frac{1}{L^{d-1}} \ln \left( \frac{L_z}{w_L} \right) \quad.
\end{equation}
Recall that in the classical limit of quantum systems the length used for counting the states for the translational entropy is the thermal de Broglie wavelength. Here, we deal with purely classical statistical mechanics, hence the use of another physical length of the system, such as $w_L$, is more appealing. In $d=2$, the exact transfer matrix results show that in geometries such as Fig.~\ref{fig: BoundaryConditionsFreeSurfaces} and~\ref{fig: BoundaryConditionsAPBC}, for large $L_z$ and large $L$ the interfacial tension can be written as $\gamma = \gamma_\infty + \Delta \gamma = \gamma_\infty - L^{-1} \ln(L_z/w_L)$, which implies that capillary wave effects are already fully accounted for through $w_L$ in Eq.~\eqref{eq10}.

\subsection{Capillary wave effects continued}
\label{sec: CapillaryWaveEffects}

For the sake of completeness, we briefly recall what is known on the finite-size effects on the interfacial tension due to capillary waves \cite{84,87,88,89,90}. Ignoring the intrinsic interfacial structure, the interface is described by a function $z=h(x)$ in $d=2$ or $z=h(x,y)$ in $d=3$, respectively, that characterizes the dividing surface between the phases with opposite order parameter. Since overhangs are forbidden, a coarse-graining as implied in Fig.~\ref{fig: TranslationalEntropySketch} is anticipated. If one assumes additionally that $|\D h(x)/\D x|$ and $|\nabla h(x,y)|$ are very small, the Hamiltonian describing the capillary wave fluctuations is \cite{100} (again in units of the thermal energy $\kb T$ and ignoring the distinction between interfacial tension $\gamma_\infty$ and interfacial stiffness \cite{100})
\begin{subequations}
\begin{align} 
\mathcal{H}_\text{cw} &=\frac{\gamma_\infty}{2} \int \D x \left|\frac{\D h}{\D x}\right|^2 &&(d=2) \label{eq11a} \\
\mathcal{H}_\text{cw} &= \frac{\gamma_\infty}{2} \int \D x \int \D y \left|\nabla h (x,y)\right|^2 &&(d=3) \;, \label{eq11b}
\end{align}
\end{subequations}
respectively. Note that here the total interface tension $\gamma_\infty$ (that results in the thermodynamic limit) is taken \cite{88,89}, rather than some renormalized quantity. Introducing Fourier transforms $h_q$ of these height variables $h(x)$ or $h(x,y)$, one finds
\begin{equation} \label{eq12}
\mathcal{H}_\text{cw} =\frac{\gamma_\infty}{2}  \frac{1}{(2 \pi)^{d-1}} \int \D^{d-1} q \; q^2 |h_q|^2
\end{equation}
and the resulting contribution to the free energy can be written in terms of path integrals
\begin{multline}\label{eq13}
\Delta F =-\kb T \ln \int D h_q \int D h_q^* \\
\exp \left(-\frac{\gamma_\infty}{2} \frac{1}{(2 \pi)^{d-1}} \int \D^{d-1} q \; q^2 |h_q|^2 \right)
\end{multline}
We now take into account that in a finite geometry with PBC in $x$, (or $x$ and $y$, respectively) directions reciprocal space is discrete, and hence Eq.~\eqref{eq13} becomes (in $d=2$)
\begin{align} 
\Delta F_\text{cw} &=-\kb T \ln \prod_\nu \int\limits_{ - \infty}^{+\infty} d h_\nu \int\limits_{ - \infty}^{+\infty} d h^*_\nu \exp \left(\frac{-\gamma_\infty}{2}\; q^2_\nu h ^*_\nu h_\nu\right) \nonumber \\
&=- \kb T \ln \prod_\nu \left(\frac{2 \pi}{\gamma_\infty q_\nu^2}\right) \label{eq14}
\end{align}
 where $q_\nu = \pm \nu \pi a/L, \, \nu=1, \ldots, N=L/a$. Of course, the term $\nu=0$ (corresponding to a uniform translation of the interface) needs to be omitted here. One can show that for large $L$ the resulting finite-size behavior is ($\Delta \gamma_\text{cw}=\Delta F_\text{cw}/L$)
\begin{equation} \label{eq15}
\Delta \gamma_\text{cw} = A + \frac{B}{L} \ln \left( \frac{L}{a} \right) + \frac{C}{L} \quad,
\end{equation}
where the regular terms in $1/L$, namely $A$ and $C/L$, are dominated by the large $q$ behavior, while the singular logarithmic term is due to small wave numbers and its prefactor $B=1/2$ agrees with transfer matrix results quoted in Eq.~\eqref{eq7}. Since the capillary wave description is no longer reliable at large $q$, however, no conclusion on the value of the leading term (A) and the coefficient $C$ of the regular finite correction $(C/L)$ can be made. The situation is worse in $d=3$, however, where in an analogous calculation no singular term due to long wavelength capillary waves can be identified. Capillary wave corrections are then expected to have the form, to leading order,
\begin{equation} \label{eq16}
\Delta \gamma_\text{cw} =\frac{\const}{L ^{d-1}}
\end{equation}
but the constant is not expected to be universal. We recall, however, that from the equipartition theorem one can conclude from Eq.~\eqref{eq12} that \cite{100}
\begin{equation} \label{eq17}
\langle |h_q |^2 \rangle =(\gamma_\infty q ^2 )^{-1}
\end{equation}
and hence Eq.~\eqref{eq3} readily follows, since (in $d=2$)
\begin{equation} \label{eq18}
w^2_L =\langle h^2 (x) \rangle - \langle h (x) \rangle^2 \propto \frac{a}{\gamma_\infty} \int\limits^{2 \pi /a}_{2 \pi/L} \frac{\D q}{q^2} \propto \frac{aL}{\gamma_\infty} \;,
\end{equation}
while in $d=3$ one finds
\begin{equation} \label{eq19}
w^2_L \propto \frac{a^2}{\gamma_\infty} \int\limits^{2 \pi/a}_{ 2 \pi/L} \frac{\D q}{q} \propto  \frac{a^2}{\gamma_\infty} \ln \left(\frac{L}{a}\right) \quad .
\end{equation}

\subsection{Domain breathing}
\label{sec: DomainBreathing}

\begin{figure}
\centering
	\includegraphics[clip=true, trim=0mm 0mm 70mm 0mm, angle=90,width=0.49 \textwidth]{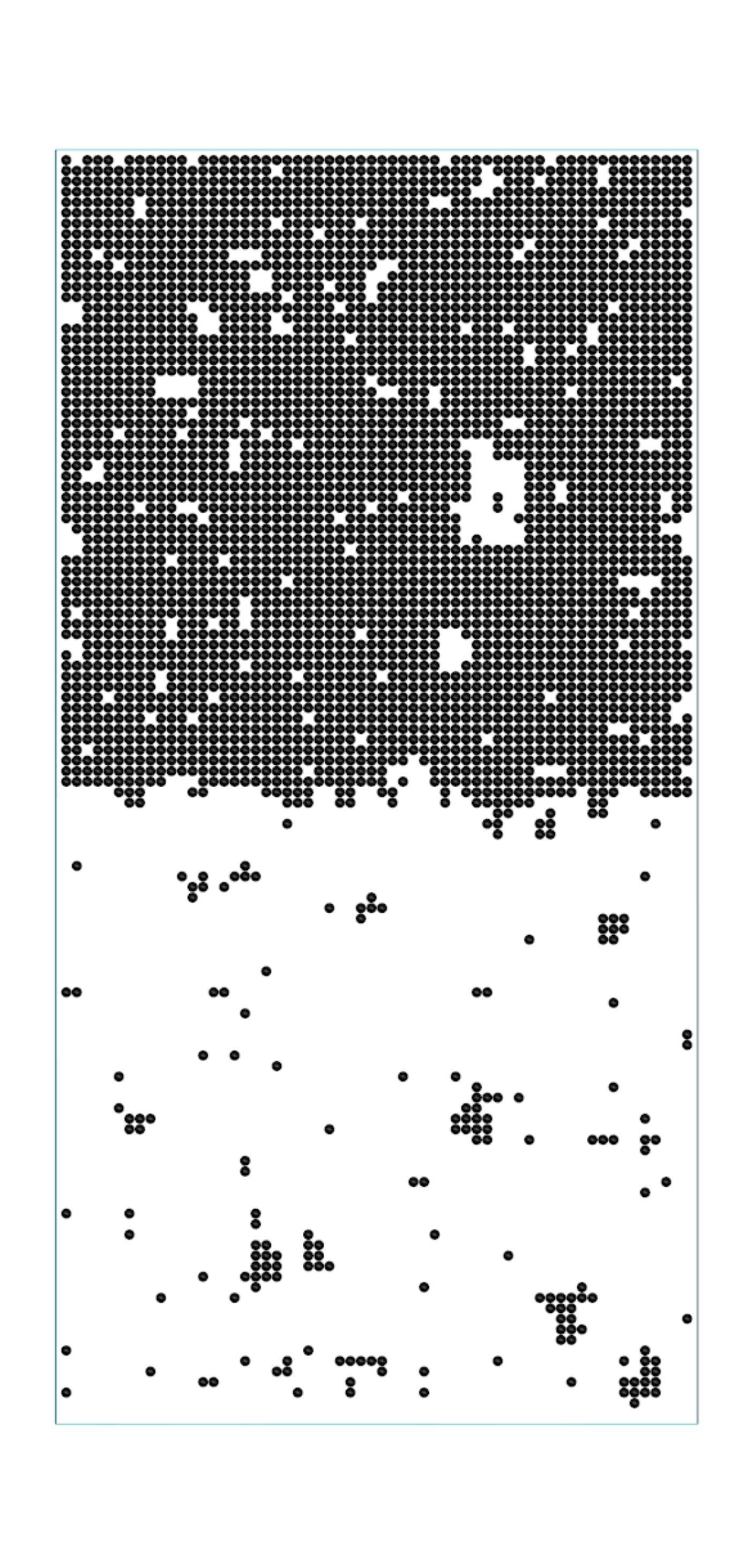}
	\includegraphics[clip=true, trim=0mm 0mm 70mm 0mm, angle=90,width=0.49 \textwidth]{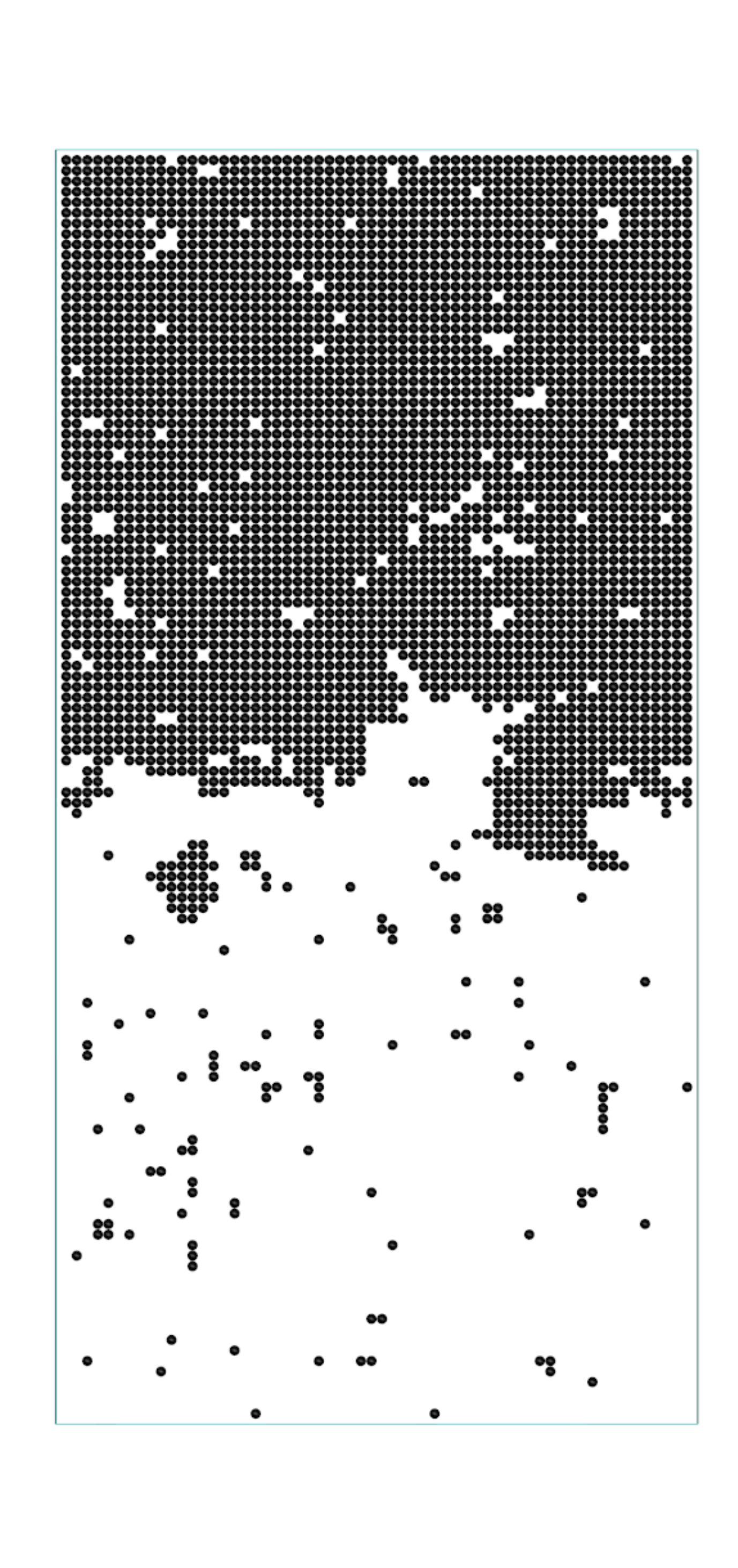}
	\includegraphics[clip=true, trim=0mm 0mm 70mm 0mm, angle=90,width=0.49 \textwidth]{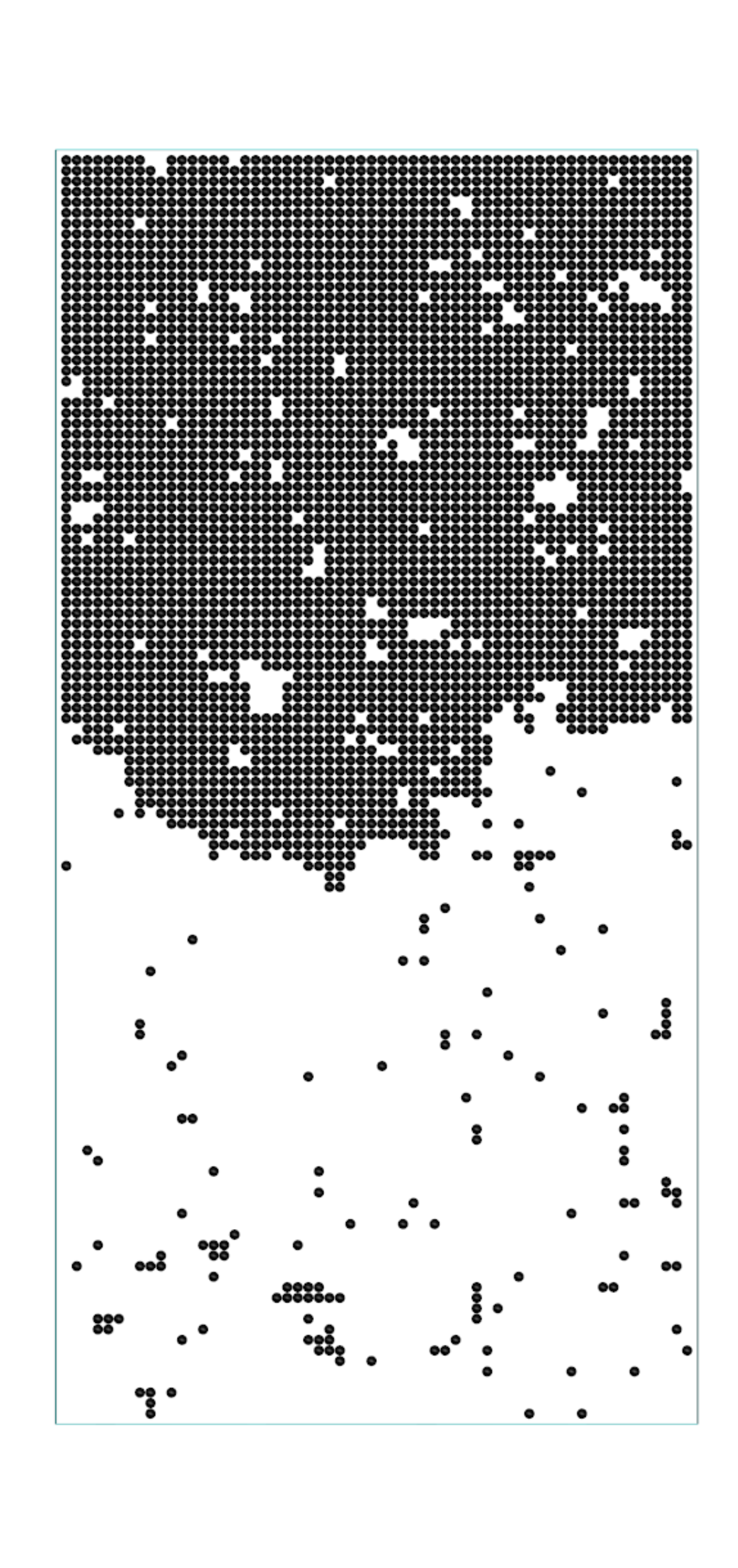}
	\includegraphics[clip=true, trim=0mm 0mm 70mm 0mm, angle=90,width=0.49 \textwidth]{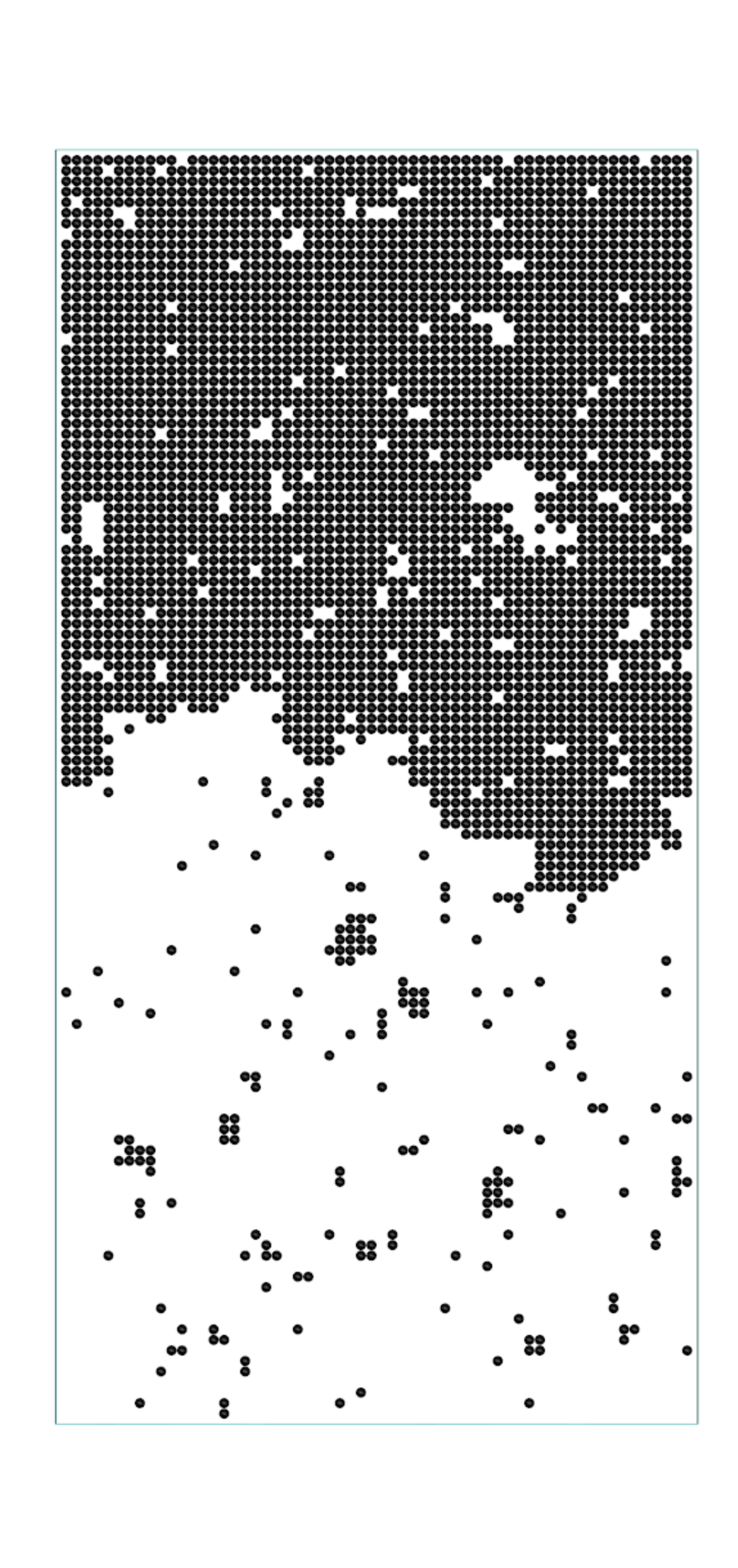}
\caption{\label{fig: DomainBreathing} Four snapshots of two-dimensional Ising systems at fixed magnetization $m=0$ to visualize domain breathing. Spontaneous fluctuations of the magnetization densities inside the two on average equal sized domains $(L_x=L_z/2=60$) have an effect on the location of the interface, when the total magnetization is conserved. All snapshots are for $\kb T/J=2.0$. APBC in $z$-direction (horizontal) and PBC in $x$-direction (vertical) are used.}
 \end{figure}

We first consider a situation with APBC, so we have a single interface, but with conserved magnetization $m=0$. Then on average we have two equally large domains, with linear dimensions $L_z/2$ in $z$-direction each, of opposite magnetization. However, the magnetization densities $m_+$, $m_-$ in both domains still can fluctuate and also the position of the interface is not fixed but can fluctuate somewhat as well. We denote this shift of the interface due to a fluctuation by $\Delta$, and note the constraint that the total magnetization in the system is strictly fixed at $m=0$, to find
\begin{equation} \label{eq20}
\begin{split}
0 &= m L^{d-1} L_z \\
&= m_+ L ^{d-1} \left(\frac{L_z}{2} - \Delta \right) + m_-L^{d-1} \left(\frac{L_z}{2} +\Delta \right)
\end{split}
\end{equation}
and hence
\begin{equation} \label{eq21}
\Delta= \frac{L_z}{2} \left(\frac{m_+ + m_-}{m_+ -m_-}\right) \approx \frac{\delta m_+ + \delta m_-}{2 m_\text{coex}}
\; \frac{L_z}{2}
\end{equation}
where we used that the fluctuations $\delta m_+ = m_+ - m_\text{coex}$, $\delta m_-=m_-+ m_\text{coex}$ are very small. From general statistical thermodynamics we know that these fluctuations of the magnetization density in the bulk obey Gaussian distributions \cite{107}
\begin{equation} \label{eq22}
P_{L, L_z/2} (\delta m) \propto \exp \left[-\frac{1}{2} \frac{(\delta m)^2 L_z L ^{d-1}}{2 \kb T \chi_\text{coex}} \right]\quad,
\end{equation}
where $\chi_\text{coex}$ is the susceptibility at the coexistence curve. Eq.~\eqref{eq22} is true both for $\delta m_+$ and $\delta m_-$, and these fluctuations in the two subvolumes of the system can occur independently of each other, so $\langle \delta m_+ \delta m_- \rangle =0$, while $ \langle \delta m^2_+ \rangle = \langle \delta m^2_- \rangle = \kb T\chi_\text{coex}/(L^{d-1} L_z/2)$. Hence we conclude from Eq.~\eqref{eq21} that
\begin{equation} \label{eq23}
\begin{split}
\left\langle \Delta ^2 \right\rangle &= \frac{L_z^2}{16 m^2_\text{coex}} \left[\left\langle\delta m_+^2 \right\rangle + \left\langle\delta m_-^2 \right\rangle \right] \\
&= \frac{\kb  T \chi_\text{coex}}{4 m^2_\text{coex}} \frac{L_z}{L^{d-1}} \quad.
\end{split}
\end{equation}
Thus, the typical length over which the interface position fluctuates is
\begin{equation} \label{eq24}
\sqrt{\langle \Delta^2 \rangle } =L_z ^{1/2} L^{-(d-1)/2} \frac{\sqrt{\kb T \chi_\text{coex}}}{2m_\text{coex}}
\end{equation}
From this motion of the interface over a width $\sqrt{\langle \Delta ^2 \rangle}$, which we call ``domain breathing'', we again get an entropy contribution, resulting in a correction of the interfacial tension
\begin{equation} \label{eq25}
\begin{split}
\Delta \gamma_\text{db} &= - \frac{1}{L ^{d-1}} \ln \left( \frac{\sqrt{\langle \Delta^2 \rangle}}{w_L}\right) \\
&= - \frac12 \frac{\ln L_z}{L ^{d-1}} + \frac{d-1}{2} \frac{\ln L}{L^{d-1}} + \frac{3-d}{2} \frac{\ln L}{L^{d-1}}+\frac{\const}{L^{d-1}} \;.
\end{split}
\end{equation}
To simplify the notation, we assume here (and in the following) that the lengths $L, L_z$ are measured in some natural units (e.g.~the lattice spacing $a$, in case of the Ising model) and hence dimensionless. Note that there is some ambiguity of interpretation possible. In our previous publication~\cite{95}, the length to normalize $\sqrt{\langle \Delta^2\rangle}$ was taken as the lattice spacing $a$, and then the capillary wave contribution $(3-d) \ln L/(2L^{d-1})$ must be added as an explicit further correction. However, when we use $w_L$ (as computed in Eq.~\eqref{eq4} or~\eqref{eq18} and~\eqref{eq19}, respectively) rather than $a$ to normalize $\sqrt{\langle \Delta^2\rangle}$, then the capillary wave effects are already fully taken care of. Fig.~\ref{fig: DomainBreathing} illustrates the occurrence of this ``domain breathing'' effect by configuration snapshots.

A special situation occurs in the case of the canonical ensemble for PBC. This is a very common situation, since then no symmetry between the coexisting phases is required, and the system exhibits translational invariance, the domains separated by the two walls can be translated along the $z$-axis as a whole. For this degree of freedom, a correction $-\ln L_z/L^{d-1}$ to the interfacial tension arises. In addition, the distance between the domain walls can fluctuate, according to the domain breathing effect, as described above, yielding an additional entropic term $-\frac{1}{2} \ln L_z/L^{d-1}$. Since there are two interfaces present in the system, the total correction $-(3/2) \ln L_z/L^{d-1}$ yields a correction of $-(3/4) \ln L_z/L^{d-1}$ per interface.

We also note that it is not necessary to fix the magnetization exactly at $m=0$ (or, in the case of a fluid that possibly lacks any symmetry between the coexisting liquid $(l)$ and the vapor $(v)$ phases, at a density $\rho=(\rho_l + \rho_v)/2)$. Rather it suffices to choose a state point where in the simulation box we have a clear slab configuration of phase coexistence. Also, in a system lacking symmetries between the coexisting phases, the distributions around $m_+$, $m_-$ are characterized by different ``susceptibilities'' $\chi^+_\text{coex}$, $\chi^-_\text{coex}$, but for the exponents of $L_z$ and $L$ in Eq.~\eqref{eq24}, this does not matter.

At this point, let us summarize the various logarithmic corrections found for the different choices of boundary conditions and ensembles: for the APBC(gc) case, we have a single interface that can freely translate (Fig.~\ref{fig: TranslationalEntropy}, Eq.~\eqref{eq10}). This yielded
	\begin{equation*}
	\Delta \gamma_{L,L_z}= - \frac{\ln L_z}{L ^{d-1}} + \frac{3-d}{2} \frac{\ln L}{L^{d-1}} + \frac{\const}{L^{d-1}} \;.
	\end{equation*}
Due to the lack of conservation laws, there is no coupling of the bulk domain fluctuations and interfacial fluctuations via the domain breathing effect in this case, unlike the APBC(c) case, for which Eq.~\eqref{eq25} implies
	\begin{equation*}
	\Delta \gamma_{L,L_z} = - \frac{1}{2} \frac{\ln L_z}{L ^{d-1}} + \frac{\ln L}{L^{d-1}} + \frac{\const}{L^{d-1}} \;.
	\end{equation*}
In the PBC(c), we have two interfaces, and we have both the above translational entropy contribution (when we translate the domains as a whole) and the domain breathing effect (considering the relative motion of the two domain walls against each other), and normalized per single interface this yields
	\begin{equation*}
	\Delta \gamma_{L,L_z} = - \frac{3}{4} \frac{\ln L_z}{L ^{d-1}} + \frac{5-d}{4}\frac{\ln L}{L^{d-1}} + \frac{\const}{L^{d-1}} \;.
	\end{equation*}
Note that by normalizing domain wall motions consistently by $w_L$ rather than by $a$, capillary wave effects are automatically included.

Taking all logarithmic finite-size corrections (due to translational entropy, domain breathing, and capillary waves) together, it makes sense to write the result for the interfacial tension in the following general form
\begin{equation} \label{eq26}
\gamma_{L,L_z} =\gamma_\infty - x_\perp \frac{\ln L_z}{L^{d-1}} + x_\parallel \frac{\ln L}{L^{d-1}} + \frac{C}{L^{d-1}}
\end{equation}
with some constant $C$ and two universal exponents $x_\perp$, $x_\parallel$ that depend on dimensionality $d$, type of boundary conditions (PBC, APBC) and statistical ensemble (grand canonical versus canonical). We present these constants $x_\perp$, $x_\parallel$ in Table~\ref{tab: ScalingConstants}.

	\begin{table}[ht]
	\begin{ruledtabular}
	\begin{tabular}{ccccc}
	$d$ & BC & ensemble & $x_\perp$ & $x_\|$ \\
	\hline
	2 & antiperiodic & grandcanonical & $1$ & $1/2$ \\
	3 & antiperiodic & grandcanonical & $1$ & $0$ \\
	2 & antiperiodic & canonical & $1/2$ & $1$ \\
	3 & antiperiodic & canonical & $1/2$ & $1$ \\
	2 & periodic & canonical & $3/4$ & $3/4$ \\
	3 & periodic & canonical & $3/4$ & $1/2$ \\
	\end{tabular}
	\end{ruledtabular}
	\caption{The universal constants $x_\perp$ and $x_\parallel$ in Eq.~\eqref{eq26} do not depend on details of the model such as particle interactions, but they rather depend on the dimensionality $d$, the boundary conditions (periodic or antiperiodic) and the ensemble (canonical or grandcanonical).}
	\label{tab: ScalingConstants}
	\end{table}

\section{Models and simulation methods}
\label{sec: ModelsAndSimulationMethods}

As stated already in Sec.~\ref{sec: PhenomenologicalTheory}, the main emphasis of this study is on the Ising model \{Eq.~\eqref{eq: IsingHamiltonian}\}, since (i) there is no source of inaccuracy due to insufficient knowledge of the conditions for which phase coexistence in the bulk occurs, symmetry requires phase coexistence to occur for $H=0$, and (ii) in the case $d=2$ the surface tension is known exactly, Eq.~\eqref{eq6}, and so the concepts described in Sec.~\ref{sec: PhenomenologicalTheory}, in particular Eq.~\eqref{eq26}, can be very stringently tested. In $d=2$, we have typically used $L=10, 20, 30$ and $40$, varying $L_z$ from $L_z=20$ to $L_z=200$ in order to test the $L_z$-dependence at fixed $L$ (Eq.~\eqref{eq10}). In addition, at fixed $L_z=60$ and $120$ runs were made varying $L$ from $L=10$ to $L=L_z$ to test the $L$-dependence in Eq.~\eqref{eq25}). In $d=3$, we have used $L=6,8,10,12,$ and 14 and varying $L_z$ from $L_z=20$ to $L_z=100$ for the test of Eq.~\eqref{eq10}, as well as using $L_z=20,40$ and 80 varying $L$ from $L=10$ to $L=40$ for the test of Eq.~\eqref{eq25}. Using the grandcanonical ensemble, all runs were performed simply using the standard single-spin flip Metropolis algorithm \cite{81}. Since the simulations are performed far below the critical point ($\kb T/J=1.2, 1.6$ and $2.0$ in $d=2$; $\kb T/J=3$ in $d=3$), the use of cluster algorithms \cite{81} would not provide any advantage. The canonical ensemble is realized via a spin exchange algorithm; choosing two spins at random from the whole simulation box, rather than choosing a pair of spins which are nearest neighbors, as in the standard spin exchange algorithm \cite{81}, we avoid slow relaxation of long wavelength magnetization fluctuations.


Special techniques are required when one wishes to sample the probability distribution $P_{L,L_z} (\rho)$, Fig.~\ref{fig: ProbDistributions}, since it varies over many orders of magnitude. Straightforward use of the Metropolis algorithm (as originally attempted \cite{39}) would not give any useful data for our purposes. While previous work \cite{45,46,49} relied on the multicanonical Monte Carlo method, we found it here more convenient to use successive umbrella sampling \cite{109} which is more straightforward to implement. We recall that from $P_{L,L_z}(\rho)$ one can extract an estimate for the interfacial tension $\gamma_{L, L_z}$ as follows \cite{39}
\begin{equation} \label{eq27}
\gamma_{L,L_z}= \frac{1}{2 L^{d-1}} \ln \left(\frac{P_{L, L_z} (\rho_\text{coex})}{P_{L, L_z} (\rho_\text{min})}\right)\quad.
\end{equation}
Here we use a notation which applies both to the lattice gas (where the density $\rho_\text{min}$ where the minimum of $P_{L, L_z}(\rho)$ occurs corresponds to a magnetization $m=0$ in the magnetic interpretation of the Ising model) and to fluids which may lack particular symmetries (then the minimum occurs for the density of the ``rectilinear diameter'', $\rho_\text{min}=\rho_d=(\rho_v + \rho_l )/2$, $\rho_v$ and $\rho_l$ being the densities of the coexisting vapor and liquid phases). The physical interpretation of Eq.~\eqref{eq27} simply is that the probability to observe a state at $\rho_\text{min}$, in comparison to the probability to observe one of the pure phases at coexistence ($\rho_v$ or $\rho_l$, respectively) is down by a factor $\exp(-2 L^{d-1} \gamma_{L, L_z})$, due to the fact that we must have 2 interfaces of area $L^{d-1}$ (Fig.~\ref{fig: BoundaryConditionsPBC}). Note that although $P_{L,L_z} (\rho)$ is generated by carrying out a sampling (multicanonical or umbrella sampling) in the grandcanonical ensemble (at magnetic field $H=0$ or chemical potential $\mu=\mu_\text{coex}$, respectively), by taking out the probability strictly at $\rho=\rho_\text{min}$ the extracted interfacial tension $\gamma_{L,L_z}$ in Eq.~\eqref{eq27} corresponds to observations sampled in a canonical ensemble.

As a second model, representative for off-lattice fluids, we study the Lennard-Jones model in $d=3$ dimensions, where point particles interact with a potential $U_{LJ} (r)$, $r$ being the distance between the particles,
\begin{equation} \label{eq28}
U_{LJ} (r) = 4 \varepsilon\left[\left(\frac{\sigma}{r}\right)^{12} - \left(\frac{\sigma}{r}\right)^6 + Y \right], \quad r< r_c \;,
\end{equation}
while $U_{LJ}(r > r_c)\equiv 0$. Here $\varepsilon$ is the strength and $\sigma$ the range of this potential, and the constant $Y$ is chosen such that $U_{LJ} (r)$ is continuous at the cutoff $r_c=2^{1/6} \cdot 2\sigma$. For this model, we choose units such that $\varepsilon=1$ and $\sigma=1$. A single temperature $T=0.78 T_c$ is used, for which $\gamma_\infty = 0.375(1)$ was already estimated in previous work \cite{110}, using Eq.~\eqref{eq27}.

In order to be able to study also other choices of boundary conditions, as shown in Fig.~\ref{fig: BoundaryConditionsFreeSurfaces} and~\ref{fig: BoundaryConditionsAPBC}, we have developed a new variant of the ensemble switch method \cite{91,92,93}. In this previous work \cite{91,92,93}, a ``mixed'' system was created from a system confined between two parallel walls and a system with no walls, to extract the excess free energy due to the walls. In the present work, we extend this method by creating a mixed system from two systems at coexistence without interfaces and a system formed from these separate systems but now having interfaces (Fig.~\ref{fig: EnsembleSwitchMethodSketch}). The two separate systems have linear dimension $L_z/2$ in $z$-direction each, and are chosen such that one of them is in the state corresponding to $+ m_\text{coex}$, the other in the state corresponding to $-m_\text{coex}$. Both systems have periodic boundary conditions individually, and hence for this state (denoted as $\kappa =0$) there are no interfaces present. The system denoted as $\kappa=1$ has exactly the same degrees of freedom as the system denoted as $\kappa =0$, namely the $N=L^{d-1}L_z$ Ising spins which may take values $S_i=\pm 1$, and we work at the same thermodynamic conditions (e.g. total magnetization fixed at $m=0$ in the canonical ensemble, and same temperature $T$). The systems denoted as $\kappa=0$ and $\kappa=1$ differ only with respect to their boundary conditions: in both halves of the system $\kappa =0$ we have PBC over a distance $L_z/2$ already, while in the system $\kappa=1$ the two halves are joined, and a single PBC over the distance $L_z$ remains (in the $z$-direction). So the difference in free energies between both systems is related to the interface tension,
\begin{equation} \label{eq29}
\gamma_{L,L_z} = \frac{F(\kappa=1) - F (\kappa=0)}{2 L^{d-1} \kb T} \quad .
\end{equation}
In order to find this free energy difference, it is useful to define a mixed system by
\begin{equation} \label{eq30}
\mathcal{H}(\kappa)= \kappa \mathcal{H}_1 + (1 - \kappa) \mathcal{H}_0 \, \quad 0 \leq \kappa \leq 1 \quad,
\end{equation}
which is a perfectly permissible Hamiltonian for a Monte Carlo simulation (although clearly such a system can never be created by an experimentalist in his laboratory).

The free energy $F(\kappa)$ of the mixed system is defined by the standard relation from the Hamiltonian,
\begin{equation} \label{eq31}
F(\kappa)=-\kb T \ln \left( \Tr \left\{\exp[-\mathcal{H}(\kappa)/\kb T]\right\}\right),
\end{equation}
but it is clear that for large $L$ the normalized free energy difference $[F(\kappa=1) - F(\kappa=0)]/\kb T$ can be huge, since we expect $\gamma_{L, L_z}$ to be of order unity. Such large free energy differences can be computed with sufficient accuracy by thermodynamic integration. In practice, the interval $0 \leq \kappa \leq 1$ is divided into $n_\kappa$ subintervals, separated by discrete values $\kappa_i$. In this work, we use $n_\kappa=1024$. Then the free energy difference $\Delta F_i=F(\kappa_{i+1} ) - F(\kappa_i)$ is obtained from a parallelized version of successive umbrella sampling, considering Monte Carlo moves $\kappa_i \rightarrow \kappa_{i+1}$ or vice versa, in addition to the sampling of the spin configuration. On each core, the system can switch between two adjacent values $\kappa_i$, $\kappa_{i+1}$ only, so one needs to use $n_\kappa$ cores. The desired free energy difference $\Delta F_i$ is simply determined by estimating the probabilities that the states with $\kappa_i$ or $\kappa_{i+1}$ are observed, $\Delta F_i=\kb T \ln [P(\kappa_i)/P(\kappa_{i+1})]$.

\begin{figure}
\centering
\includegraphics[clip=true, trim=0mm 0mm 0mm 0mm, angle=-90,width=0.49 \textwidth]{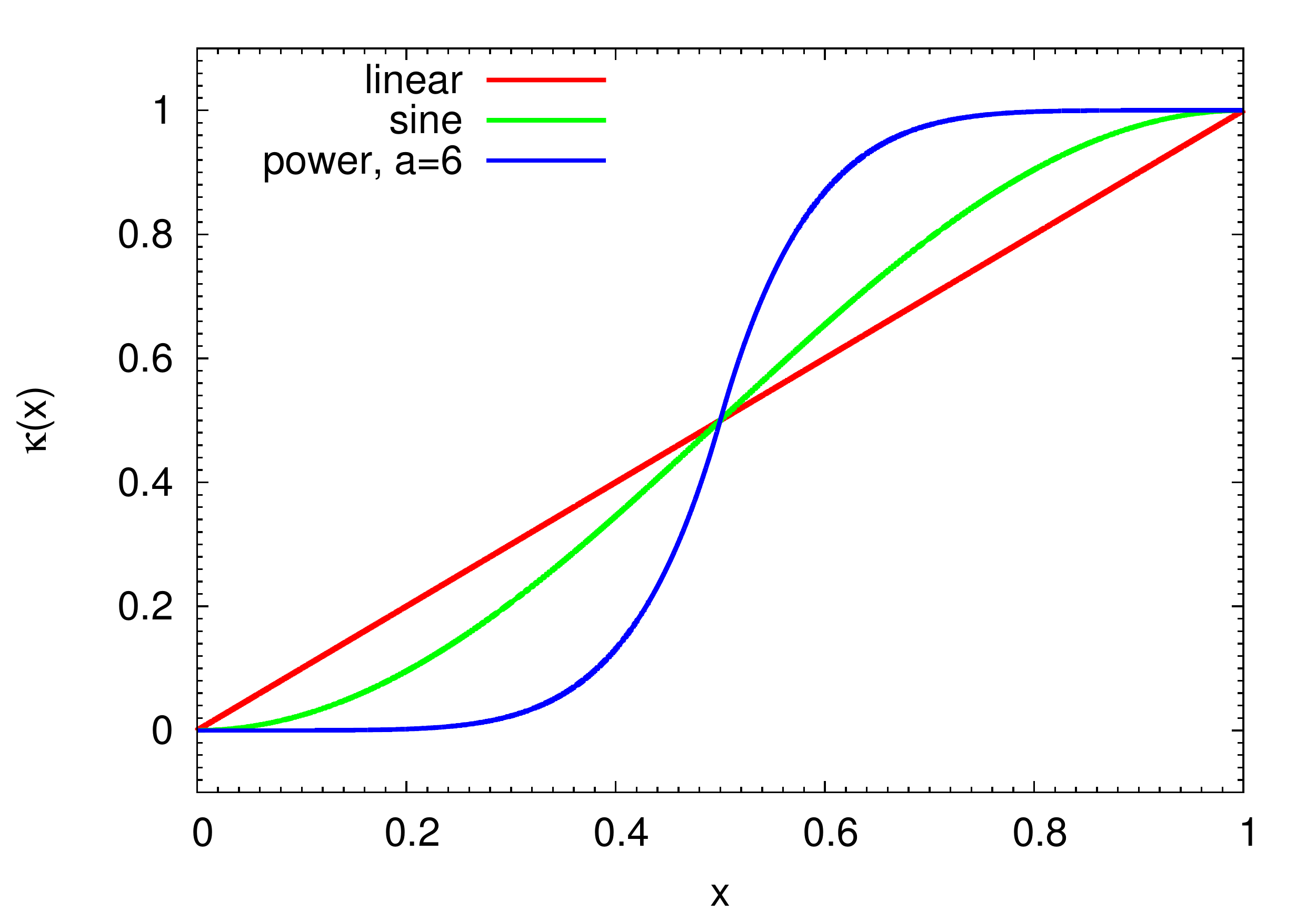}
\caption{\label{fig: SketchKappaFunctions} Various choices for choosing the set of points $\{\kappa_i\}$. Instead of an equidistant set of points, it is useful to increase the density of points near $\kappa=0$ and $\kappa=1$ by choosing for example $\kappa_i(x) = \sin^2(\pi x/2)$ [labeled sine] or $\kappa_i(x)=(2x)^a/2$ for $x<0.5$ and $\kappa_i(x)=1- (2(1-x))^a/2$ for $x\geq 0.5$ [labeled power], where $x=i/n_\kappa$, $n_\kappa$ being the number of points one intends to use.}
  \end{figure}

An important technical aspect is that the set of points $\{\kappa_i\}$ need not be chosen equidistantly in the interval from zero to unity, but the location of these points can be chosen in a way which optimizes the accuracy of the thermodynamic integration. For the Ising model we have found it useful to choose $\kappa_i=\sin^2 (\pi i / (2n_\kappa))$. Note that this function yields more points $\kappa_i$ near $\kappa=0$ and $\kappa=1$, and this clearly is useful since the states for intermediate values of $\kappa$ only are needed for the thermodynamic integration, but have no direct physical significance. Figure~\ref{fig: SketchKappaFunctions} shows various choices for the mapping $i\to\kappa_i$.

\begin{figure}
\centering
\subfigure[\label{fig: betaFvsKappaNearKappa1_Ising2d}]{\includegraphics[clip=true, trim=0mm 0mm 0mm 0mm, angle=-90,width=0.49 \textwidth]{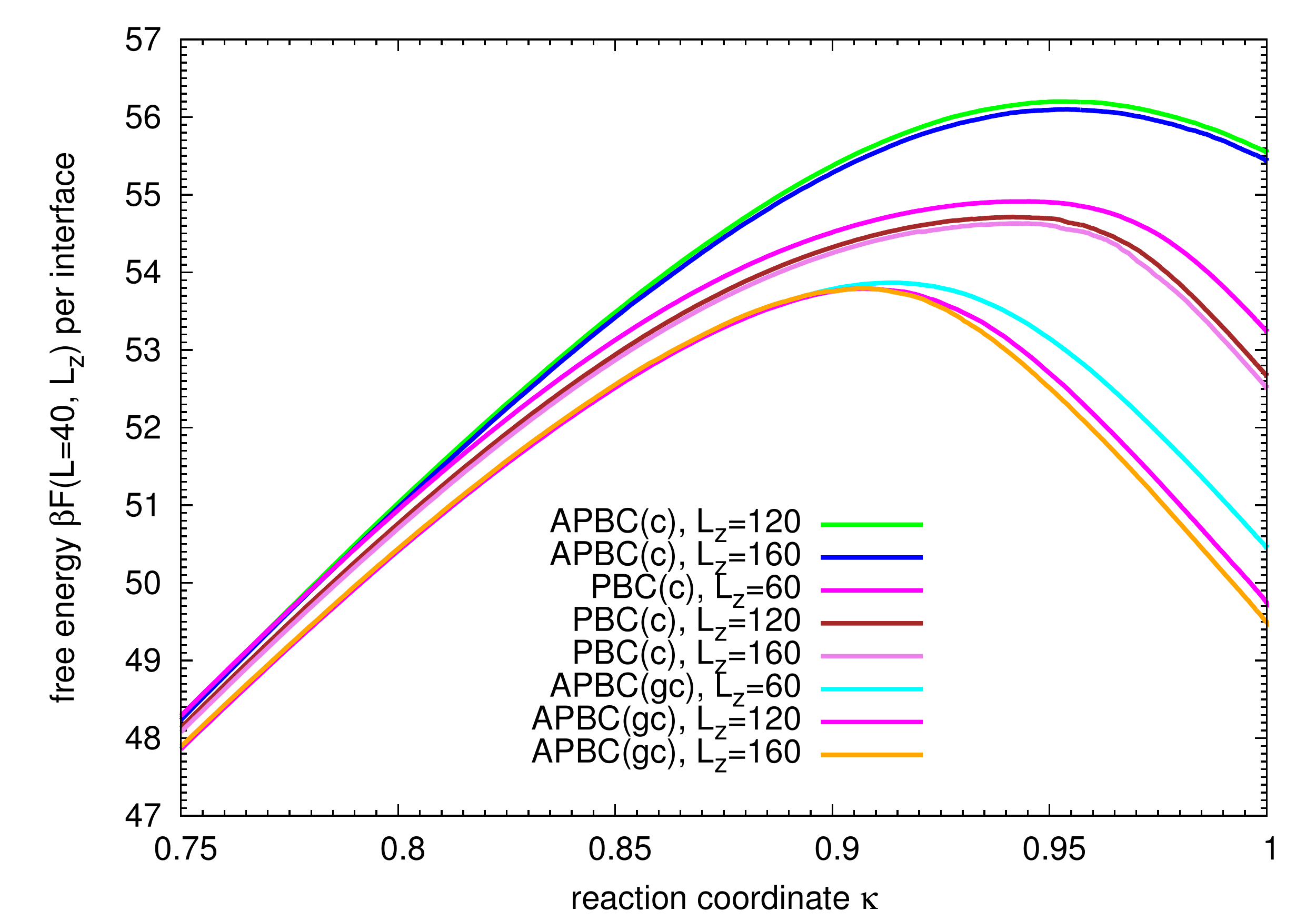}}
\subfigure[\label{fig: betaFvsKappaNearKappa1_Ising3d}]{\includegraphics[clip=true, trim=0mm 0mm 0mm 0mm, angle=-90,width=0.49 \textwidth]{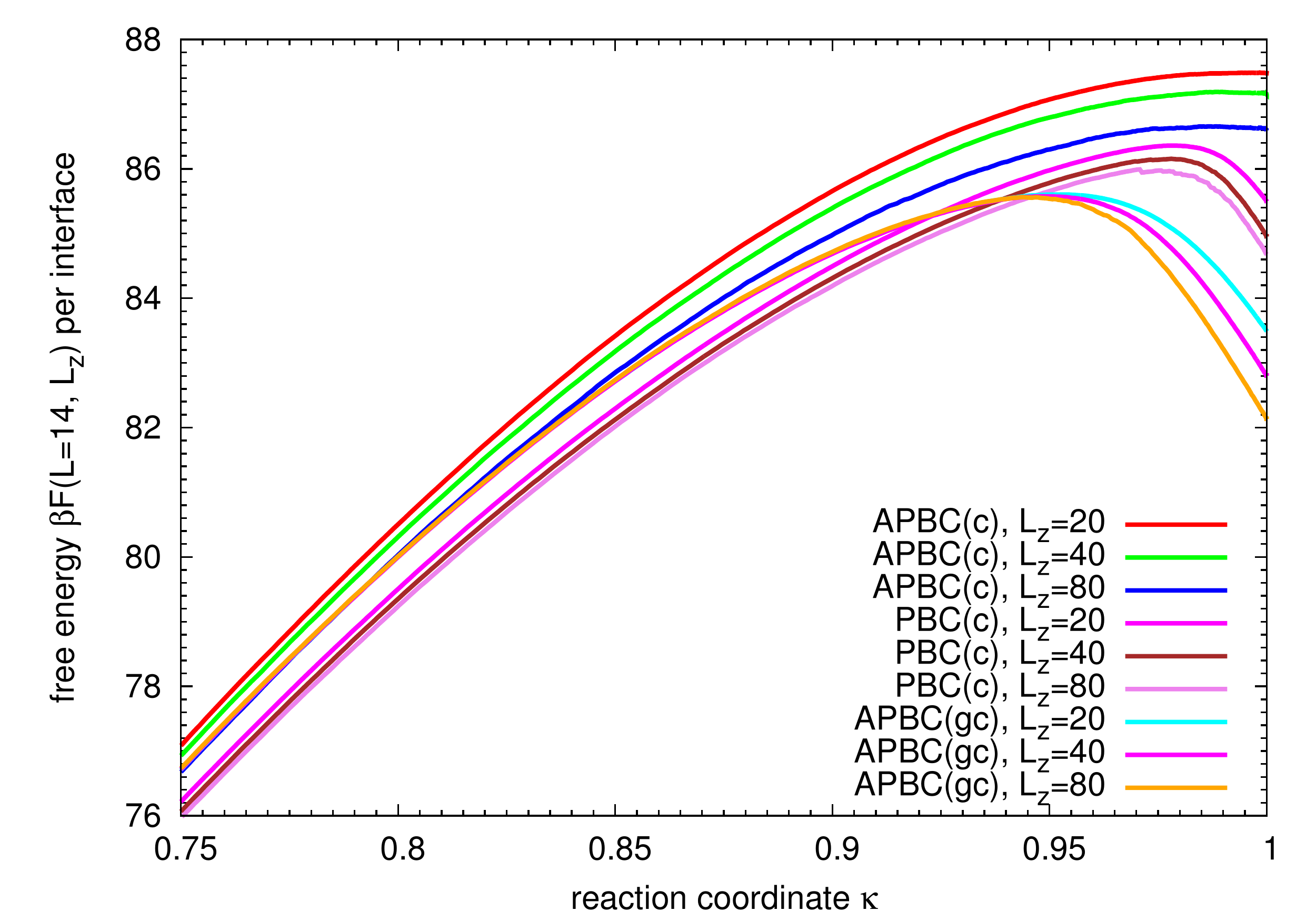}}
\caption{\label{fig: betaFvsKappaNearKappa1} Free energy difference $\Delta F ({\kappa})$ versus $\kappa$, as obtained from the ensemble switch method for $L \times L_z$ Ising systems. (a) shows data for $d=2$ and $L=40$, (b) shows data for $d=3$ and $L=14$. Different choices of boundary conditions (PBC, APBC) and ensembles (grandcanonical (gc), canonical (c) are compared for several $L_z$. Note that only a small section of the whole variation of $\kappa$ and of $\Delta F (\kappa)$ is shown, to display the finite-size effects on $\Delta F(\kappa)$ clearly.}
 \end{figure}

A typical example of the free energy function $\Delta F(\kappa)$ is given in Fig.~\ref{fig: betaFvsKappaNearKappa1}, comparing for the $d=2$ and $3$ Ising model three cases, namely APBC in the canonical and grandcanonical ensemble, as well as the PBC case (canonical ensembles). One sees that in general, the variation with $\kappa$ is slightly non-monotonic. However, since the height of this maximum of $\Delta F(\kappa)$ exceeds the final result ($\Delta F (\kappa=1))$ only by at most a few $\kb T $ (which is the unit of the ordinate scale), we do not think that entropic barriers for intermediate values of $\kappa$ provide a problem here. Of course, this aspect needs to be carefully checked for other models.

We have verified for the Ising model that this method, with the choice of PBC as indicated in Fig.~\ref{fig: EnsembleSwitchMethodSketch} yields results that are completely equivalent to the standard method of Eq.~\eqref{eq27}, as expected. But the advantage of the ensemble switch method (Fig.~\ref{fig: EnsembleSwitchMethodSketch}) is that it is not restricted to simple Ising systems, but can be applied to cases such as liquid-solid interfaces, for which an approach such as Eq.~\eqref{eq27} is difficult to apply: In fact one cannot easily construct convenient reversible paths connecting the two pure phases (liquid and crystal in this case) in a simulation of a single system, where just the volume fraction of the crystal is continuously increased, unlike the case of the Ising model, where starting out at $m=-m_\text{coex}$ the volume fraction of the state with $m= +m_\text{coex}$ is gradually increased and hence $P_{L, L_z}(m)$ is sampled (Fig.~\ref{fig: ProbDistributions}). At this point, we mention that also in the Ising model entropic barriers associated with the droplet evaporation-condensation transition and the transition from circular droplets (in $d=2$) to slabs, in principle, are also a problem when one aims at very high accuracy \cite{111}, but for the data in the present paper this problem was not yet important; nevertheless it is useful to have an alternative method. Moreover, the ensemble switch method can also straightforwardly be applied when we use APBC in the $z$-direction: then the state with $\kappa=1$ has a single interface rather than two interfaces. In the APBC case, both canonical and grandcanonical ensembles can be implemented. Of course, the limiting behavior for $L \rightarrow \infty$ and $L_z \rightarrow \infty$ always must yield the same interfacial tensions, but since the nature of the finite-size corrections differ, it is useful to carry out simulations in different ensembles and or different choices of boundary conditions, and verify that in practice one indeed converges to the same result. This will be the strategy that we will follow in the next section.

For the computations presented in this paper, the total computing effort was of the order of 40~million single core hours of the Interlagos Opteron 6272 processor at the high-performance computer Mogon of the University of Mainz.

We emphasize that additional methods to estimate interfacial tensions from simulations, of course, exist. E.g.~for off-lattice fluids a popular approach is based on the anisotropy of the pressure tensor $p_{\alpha \beta} (z)$ ($ \alpha,\beta=x,y,z)$ across an interface \cite{16,41},
\begin{equation} \label{eq32}
\gamma_{L,L_z} =\frac{1}{2} \int\limits^{L_z/2}_{-L_z/2} \D z \left[p_{zz} (z) - \frac{p_{xx} (z) + p_{yy} (z)}{2} \right]
\end{equation}
where we have assumed a system with linear dimension $L_z$ and PBC in all directions, so that two interfaces contribute. Such simulations normally are done in the canonical ensemble, and we expect that the finite-size effects are of the same character as for the method based on Eq.~\eqref{eq27}. For temperatures close to the critical temperature, Eq.~\eqref{eq32} is computationally inconvenient, since the integrand is very small, and very accurate sampling is required. We expect that Eq.~\eqref{eq32} has an advantage at rather low temperatures, where the grandcanonical sampling of $P_{L, L_z} (\rho)$ becomes less efficient. Note, however, that for computing the pressure tensor $p_{\alpha \beta}(z)$ from the virial theorem one should avoid the sharp cutoff of the potential, as done in Eq.~\eqref{eq28}, and apply a smoothened cutoff
to avoid jumps of the force at $r=r_c$.

A difficult issue are finite-size effects associated with the use of Eq.~\eqref{eq4} or Eq.~\eqref{eq17}, respectively: one either observes the dependence of $\langle|h_q|^2\rangle$ on $q^2$ (Eq.~\eqref{eq17}) or of $w^2_L$ on $\ln L$ \{Eq.~\eqref{eq4}\} and estimates $\gamma_\infty$ from fitting the prefactor. Finite-size effects make the set of possible wave numbers $q$ discrete, of course: in addition one must note that Eq.~\eqref{eq17} is believed to hold in the long wave length limit only, while at shorter wave lengths (corresponding to large $q$) systematic deviations are expected (sometimes a wave vector-dependent interfacial tension $\gamma(q)$ is discussed \cite{26,38}). However, this problem is out of focus here.

\section{Numerical Results for Finite-Size Effects on Interfacial Tensions}
\label{sec: NumericalResults}

\subsection{Two-Dimensional Ising Model}
\label{sec: TwoDimensionalIsingModel}

\begin{figure}
\centering
\includegraphics[clip=true, trim=0mm 0mm 0mm 0mm, angle=-90,width=0.49 \textwidth]{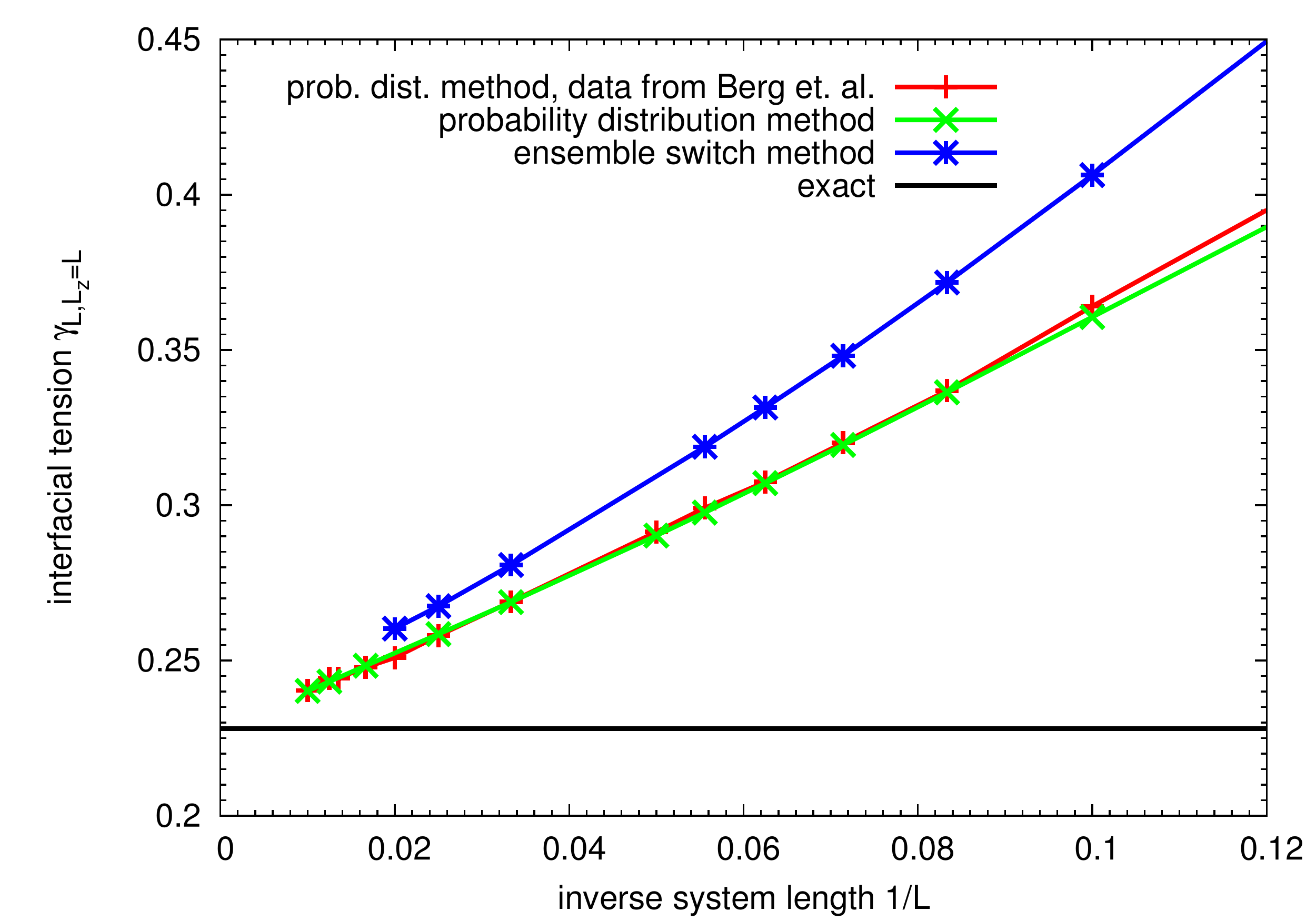}
\caption{\label{fig: Ising2d_Scaling_Ratio1} Estimates for the interfacial tension $\gamma_{L,L}$ of the $d=2$ Ising model at $\kb T/J=2.0$ plotted vs. $1/L$. Two sets of data from the probability distribution method are included (one from Ref. \cite{45}, one from the present work), which agree within statistical errors with each other (the statistical error is smaller than the size of the symbols throughout). The third set of data is due to the ensemble switch method, and has in this case slightly larger finite-size effects than the method based on Eq.~\eqref{eq27}.}
 \end{figure}

As a starting point of the discussion, we use data for $L \times L$ systems with PBC obtained with the help of Eq.~\eqref{eq27}, including both the previous results by Berg et al.~\cite{45}, and results taken by us including also additional choices for $L$, and compare them to the results from the ensemble switch method for the PBC case. The traditional use of such data is to plot the estimates for $\gamma_{L}$ linearly versus $1/L$ and try an extrapolation towards $1/L \rightarrow 0$ (Fig.~\ref{fig: Ising2d_Scaling_Ratio1}). Indeed such an extrapolation seems to be compatible with the exact result (from Eq.~\eqref{eq6} \cite{105}), highlighted by a horizontal straight line, but one can also clearly recognize the problems of the approach: (i) even for relatively large $L$, such as $L=50$, the relative deviation is still of the order of 10\%. (ii) Over the whole range of $1/L$, there is a distinct curvature of the data visible, indicating that it is unclear whether or not the asymptotic regime of the extrapolation has actually been reached. In cases of real interest, of course, the exact answer is not known beforehand, and it is  also very difficult (and may need orders of magnitude more computational resources) to obtain data of the same statistical quality as shown in Fig.~\ref{fig: Ising2d_Scaling_Ratio1}. Thus, in general it will be very helpful to understand the origin of the finite-size effects, and - if possible - to combine different variants of the method where the finite-size effects differ, but the resulting estimate for $\gamma_\infty$ must be the same.

\begin{figure}
\centering
\subfigure[\label{fig: Ising2d_ScalingZ_All}]{\includegraphics[clip=true, trim=0mm 0mm 0mm 0mm, angle=-90,width=0.49 \textwidth]{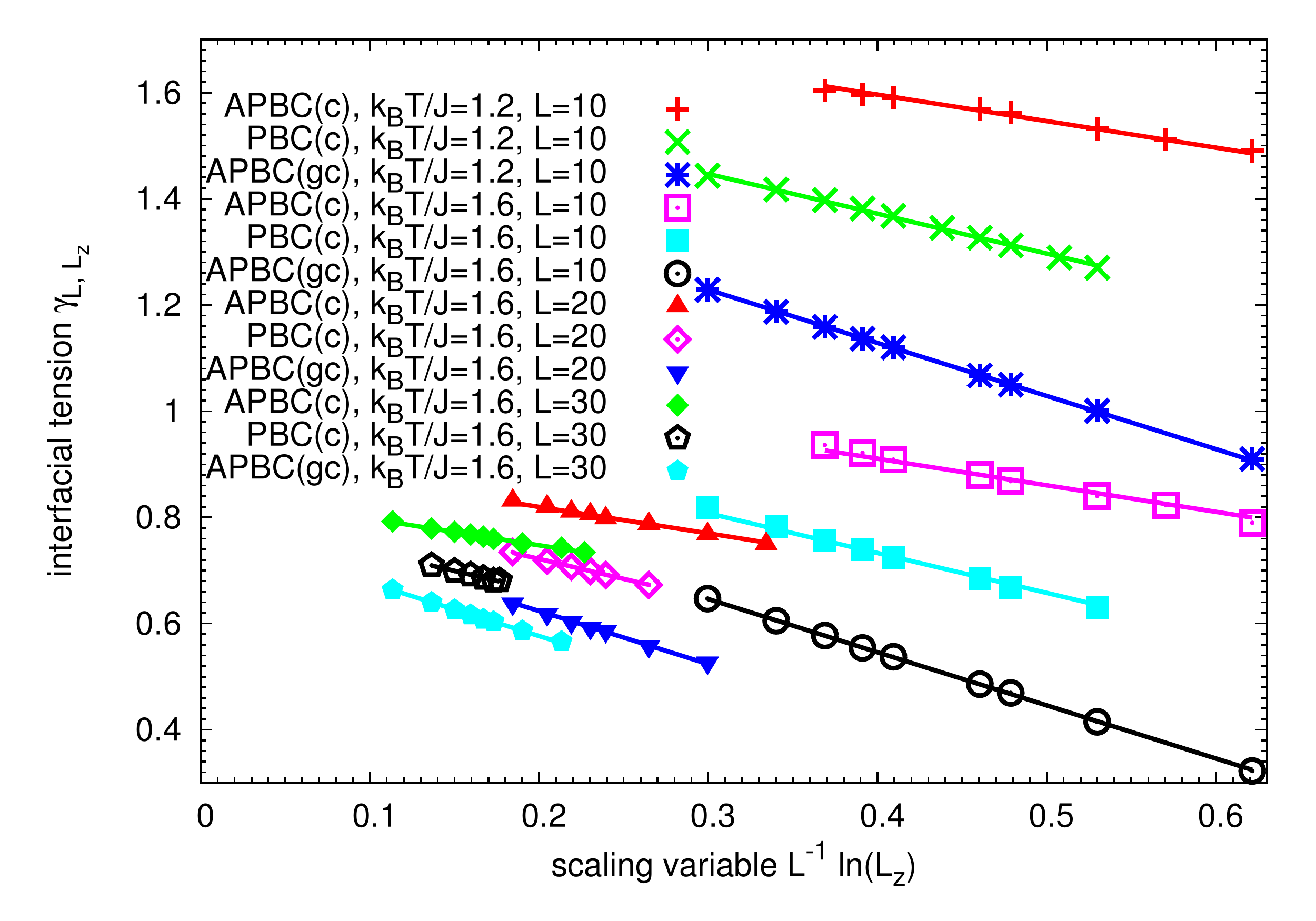}}
\subfigure[\label{fig: Ising2d_ScalingZ_PBCcan}]{\includegraphics[clip=true, trim=0mm 0mm 0mm 0mm, angle=-90,width=0.49 \textwidth]{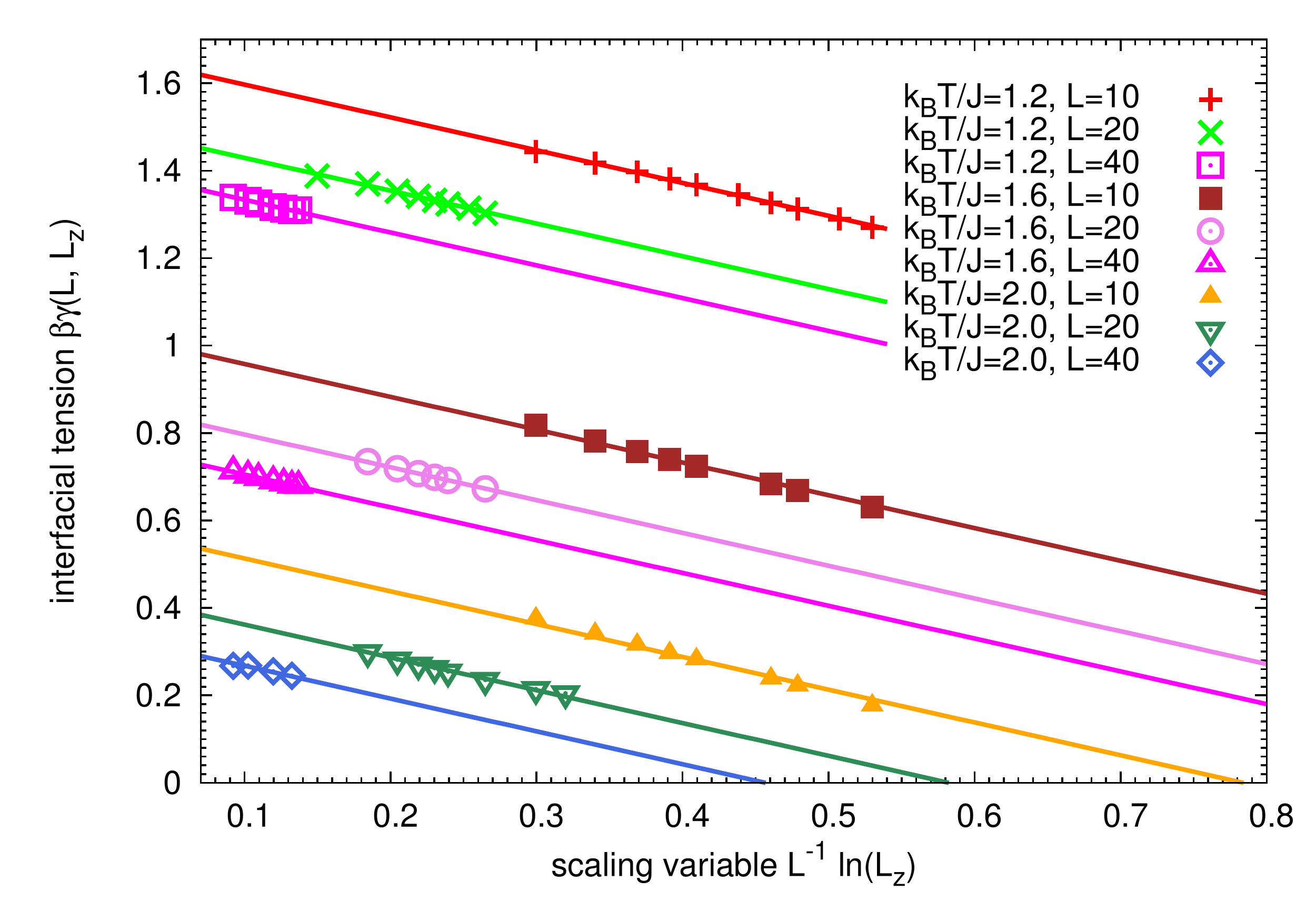}}
\caption{\label{fig: Ising2d_ScalingZ}a) Interfacial tension $\gamma_{L,L_z}$ for the $d=2$ Ising model plotted at fixed $L$ versus $L ^{-1}\ln (L_z)$, for the cases APBC(c), PBC(c) and APBC(gc). The upper set of data refers to $\kb T/J=1.2$, $L=10$, the lower set of data refers to $\kb T/J=1.6$, $L=10,20$ and $30$, as indicated. The straight lines shown have the theoretical slopes $x_\perp=1/2$, 3/4, and 1, respectively. b) Same as a), but showing only the PBC(c) case, for the three temperatures $\kb T/J=1.2$, $1.6$ and $2.0$, respectively. Always three choices of $L$ are shown namely $L=$10, 20 and 40 (from top to bottom). The straight lines illustrate the theoretical slope $x_\perp=3/4$ throughout.}
 \end{figure}

In order to identify the sources of the various finite-size effects in the problem, it is useful to choose $L_z$ different from $L$ and vary $L_z$ at fixed $L$: Executing this with the ensemble switch method for the three different choices APBC(gc), APBC(c) and PBC(c), we see from Eq.~\eqref{eq26} that we must get a result of the form
\begin{equation} \label{eq33}
\gamma_{L, L_z}=\const - x_\perp \frac{\ln L_z}{L},
\end{equation}
where all the terms depending on $L$ only (and $\gamma_\infty$) have been combined in the constant on the right-hand side of this equation, and the prefactor $x_{\perp}$ of the $(1/L)\ln L_z$ term is 1/2, 3/4 or 1, for the three choices APBC(c), PBC(c) and APBC(gc), respectively (cf. Table~\ref{tab: ScalingConstants}). Figure~\ref{fig: Ising2d_ScalingZ_All} verifies this behavior, focusing on two examples, namely $\kb T/J=1.2$, $L=10$ and $\kb T/J=1.6$, $L=10, 20$ and $30$. The straight lines have precisely these theoretical values for $x_{\perp}$, and fit the simulated data rather perfectly. We recall that in the case APBC(gc) where we have a single mobile interface, we test the simple translational entropy of the interface $x_\perp=1$, while in the case APBC(c) we just test the ``domain breathing'' contribution to the interface $(x_\perp=1/2)$. In the PBC(c) case, two interfaces are present, and both these mechanisms contribute once, yielding $x_\perp=(1 + 1/2)/2=3/4$ per interface. Fig.~\ref{fig: Ising2d_ScalingZ_PBCcan} verifies that the latter exponent indeed is found at all temperatures and all $L$.

\begin{figure*}
\centering
\subfigure[\label{fig: Ising2d_ScalingX_T2-0_L}]{\includegraphics[clip=true, trim=0mm 0mm 0mm 0mm, angle=-90,width=0.49 \textwidth]{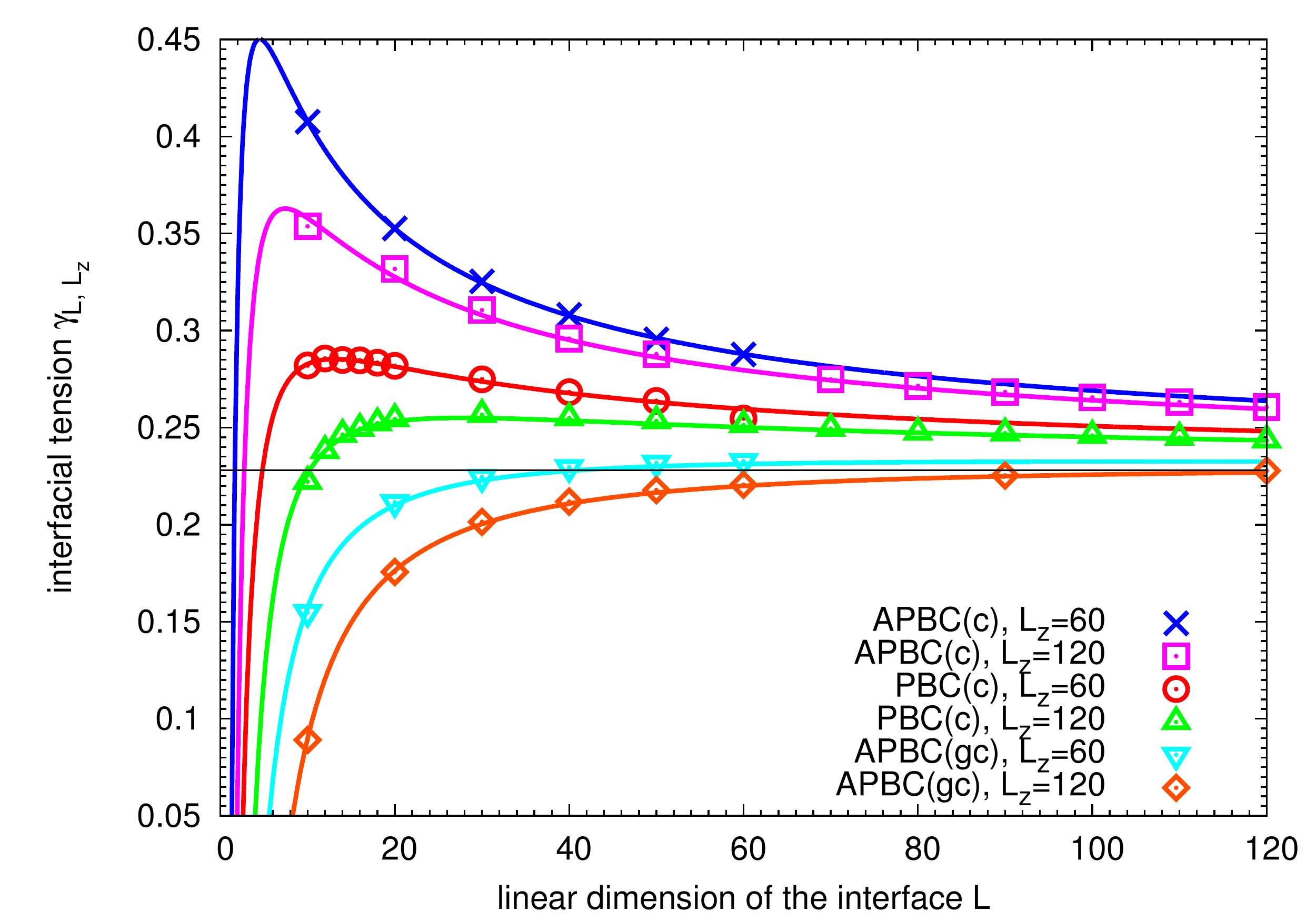}}
\subfigure[\label{fig: Ising2d_ScalingX_T1-2_sublogs_1dL}]{\includegraphics[clip=true, trim=0mm 0mm 0mm 0mm, angle=-90,width=0.49 \textwidth]{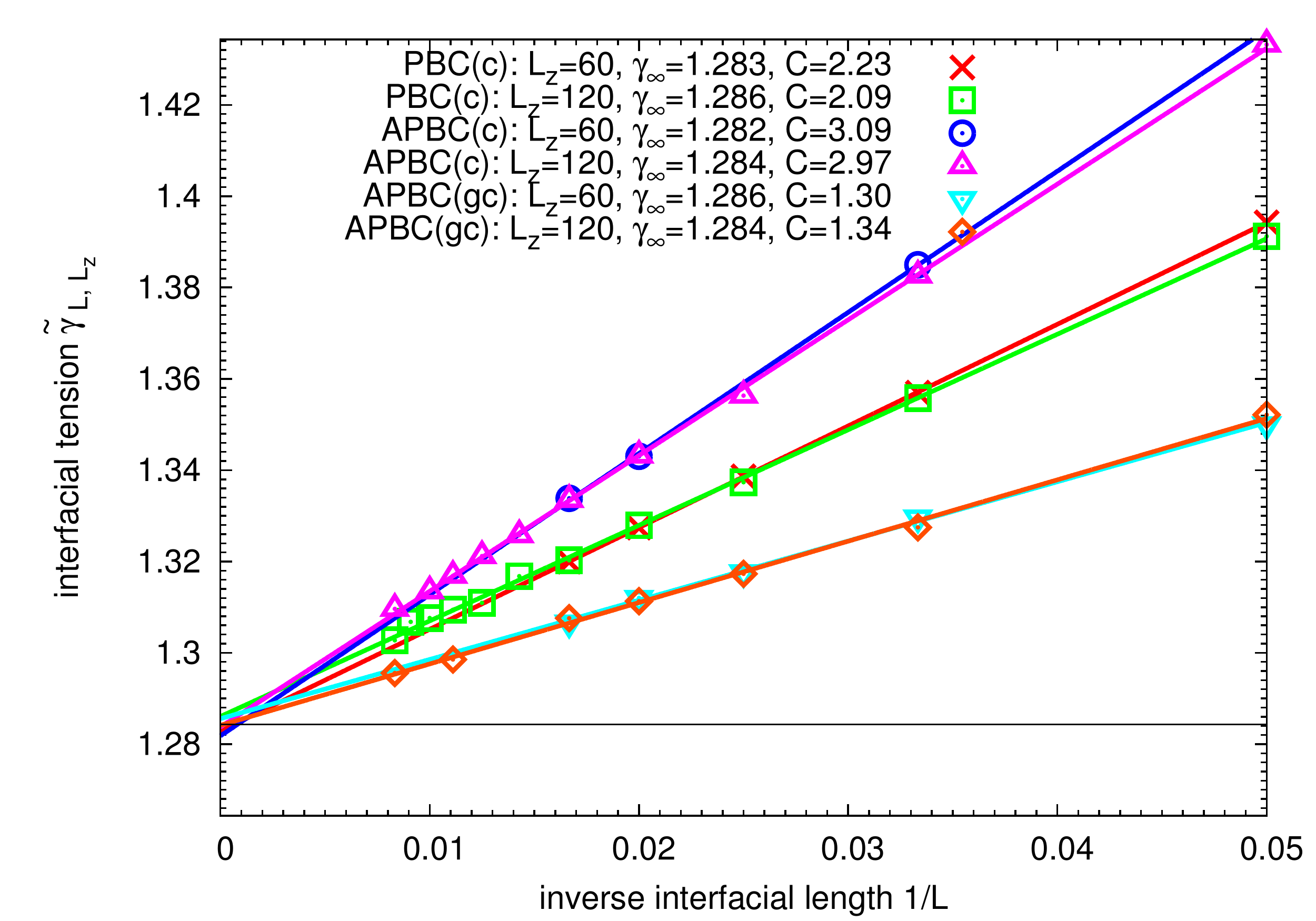}}
\subfigure[\label{fig: Ising2d_ScalingX_T1-6_sublogs_1dL}]{\includegraphics[clip=true, trim=0mm 0mm 0mm 0mm, angle=-90,width=0.49 \textwidth]{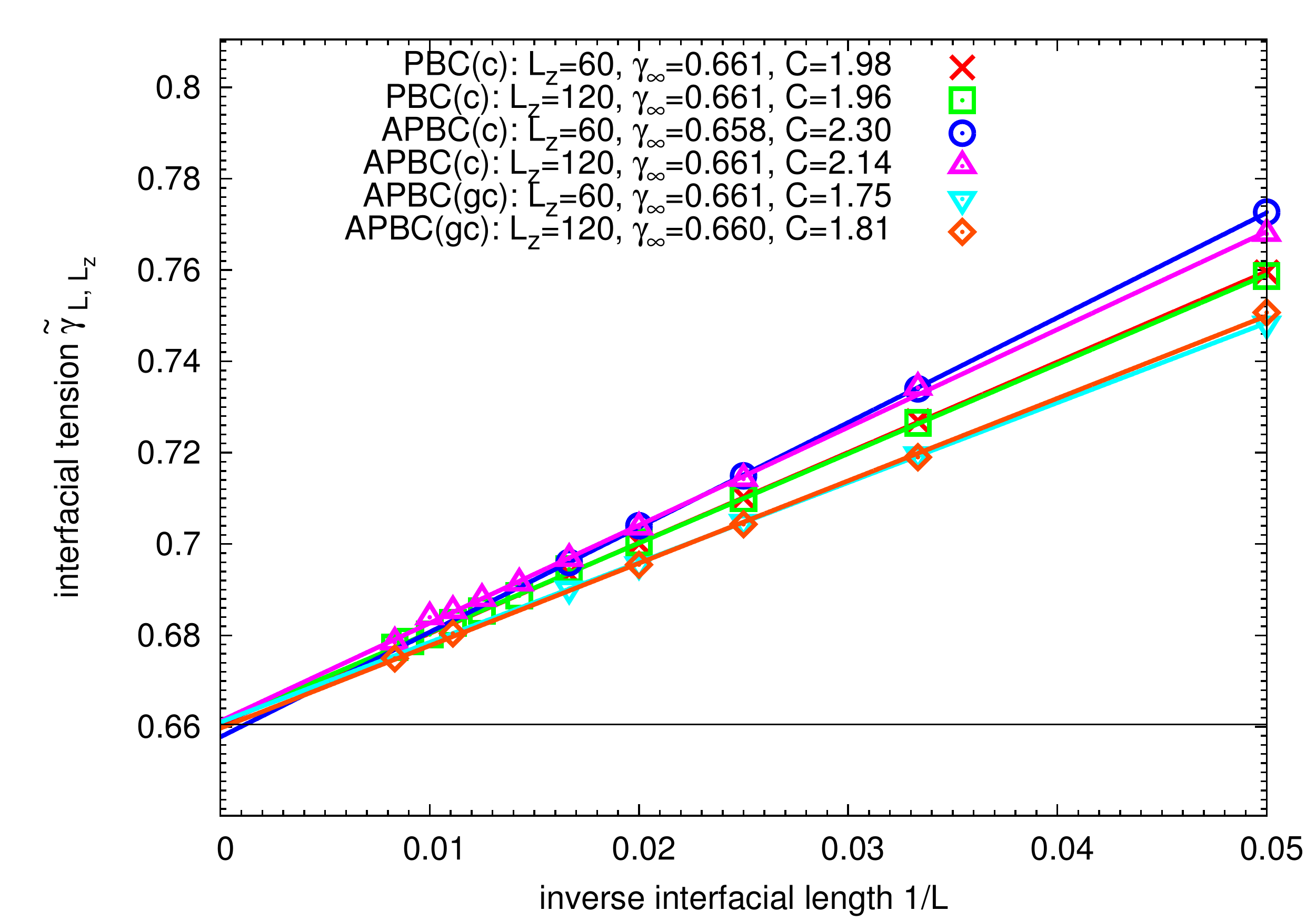}}
\subfigure[\label{fig: Ising2d_ScalingX_T2-0_sublogs_1dL}]{\includegraphics[clip=true, trim=0mm 0mm 0mm 0mm, angle=-90,width=0.49 \textwidth]{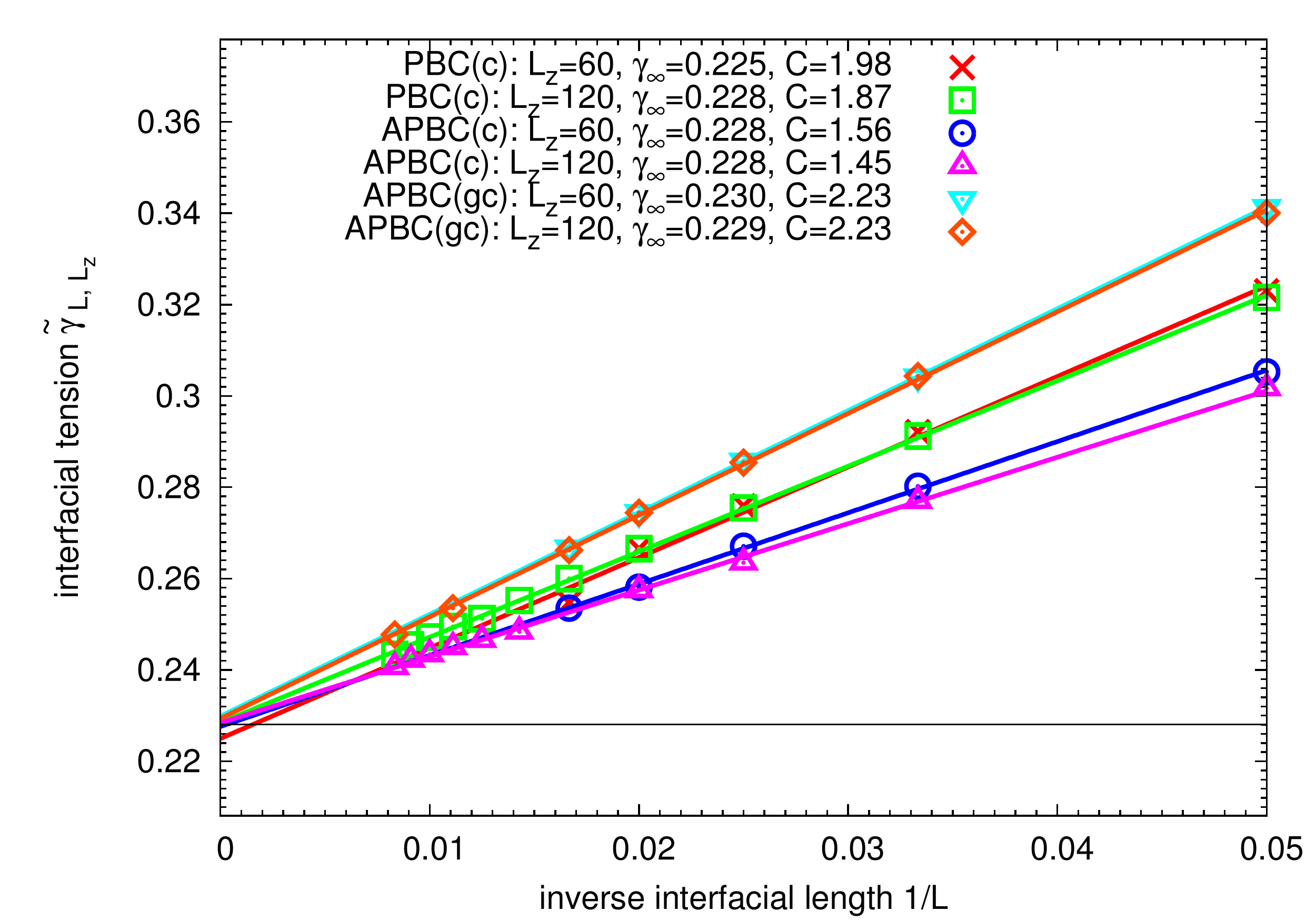}}
\caption{\label{fig: Ising2d_ScalingX} (a) Interfacial tension $\gamma_{L,L_z}$ plotted vs. $L$ for the $d=2$ Ising model at $\kb T/J=1.2$ for two choices of $L_z$, $L_z=60$ and $L_z=120$; the horizontal straight line shows the known value of $\gamma_\infty$ (Eq.~\eqref{eq6} \cite{105}), while the curves are fits of Eq.~\eqref{eq26} to the data (symbols), for the cases APBC(c), top set of curves; PBC(c), middle set of curves; APBC(gc), bottom set of curves. In each set $L_z$ increases from top to bottom. The theoretical values $x_\perp$, $x_\parallel$ from Table~\ref{tab: ScalingConstants} were used in the fit. (b) Reduced interfacial tension $\tilde{\gamma}$ \{Eq.~\eqref{eq34}\} plotted vs. $1/L$, for $\kb T/J=1.2$, and three choices of boundary conditions and/or ensembles, as indicated (PBC(c), APBC(c), and APBC(gc)). In each case two choices of $L_z$ are included, $L_z=60$ and $L_z=120$. Symbols represent the simulation results, and straight lines show the fits $\tilde{\gamma}=\gamma_\infty + C/L$; the fitted values $\gamma_\infty$, $C$ are quoted in the figure. The horizontal straight line shows the known exact result, $\gamma_\infty=1.284$ (from Eq.~\eqref{eq6} \cite{105}). (c) Same as (b), but for $\kb T/J=1.6$. Here the exact result is $\gamma_{\infty}=0.660$ \cite{106}. (d) Same as (b), but for $\kb T/J=2.0$. Here the exact result is $\gamma_\infty=0.228$ \cite{105}.}
 \end{figure*}

Of course, varying $L_z$ at fixed finite $L$ does not yield the desired information on $\gamma_\infty$; thus both $L$ and $L_z$  need to be varied and the limit that both $L$ and $L_z$ tend to infinity needs to be considered. As a first step to also test that the quoted results for $x_\parallel$ (Table~\ref{tab: ScalingConstants}) are compatible with the simulation results as well, we have fitted $\gamma_{L, L_z}$ to Eq.~\eqref{eq26}, using the theoretical values for $x_\parallel$, $x_\perp$ and $\gamma_\infty$ so that a single fit parameter remains, namely the coefficient $C$ of the $C/L$ term in Eq.~\eqref{eq26}. Fig.~\ref{fig: Ising2d_ScalingX_T2-0_L} shows that indeed an excellent fit of the data results, giving further credence to our assertion that the finite-size effects are under control. However, in the general case $\gamma_\infty$ is not known in beforehand, of course, but rather should be an output of the computation. Then a very natural strategy is to subtract the theoretical contributions $[x_\parallel \ln (L) - x_\perp \ln (L_z)]/L$ from $\gamma_{L, L_z}$, so that Eq.~\eqref{eq26} reduces to (in $d=2$)
\begin{equation} \label{eq34}
\tilde{\gamma} \equiv \gamma_{L, L_z} + \frac{x_\perp \ln L_z-x_\parallel \ln L}{L}=\gamma_\infty + \frac{C}{L}
\end{equation}
and estimate both constants $\gamma_\infty$ and $C$ from a fit of Eq.~\eqref{eq34} to the data. The results of this procedure are shown in Fig.~\ref{fig: Ising2d_ScalingX}. It is seen that the theoretical values $\gamma_\infty (T=1.2)=1.284$, $\gamma_\infty (T=1.6) =0.660$ and $\gamma_\infty (T=2.0)=0.228$ are almost perfectly reproduced! We also note that the constant $C$, which is expected to depend on both temperature and boundary conditions and the type of ensemble, since not the same fluctuations are probed, takes in each case roughly the same value for both choices of $L_z$: in the asymptotic limit, this parameter $C$ should no longer depend on $L_z$ at all, and the fact that this is not strictly true indicates that presumably there is some residual effect of higher order corrections, that were neglected in our analysis. When we try to improve the estimation of this parameter $C$ by imposing the theoretical value of $\gamma_\infty$ in the analysis, the differences between the two estimates for $C$ obtained are still slightly affected by statistical errors. Nevertheless, we judge the quality of the straight line fits in Figs.~\ref{fig: Ising2d_ScalingX}, as rather gratifying. In particular, the coincidence of the estimates for $\gamma_\infty$ for the 6 cases shown at every temperature shows that the possibility of the ensemble switch method to apply it for different boundary conditions (and/or ensemble) is most valuable for ensuring that the desired accuracy really has been reached.

From the fits in Figs.~\ref{fig: Ising2d_ScalingX_T1-2_sublogs_1dL}, \ref{fig: Ising2d_ScalingX_T1-6_sublogs_1dL} and \ref{fig: Ising2d_ScalingX_T2-0_sublogs_1dL}, we see that the constant $C$ is of order unity but temperature-dependent, and it is of interest, of course, to ask where this temperature dependence comes from. The easiest case to discuss is the case of APBC(gc), where we have argued that the singular size effects solely reflect the translational entropy contribution, Eq.~\eqref{eq10}. The capillary wave effects are already included if for the ``counting'' of states where the interface can be placed (Fig.~\ref{fig: TranslationalEntropySketch}), the length $L_z$ is measured in units of $w_L$. Of course, an additional regular contribution $c/L$ with some coefficient $c$ can also occur; this is already seen from Eqs.~\eqref{eq3}, \eqref{eq4}, which in $d=2$ can be written as $\xi_\parallel = A w_L \exp(\gamma_\infty L)$, where $A$ is another constant, and putting (in the spirit of Eq.~\eqref{eq9}) $\xi_\parallel = L_{z,0}$, where $\gamma_{L,L_z} = \gamma_\infty - \frac{1}{L} \ln(L_z/w_L) + c/L$ vanishes, we conclude $c=\ln A$. However, another contribution to this regular term comes from the prefactor in the relation $w_L\propto L^{1/2}$ in Eq.~\eqref{eq4}. In the $d=2$ Ising model it is known exactly \cite{100} that $w_L^2/L = (2 \sinh(\gamma_\infty))^{-1} \equiv l_0$ (recall that lengths are measured in units of the lattice spacing $a$). Using Eq.~\eqref{eq6} to evaluate this term for the three temperatures $\kb T/J=1.2, 1.6$ and $2.0$ considered in Fig.~\ref{fig: Ising2d_ScalingX}, we find that the remaining constant $c$, as defined above, is almost temperature independent (namely 1.94, 1.98 and 1.99, respectively, for the three mentioned temperatures). So the increase of the parameter $C$ with temperature in Fig.~\ref{fig: Ising2d_ScalingX} simply reflects the increase of the length $l_0$ (which also is measured in units of the lattice spacing and hence dimensionless) with temperature, since $C=(\ln (l_0) + c)/2$.

\subsection{Three-Dimensional Ising Model}
\label{sec: ThreeDimensionalIsingModel}

Since the computational effort in $d=3$ is substantially larger, we restrict attention here to a thorough study of a single temperature only, $\kb T/J=3.0$, where the correlation length in the bulk still is very small (recall that the critical temperature occurs at about $\kb T_c/J \approx 4.51$ \cite{81}) but this temperature is sufficiently distant from the roughening transition temperature $\kb T_R / J \approx 2.45$ \cite{114}, and hence the anisotropy effects on the interfacial free energy of flat interfaces are already small \cite{44,47,115}.

\begin{figure}
\centering
\subfigure[\label{fig: Ising3d_Scaling_ScalingZ}]{\includegraphics[clip=true, trim=0mm 0mm 0mm 0mm, angle=-90,width=0.49 \textwidth]{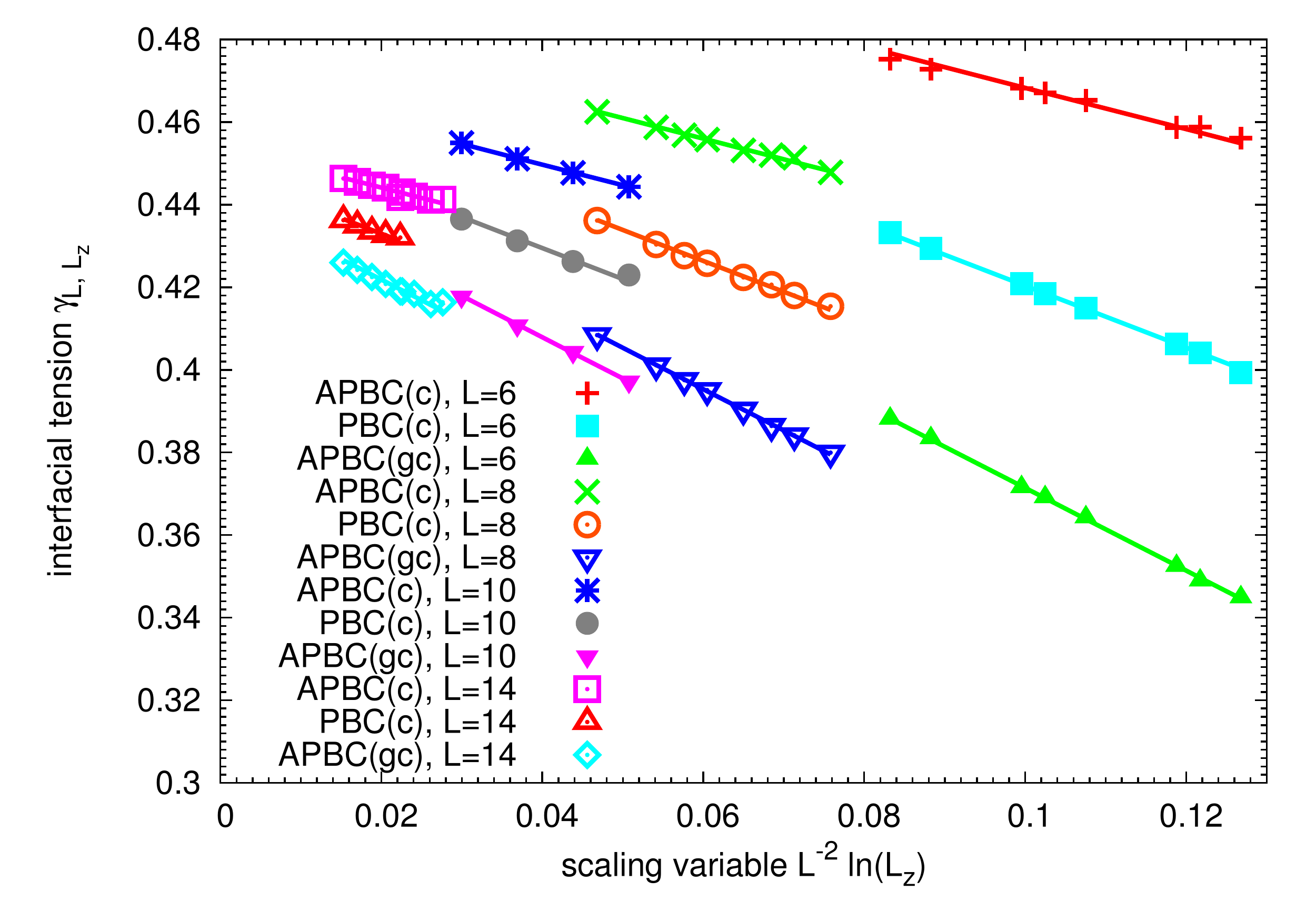}}
\subfigure[\label{fig: Ising3d_Scaling_ScalingX}]{\includegraphics[clip=true, trim=0mm 0mm 0mm 0mm, angle=-90,width=0.49 \textwidth]{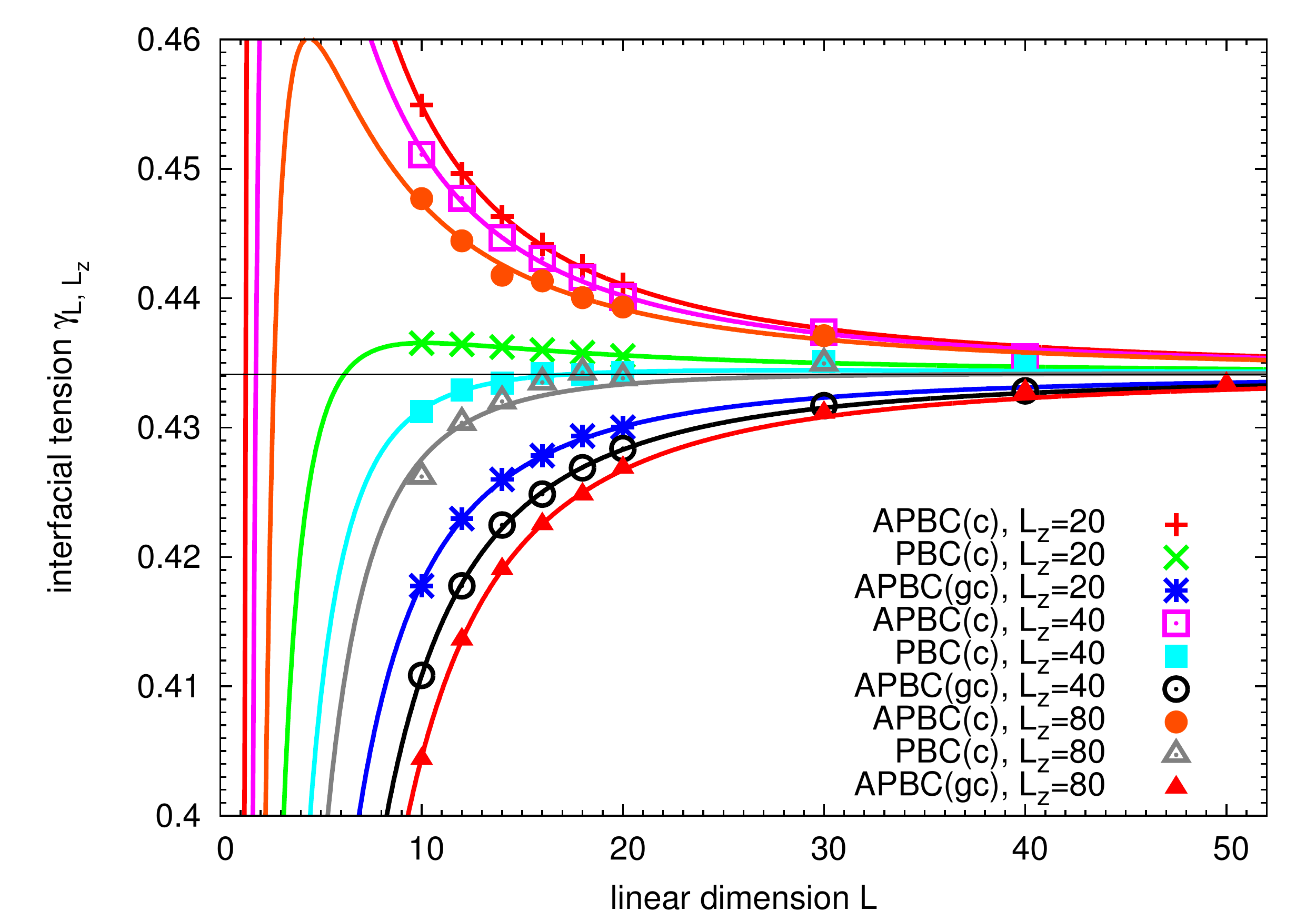}}
\caption{\label{fig: Ising3d_Scaling} (a) Interfacial tension $\gamma_{L, L_z}$ for the $d=3$ Ising model at $\kb T/J =3.0$ is plotted vs. the variable $(1/L^2) \ln (L_z)$, using PBC and the canonical ensemble. Several choices of $L$ are included, as indicated. The straight lines show the theoretical exponent $x_\perp=3/4$ (resulting from the entropy of interface translation and domain breathing). (b) Interfacial tension $\gamma_{L, L_z}$ for the $d=3$ Ising model at $\kb T/J=3.0$ plotted vs. $L$ for three choices of $L_z$, $L_z$=20, 40 and 80. The horizontal straight line shows the previous result $\gamma_{\infty}$=0.434 due to Hasenbusch and Pinn \cite{48}, while the curves are fits of Eq.~\eqref{eq26} to the data (symbols) for the cases APBC(c), top set of curves; PBC(c), middle set; APBC(gc), bottom set. In each set, $L_z$ increases from top to bottom. The theoretical values of $x_\perp$, $x_\parallel$ from Table~\ref{tab: ScalingConstants} were used, so each curve contains a single adjusted constant (the prefactor of the $1/L^2$ term) only.}
 \end{figure}

\begin{figure}
\centering
\subfigure[\label{fig: Ising3d_ScalingSublogs_Param2}]{\includegraphics[clip=true, trim=0mm 0mm 0mm 0mm, angle=-90,width=0.49 \textwidth]{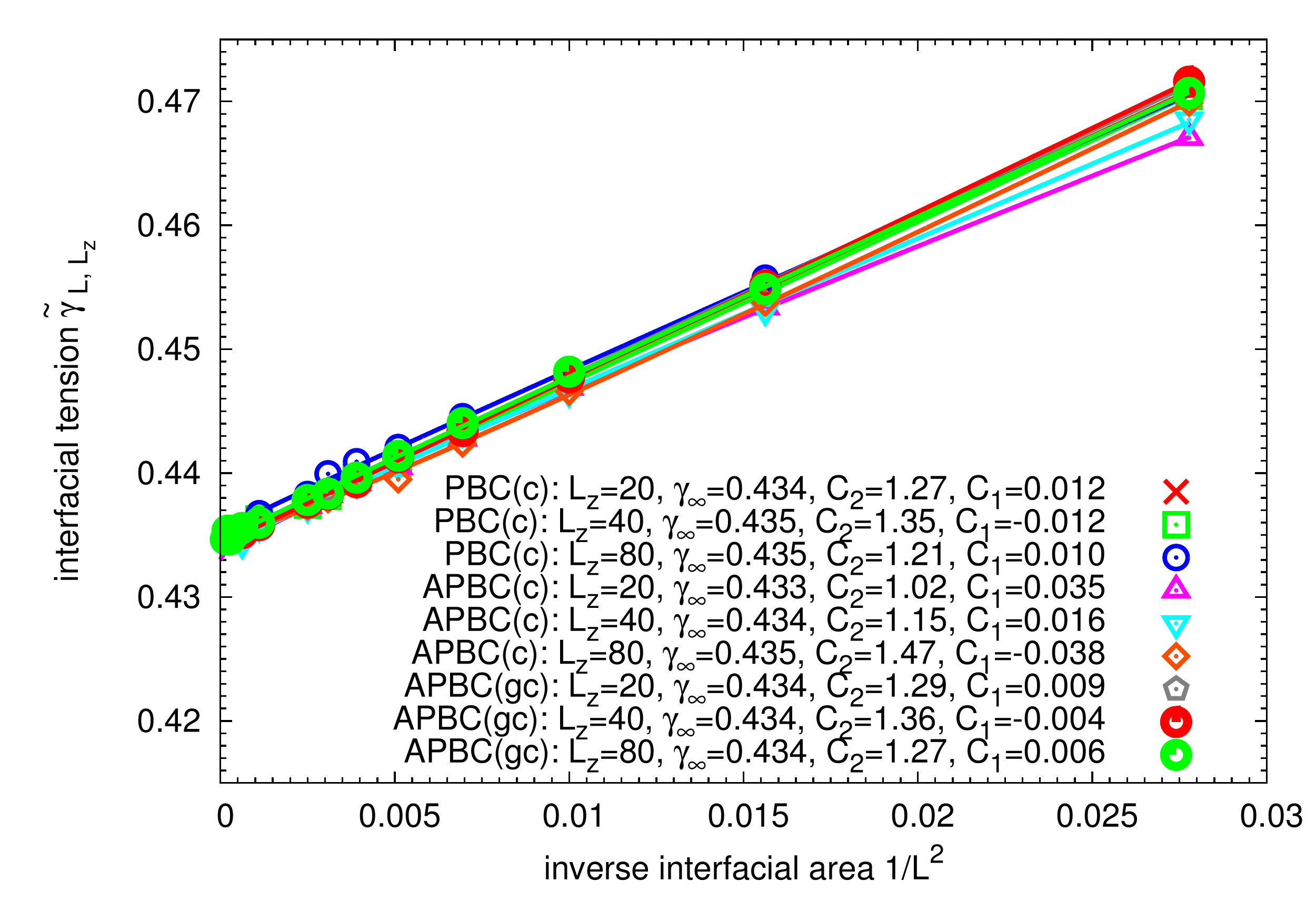}}
\subfigure[\label{fig: Ising3d_ScalingSublogs_Param1}]{\includegraphics[clip=true, trim=0mm 0mm 0mm 0mm, angle=-90,width=0.49 \textwidth]{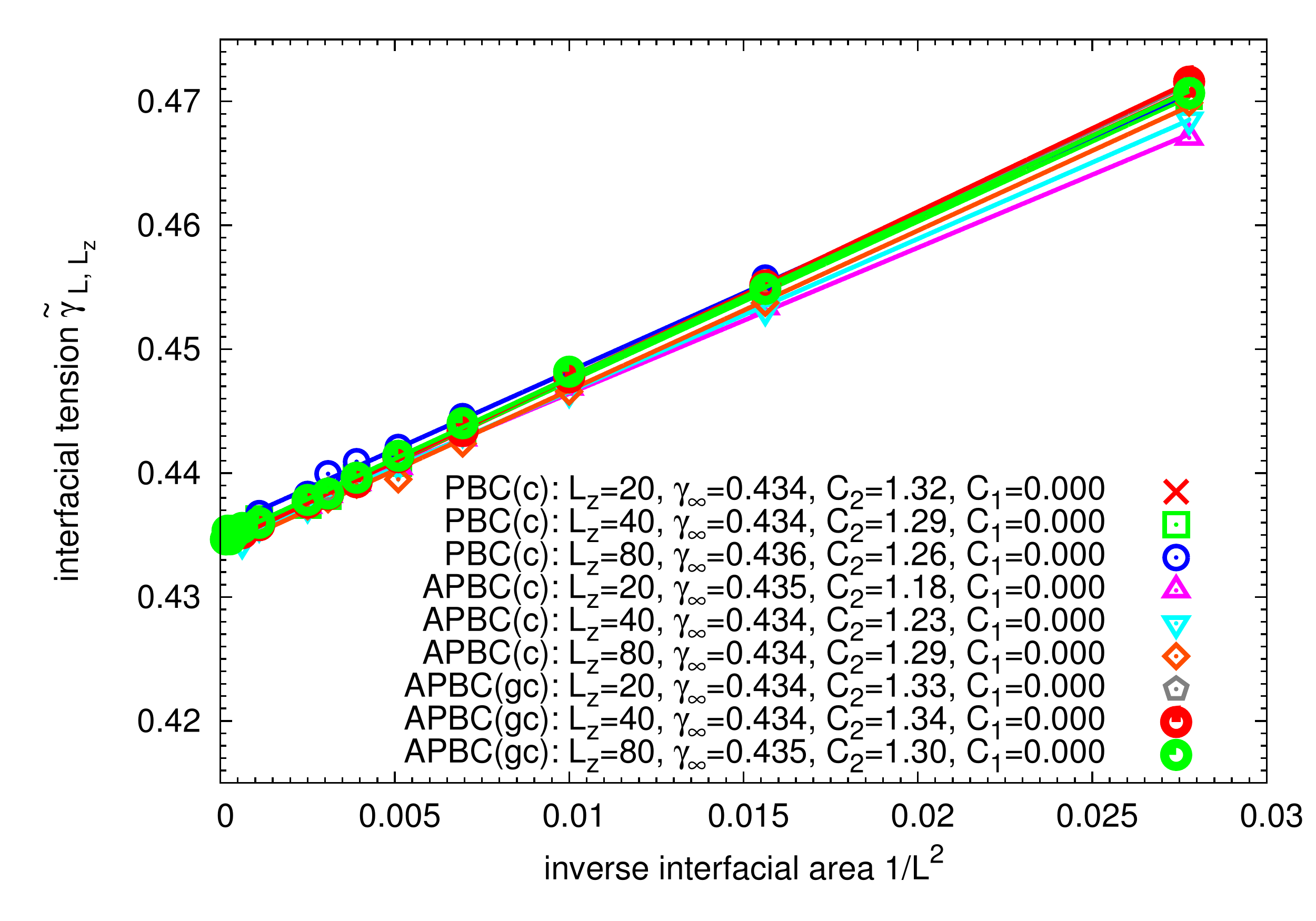}}
\caption{\label{fig: Ising3d_ScalingSublogs} Reduced interfacial tension $\tilde{\gamma}$ \{Eq.~\eqref{eq35}\} plotted vs. $1/L^2$ for the $d=3$ Ising model, using three chocies of $L_z$ and three choices of boundary conditions and/or ensembles, as indicated. Case (a) includes the parameter $C_1$, in the fit, while case (b) requires $C_1=0$. The resulting estimates for the parameters $\gamma_\infty$ and $C_2$ are quoted in the figure.}
   \end{figure}

Again, we begin by asserting that the effects demonstrated to be important in the $d=2$ case, such as the translational entropy of the interface and ``domain breathing'' fluctuations, have a significant impact in three dimensions, too. Fig.~\ref{fig: Ising3d_Scaling_ScalingZ} is the counterpart of Fig.~\ref{fig: Ising2d_ScalingZ}, demonstrating the presence of a correction $-x_\perp (1/L^2) \ln (L_z)$, due to the translational entropy of the interface(s) and domain breathing, when $L_z$ is varied at fixed $L$. Fig.~\ref{fig: Ising3d_Scaling_ScalingX} is the counterpart of Fig.~\ref{fig: Ising2d_ScalingX_T2-0_L}, where we fit the data to the full Eq.~\eqref{eq26} when $L$ is varied for several choices of $L_z$, using the known value \cite{48} $\gamma_\infty=0.434$ and the theoretical values of $x_\perp$, $x_\parallel$ from Table~\ref{tab: ScalingConstants}, so that a single parameter (the prefactor of the $1/L^2$ term in Eq.~\eqref{eq26}) is fitted. As in the case of $d=2$ the quality of the fit is excellent. Thus, in order to estimate $\gamma_\infty$, we proceed in analogy with Eq.~\eqref{eq34}, reducing the data with the known theoretical corrections (using Eq.~\eqref{eq26} and Table~\ref{tab: ScalingConstants})
\begin{equation} \label{eq35}
\widetilde{\gamma} \equiv \gamma_{L,L_z} + \frac{x_\perp \ln L_z - x_\parallel \ln L}{L^2} = \gamma_\infty + \frac{C_1}{L} +\frac{C_2}{L^2}
\end{equation}
Here we have made an important phenomenological modification, not suggested by our theoretical considerations of Sec.~\ref{sec: PhenomenologicalTheory}: there must be the theoretically expected term of order $1/L^2$, which is strictly required because the arguments of the logarithms in Eq.~\eqref{eq26} must have the form $\ln (L_z/ l')$, $\ln (L/l'')$ with some lengths $l '$, $l''$, to make the arguments dimensionless, and so the unspecified constant in the last term on the right hand side of Eq.~\eqref{eq26} must contain a factor $x_\perp \ln l' - x_\parallel \ln l ''$. We have written this theoretically expected term then in the form $C_2 / L^2$, where $C_2$ is some effective parameter. However, in addition we have allowed for a term $C_1/L$, where $C_1$ is another (hypothetical) effective parameter. Fig.~\ref{fig: Ising3d_ScalingSublogs_Param2} shows the result of such an analysis: we see that the parameter $C_1$, if it exists, is very small (of order 10$^{-2}$ lattice spacings), while the parameter $C_2$ is of order unity (and almost independent of $L_z$: the weak variation of this parameter with $L_z$ is surely due to residual statistical errors, and possible higher order corrections which were disregarded from the start). The value of $\gamma_\infty$ estimated from such a fit is in excellent agreement with the value known from a completely different method \cite{47}. Thus, it is tempting to require that the parameter $C_1$, that was introduced phenomenologically in Eq.~\eqref{eq35}, actually must be zero. Fig.~\ref{fig: Ising3d_ScalingSublogs_Param1} shows that the data are fully compatible with this assumption, the random spread in the estimates for $\gamma_\infty$ and $C_2$ is now distinctly smaller than before, and no evidence for some systematic error is detected. We also emphasize that for $L=10$ the deviation of $\tilde{\gamma}$ still is about 3\%, for $L=20$ it is almost 1\%, and so it is clear that finite-size extrapolations are needed for a very precise estimate.

In fact, the non-existence of a term $C_1/L$ in Eq.~\eqref{eq35} is desirable in view of a completely different argument. Consider the situation that in the directions parallel to the interface we do not use a PBC but rather use free boundaries. Then we expect that the interfacial tension must contain a correction of order $2 \gamma_\text{line}/L$ where $\gamma_\text{line}$ is the line tension \cite{16,116,117} of the contact line of the interface at such a boundary. This geometry in fact has been suggested (and used) to obtain estimates for the line tension \cite{118,119}. Such an approach would not make sense if it would be spoiled by ``intrinsic'' finite-size effects that are of the same order (see also \cite{120}).

In view of this conclusion that the parameter $C_1$ for the $d=3$ Ising model does not exist, the reader may wonder why we present this discussion in such detail. However, as we shall see in the next section, the situation may be more subtle: previous work on LJ fluids and LJ mixtures \cite{110} in fact assumed that the leading corrections are of order $1/L$.

\subsection{The Lennard-Jones Fluid}
\label{sec: LennardJonesFluid}

We now study the interfacial tension of a generic off-lattice system, namely the (truncated and shifted) Lennard Jones fluid of point particles with a pairwise interaction potential $U(r)$ as defined in Eq.~\eqref{eq28}. It is known that this model has a vapor-liquid phase separation for temperature $T$ below the critical temperature of $\kb T_c/\varepsilon=0.999$ \cite{57}. Here we shall only analyze data at temperature $\kb T/\varepsilon = 0.78$. For this temperature Eq.~\eqref{eq27} was already used previously \cite{110} to estimate $\gamma_\infty= 0.375(1)$ (choosing units $\varepsilon=1$ and $\sigma=1$, as mentioned in Sec.~\ref{sec: ModelsAndSimulationMethods}.

For the off-lattice LJ fluid, an analogue of the APBC is not known. Therefore, we restrict attention to the PBC(c) case. We apply here only the ensemble switch method, using standard local displacements as the elementary Monte Carlo move for the particles \cite{80,81}.

\begin{figure}
\centering
\includegraphics[clip=true, trim=5mm 5mm 5mm 5mm, angle=-90,width=0.49 \textwidth]{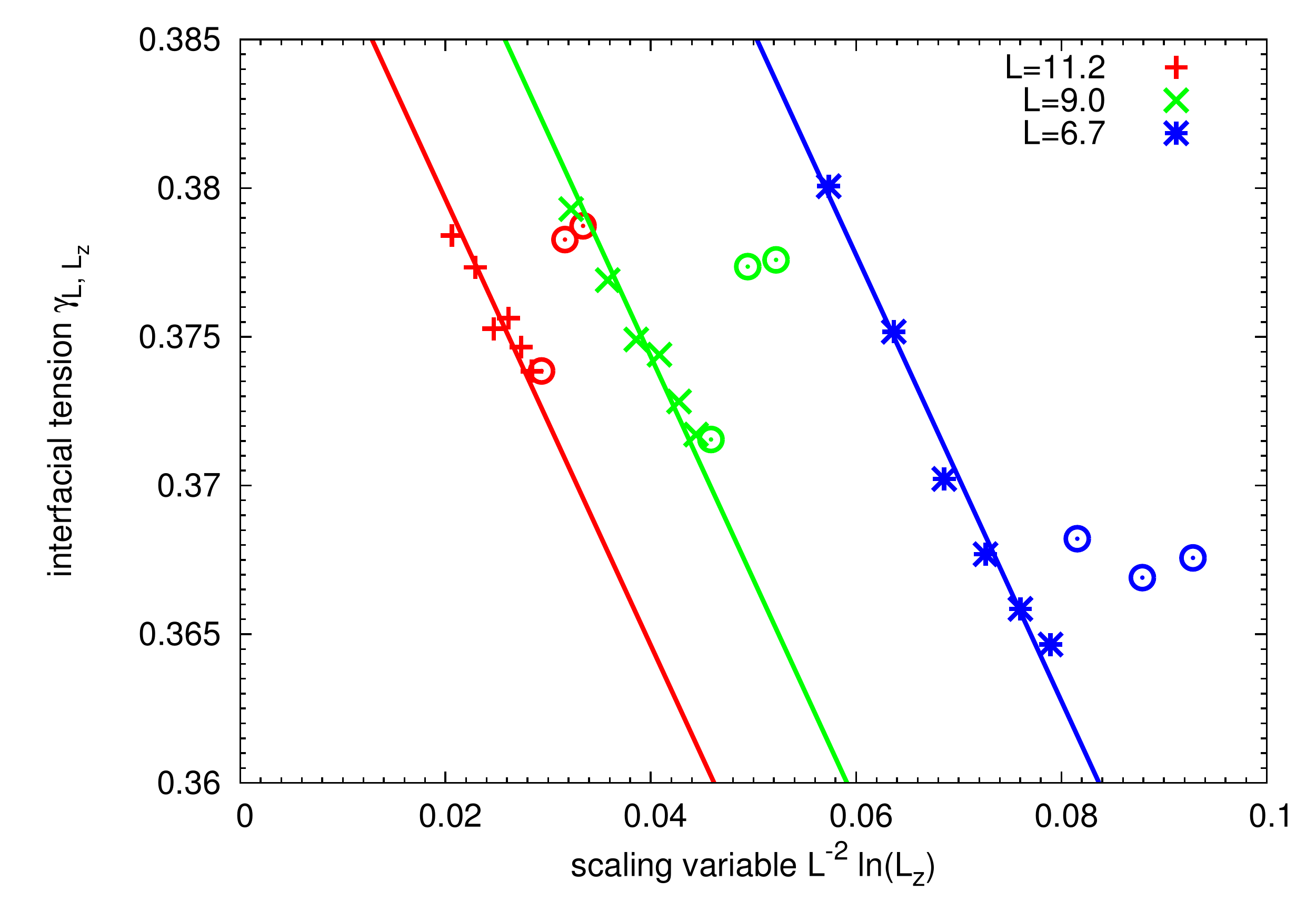}
\caption{\label{fig: LJSCT0-78_ScalingZ} Interfacial tension of the LJ fluid at $\kb T/\varepsilon=0.78$ plotted vs. the scaling
variable $L^{-2} \ln (L_z)$, for three choices of the cross-sectional area $A=L^2$, as indicated. The slope of the straight lines is again the theoretical value, $x_\perp=3/4$. The circles show preliminary data for rather large $L_z$ with insufficient statistics (see text).}
  \end{figure}

We proceed as in the last subsection, testing first the variation of $\gamma_{L, L_z}$ with $L_z$, for several choices of cross sectional area $A=L^2$ (Fig.~\ref{fig: LJSCT0-78_ScalingZ}). Indeed the predicted logarithmic variation (again due to the translational entropy of the interface and the domain breathing effect) is verified. But we have to make a caveat: Due to the use of a local algorithm for moving particles (unlike the Ising model, where in the conserved case spins at arbitrary distances from each other were interchanged) the relaxation of the particle configurations is very slow. Basically, in order to actually observe the logarithmic contributions quantitatively correct, the simulation runs must be long enough that the interface in Fig.~\ref{fig: TranslationalEntropySketch} can explore the available volume. If the runs are too short, and the liquid domain diffuses only over a length $L_\text{diff} \ll L_z$, we expect that the contribution to the entropy that is ``measured'' by such a simulation is only $-L^{-2} \ln (L_\text{diff})$ rather than $-L^{-2} \ln (L_z)$. Since diffusive displacements only increase with the square root of time, we expect that a simulation time $\tau_\text{sim} \approx L^2_z/D$ would be needed to observe the correct entropic effect on the interfacial tension where $D$ is the effective domain diffusion constant. Since the diffusion constant $D$ with which the liquid domain can move in the simulation box is expected to be very small, for our local Monte Carlo algorithm, for large $L_z$ the simulation time will not suffice to sample the full equilibrium result, and we rather observe a result which is independent of $L_z$ but depends on the simulation time $\tau_\text{sim}$ via the equation $\tau_\text{sim} \approx L^2_\text{diff}/D.$ So we see the correct logarithmic variation only for $L_z < L_\text{diff}$ in Fig.~\ref{fig: LJSCT0-78_ScalingZ}, while for $L_z > L_\text{diff}$ there is no longer a systematic decrease of $\gamma_{L, L_z}$ with $L_z$ (data from too short runs are shown by circles), rather the data fluctuate randomly around a value that was dictated by the choice of $\tau_\text{sim}$.

In view of this problem, it is in fact desirable to use also grandcanonical particle insertion and deletion moves for the Lennard-Jones fluid as well, as we did in the Ising model. A simulation in the canonical ensemble then is realized by trial moves where one attempts both to randomly delete a particle somewhere in the box and also insert a particle at a randomly selected position simultaneously. The trial move is accepted and executed only if both parts of the move together are accepted in the Metropolis test. It is clear that such nonlocal displacements of particles will fulfill detailed balance and have a reasonably high acceptance probability at the temperatures where grandcanonical ensemble simulations of the considered model are still feasible. For the LJ fluid studied here, this is the case for $\kb T/\varepsilon=0.78$.

\begin{figure}
\centering
\includegraphics[clip=true, trim=5mm 5mm 5mm 5mm, angle=-90,width=0.49 \textwidth]{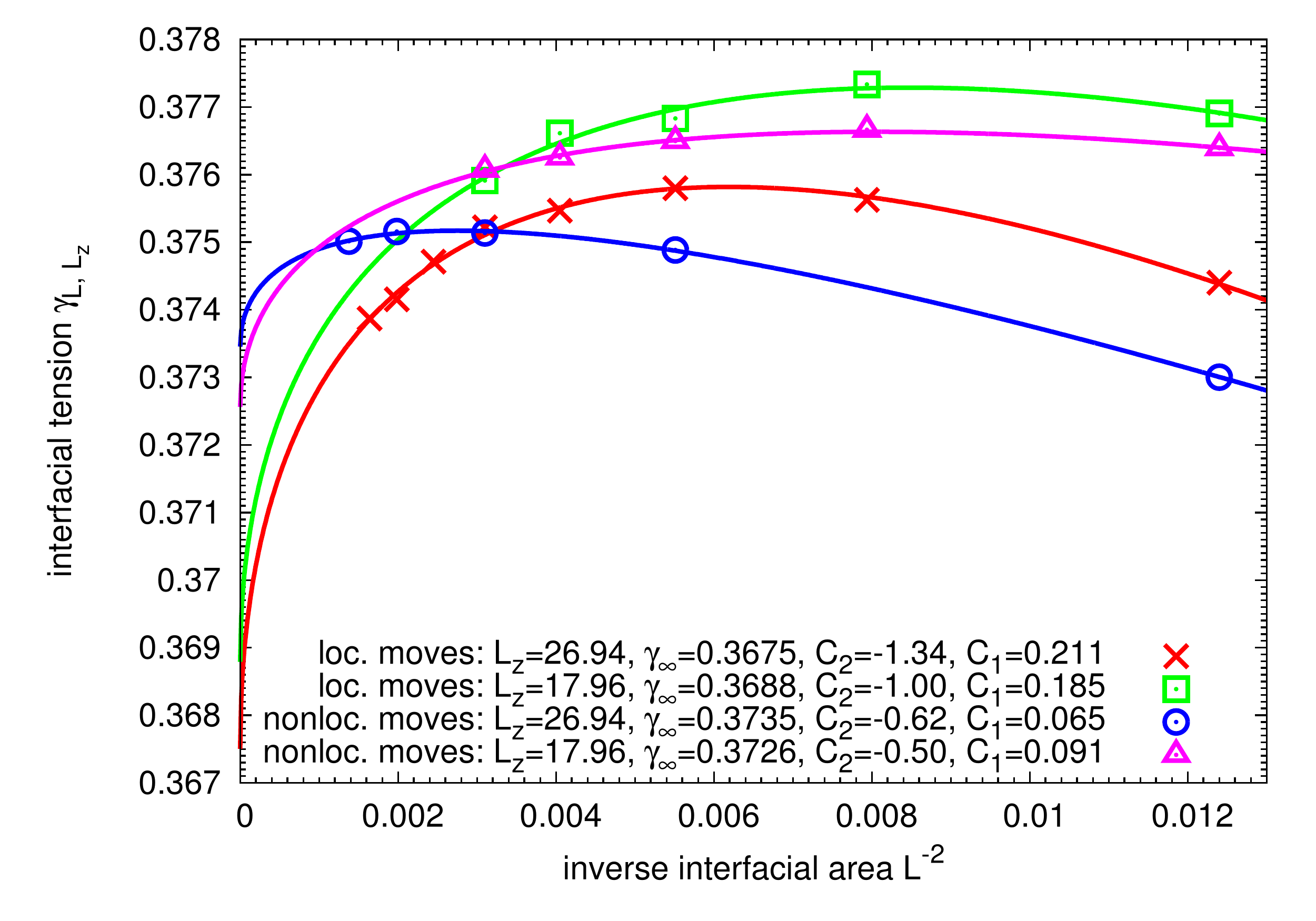}
\caption{\label{fig: LJSCT0-78_ScalingXsublogsforgotten} Interfacial tension of the LJ fluid at $\kb T/\varepsilon=0.78$ plotted vs. $L^{-2}$, using either local or nonlocal moves, for fixed $L_z$, as indicated. The logarithmic corrections have not been subtracted. The lines are fits of the form as in Eq.~\eqref{eq35}.}
  \end{figure}

\begin{figure}
\centering
\subfigure[\label{fig: LJSCT0-78_ScalingX_Param2}]{\includegraphics[clip=true, trim=5mm 5mm 5mm 5mm, angle=-90,width=0.49 \textwidth]{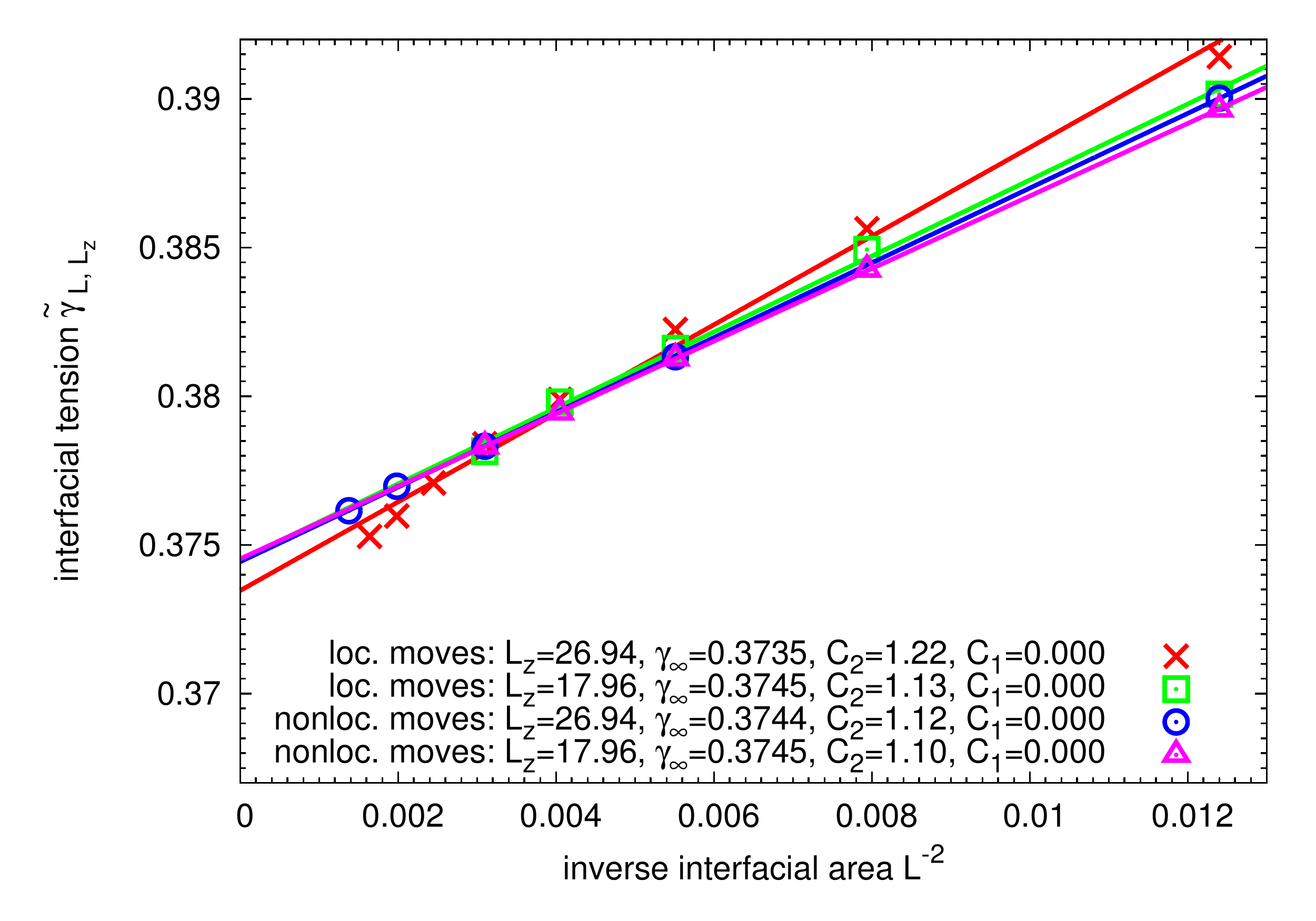}}
\subfigure[\label{fig: LJSCT0-78_ScalingX_Param3}]{\includegraphics[clip=true, trim=5mm 5mm 5mm 5mm, angle=-90,width=0.49 \textwidth]{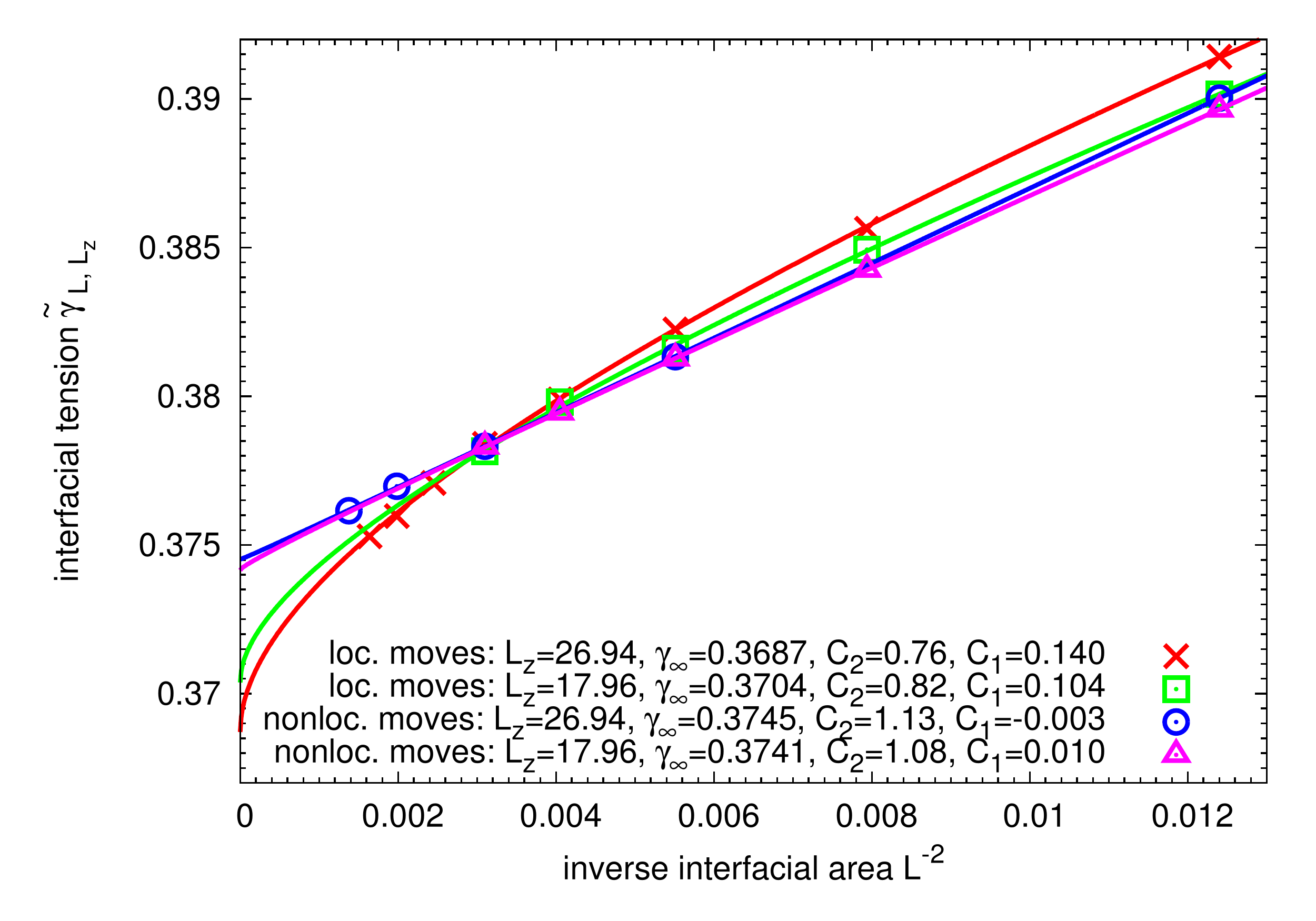}}
\caption{\label{fig: LJSCT0-78_ScalingX} Reduced interfacial tension $\tilde{\gamma}$ [Eq.~\eqref{eq35}] of the Lennard Jones fluid at $\kb T/\varepsilon=0.78$ plotted vs. $L^{-2}$. In case a) the parameter $C_1$ was forced to be zero, while in case b) both parameters $C_1$, $C_2$ were fitted. The results of the fits are quoted in the figure. Note that both data dare included where only local moves of the particle were permitted, as well as data where randomly chosen particles were removed from their position and re-inserted at a randomly chosen position anywhere in the box.}
 \end{figure}

Figure~\ref{fig: LJSCT0-78_ScalingXsublogsforgotten} plots then data for $\gamma_{L,L_z}$ at two fixed choices for $L_z$ versus $L^{-2}$, comparing results obtained using only local moves (which we believe are insufficiently equilibrated) with the results based on the nonlocal moves. One can see two features: 

	\begin{itemize}
	\item[{(i)}] The data based on the local moves are systematically off, but they are not visibly irregular, and so without the availability of the better data based on the nonlocal algorithm, it would not be obvious that the data based on local moves are unreliable.
	\item[{(ii)}] Fitting either set of data in the traditional way, i.e.~assuming a variation $\gamma_{L,L_z} = \gamma_\infty + C'_1/L + C'_2/L^2$, both fit parameters $C'_1$ and $C'_2$ clearly are nonzero, as is visually obvious from the curvature of this plot. The constant $C'_1$ is larger for the unreliable data. Omitting data for smaller values of $L$, one can get off with the simpler variation $\gamma_{L,L_z} = \gamma_\infty + C'_1/L$ as done in the literature \cite{110}, but we now know that such a fit is meaningless, a parameter $C'_1$ should not occur, and hence the resulting estimate for $\gamma_\infty$ would be inaccurate.
	\end{itemize}
Of course, a naive data analysis as shown in Fig.~\ref{fig: LJSCT0-78_ScalingXsublogsforgotten} ignores all the knowledge on the logarithmic corrections derived in the present paper. In fact, if we use this knowledge, subtracting the logarithmic correction via Eq.~\eqref{eq35}, and fit only the reduced surface tension $\widetilde{\gamma}$, as we did in the $d=3$ Ising model, the picture becomes much clearer (Fig.~\ref{fig: LJSCT0-78_ScalingX}). The reliable nonlocal data yield very small values for $C_1$ again, hence giving evidence that $C_1=0$, and if we require $C_1=0$ from the outset, a very good fit with $\gamma_\infty \approx 0.3745\pm0.0005$ is in fact obtained (Fig.~\ref{fig: LJSCT0-78_ScalingX_Param2}). The less reliable data based on the local algorithm are in fact compatible with this conclusion, if we omit the data for $L_z=26.94$ for the two largest choices of $L$, which fall systematically below the straight lines in Fig.~\ref{fig: LJSCT0-78_ScalingX_Param2}. In Fig.~\ref{fig: LJSCT0-78_ScalingX_Param3}, where $C_1$ was not forced to be zero, a systematically too small value for $\gamma_\infty$ would result from the unreliable data with the local algorithm, but it is clear that this is an artifact due to the combined effect of unreliable data and an inappropriate fitting formula (allowing for a nonzero $C_1$).

We have given this detailed discussion to show that in cases of practical interest, the knowledge of the logarithmic corrections indeed is very valuable to extract reliable estimates for $\gamma_\infty$; but high quality well equilibrated ``raw data'' for $\gamma_{L,L_z}$ are an indispensable input in the analysis.

\begin{figure}
\centering
\includegraphics[clip=true, trim=5mm 5mm 5mm 0mm, angle=-90,width=0.49 \textwidth]{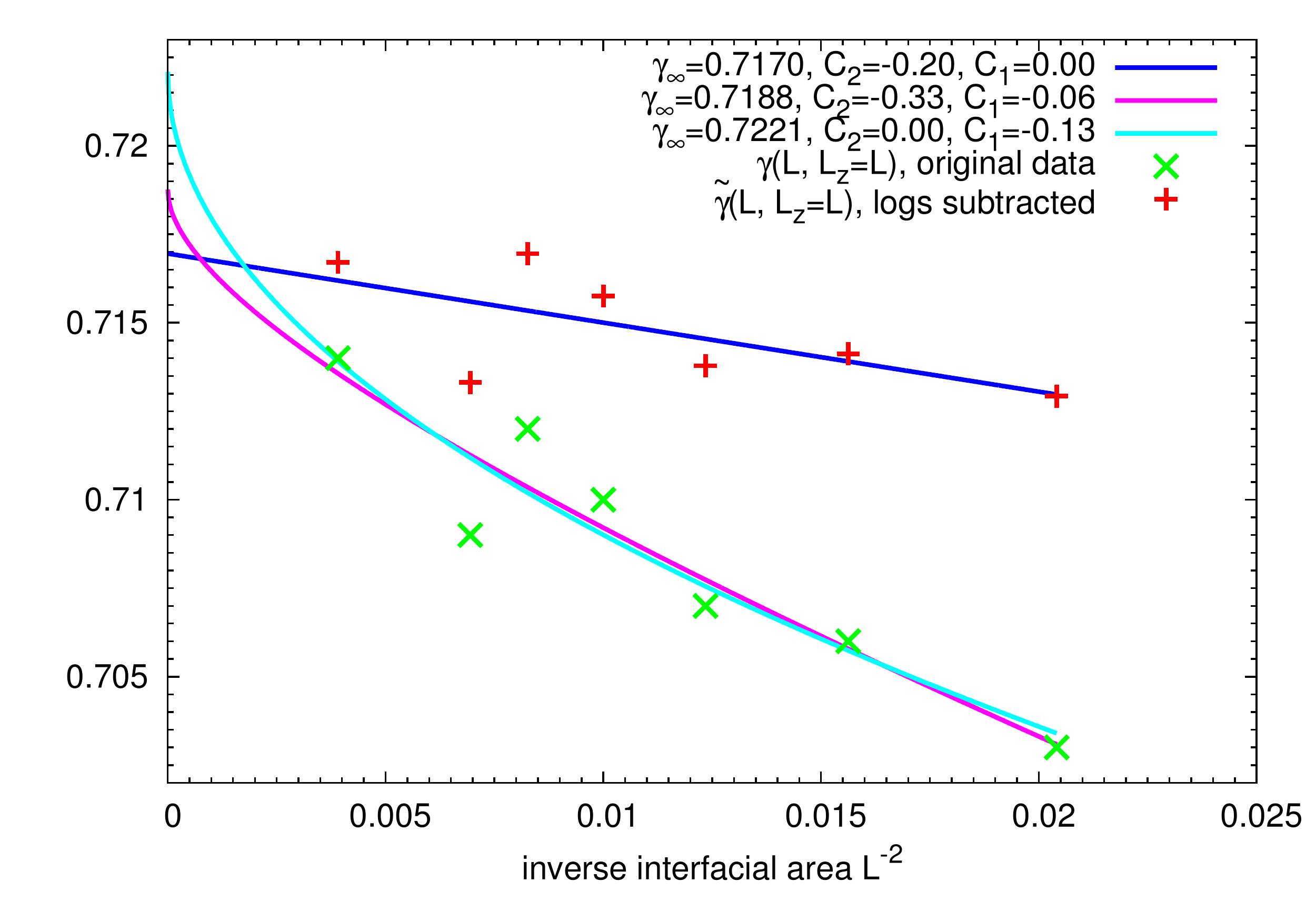}
\caption{\label{fig: SymmLJ_DataFromDas} Interfacial tension of the symmetric LJ fluid at $\kb T/\varepsilon=1.0$ plotted vs. $L^{-2}$. Cubic boxes of volume $L^3$ have been used. The lines are fits of the form as in Eq.~\eqref{eq35}. The original data is from~\cite{110}, in which a fit of the form $A+B/L$ (corresponding to fit with $C_2=0$ and $C_1=-0.13$) suggests $\gamma_\infty\approx 0.722$. After subtracting the logarithmic contributions, the result is rather $\gamma_\infty \approx 0.717$.}
  \end{figure}

As a final example, we present a re-analysis of the data for the symmetrical binary (AB) Lennard-Jones mixture at $\kb T/\varepsilon=1.0$ presented in \cite{110}. The original data (resulting from semi-grandcanonical exchange moves between the particles) were extrapolated against $1/L$, yielding $\gamma_\infty\approx 0.722$. Using again Eq.~\eqref{eq35}, we see that the data are compatible with the absence of a term $C_1/L$ as well (Fig.~\ref{fig: SymmLJ_DataFromDas}), and the final estimate for $\gamma_\infty$ ($\approx 0.717$) is only slightly off from the original estimate.

\section{Summary \& Outlook}
\label{sec: Conclusion}

\begin{figure}
\centering
\subfigure[\label{fig: BoundaryConditionsFreeSurfaces}]{\includegraphics[clip=true, trim=3mm 3mm 3mm 3mm, angle=0,width= 0.4 \columnwidth]{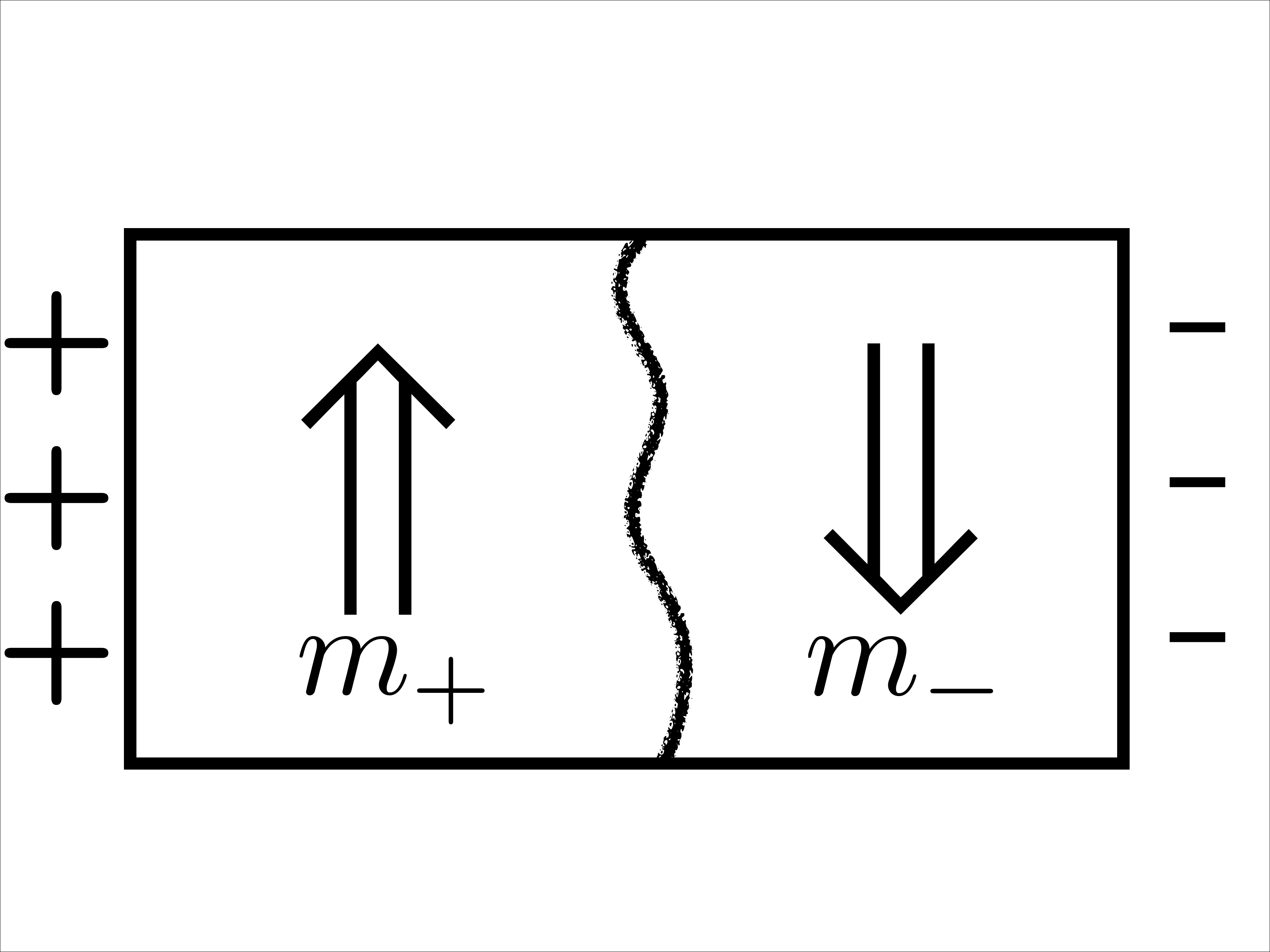}}
\subfigure[\label{fig: BoundaryConditionsAPBC}]{\includegraphics[clip=true, trim=3mm 3mm 3mm 3mm, angle=0,width= 0.4 \columnwidth]{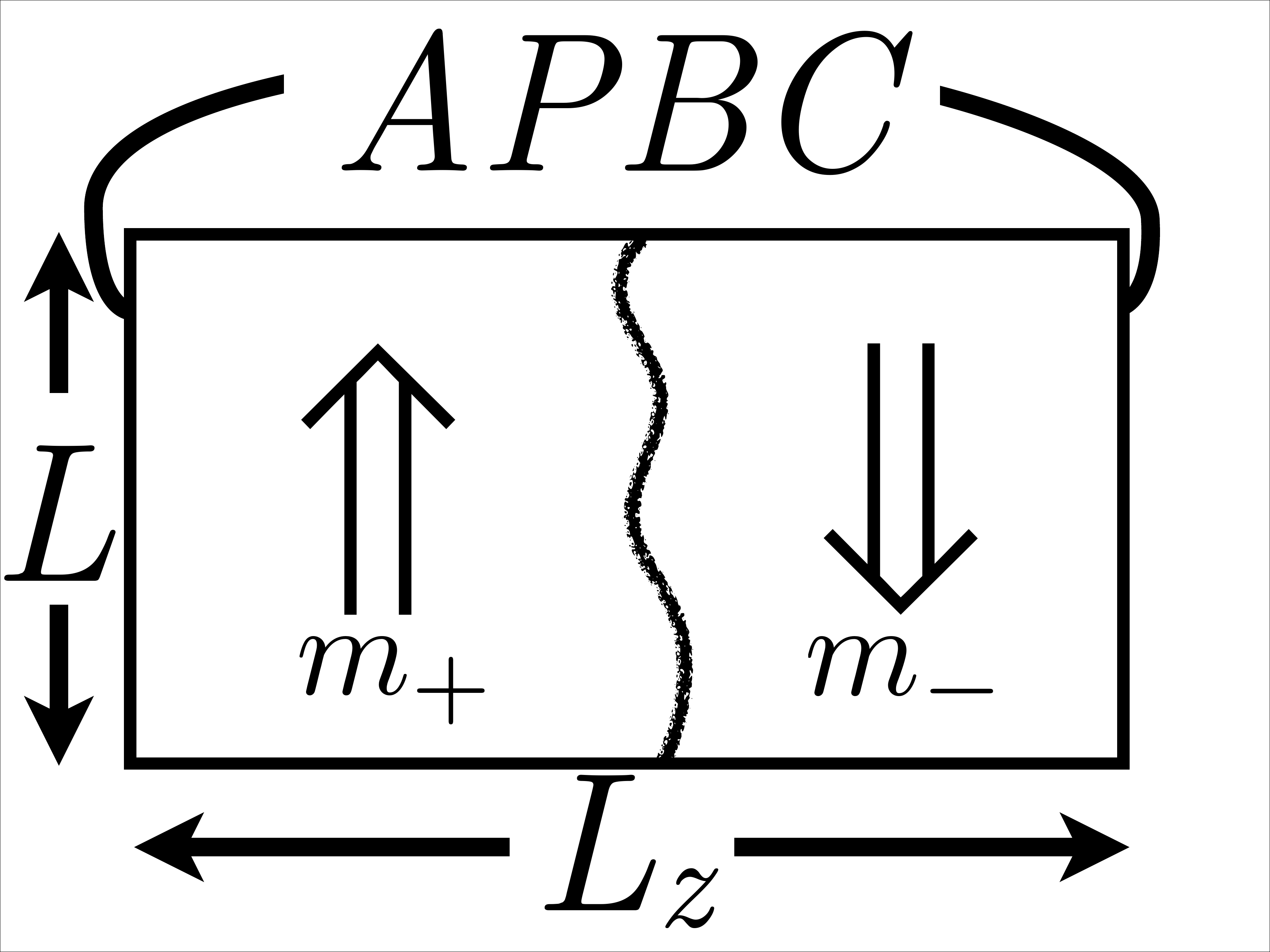}}
\subfigure[\label{fig: BoundaryConditionsPBC}]{\includegraphics[clip=true, trim=3mm 3mm 3mm 3mm, angle=0,width= 0.4 \columnwidth]{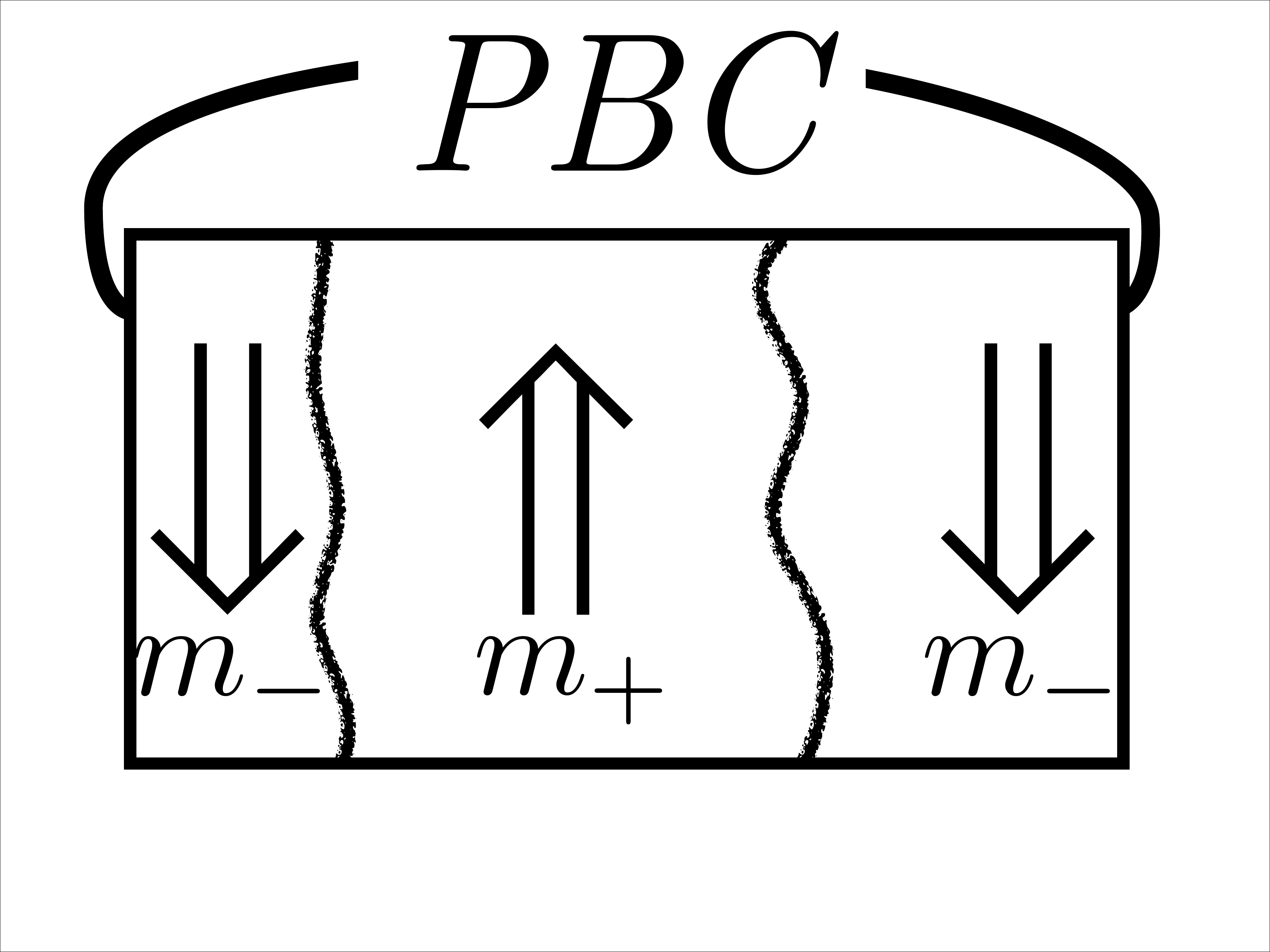}}
\caption{\label{fig: BoundaryConditions} Useful boundary conditions to study interfaces in Ising models, using a simulation box of linear dimension(s) $L$ parallel to the interface(s) and $L_z$ in the perpendicular direction. In the parallel direction(s), periodic boundary conditions (PBC) are used in all cases. Interfaces are schematically indicated by thick wavy lines, and the double arrows indicate the sign of the magnetization in the coexisting domains $(m_+, m_-)$. Note that $\langle m_+ \rangle =m_\text{coex}$, $\langle m_-\rangle = -m_\text{coex}$, $m_\text{coex}$ being the spontaneous magnetization of the Ising model in the thermodynamic limit, while in finite simulation boxes $m_+$, $m_-$ may fluctuate around these average values. Case (a) assumes free surfaces at the first $(n=1)$ and last $(n=L_z)$ layer, with a fixed spin boundary condition for all spins in the adjacent layers, $S_i= +1$ for all spins in the layer $n=0$, $S_i=-1$ in the layer $n=L_z+1$. Case (b) indicates the antiperiodic boundary condition (APBC) and case (c) uses also a periodic boundary condition in $z$-direction.}
\end{figure}

In this paper we have discussed the estimation of interfacial free energies associated with planar interfaces between
coexisting phases in thermal equilibrium, emphasizing the need to carefully address the finite-size effects when one employs a computer simulation approach. We have focused on the use of a simulation geometry where the linear dimension $(L)$ of the simulation box in the direction(s) parallel to the interface differs from the linear dimension perpendicular to the interface $(L_z)$. Using periodic boundary conditions in all (two or three) space directions, the situation that is normally considered (Fig.~\ref{fig: BoundaryConditionsPBC}) is a ``slab geometry'', where (for a fluid system) a domain of the liquid phase is separated by two planar interfaces (that are connected in themselves via the periodic boundary conditions (PBC) in the direction(s) parallel to the interface) from the vapor phase (the two vapor regions on the left and on the right of the liquid slab are connected by the periodic boundary condition in $z$-direction). This choice of geometry also applies to other systems (fluid binary mixture, Ising magnets, etc). For systems exhibiting a strict symmetry between both coexisting phases (Ising model, symmetrical binary Lennard-Jones mixture, etc.) also a simpler choice with a single interface is useful, where the boundary condition in the $z$-direction is antiperodic (APBC) rather than periodic. While for the situation with the PBC in $z$-direction we consider here only the canonical ensemble (conserved density of the fluid, or conserved relative concentration of the binary mixture, or conserved magnetization in the Ising magnet, respectively), for the systems with APBC it is instructive to study both the case of the canonical (c) ensemble and the case of the grandcanonical (gc) ensemble (where the respective order parameter, i.e., density, concentration, or magnetization, respectively, is not conserved, and the variable that is thermodynamically conjugate to this order parameter is fixed at the value that is appropriate for bulk two-phase coexistence). While in this APBC(c) case the position of the interface on average is fixed (by the chosen value of the order parameter), for the APBC(gc) case it is not, and the statistical fluctuation associated with this degree of freedom needs to be carefully considered. As discussed in Sec.~\ref{sec: SystemGeometryAndBoundaryConditions}, this translational degree of freedom of the interface gives rise to an entropic correction to the interfacial tension. Likewise, in the PBC case the liquid slab can be translated in the system as a whole, and this also shows up as a logarithmic correction.

But additional corrections arise as a consequence of the coupling between fluctuations of the bulk order parameter in the coexisting domains and the interface location (created by the constraint that there cannot be a net fluctuation of the total order parameter in the canonical ensemble, and so the individual fluctuations of the order parameter densities in both domains must be compensated by a suitable interface displacement). This so-called ``domain breathing'' effect causes an entropic correction for both the PBC and APBC(c) cases. We have given detailed evidence for these effects both in the case of the two-dimensional $(d=2)$ and three-dimensional $(d=3)$ Ising model. Note that for the $d=2$ Ising model capillary-wave type fluctuations of the interface affect these interfacial entropy corrections strongly as well, since the root mean squared interfacial width $\sqrt{\langle w_L^2 \rangle}$ scales like $L^{1/2}$ \{Eq.~\eqref{eq4}\}, and via the normalization of the translational entropy this gives rise to an additional $\ln (L)/(2L)$ correction to the interfacial tension.

\begin{figure}
\centering
\subfigure[\label{fig: ProbDistributionsFixedL}]{\includegraphics[clip=true, trim=0mm 0mm 0mm 0mm, angle=-90,width=0.49 \textwidth]{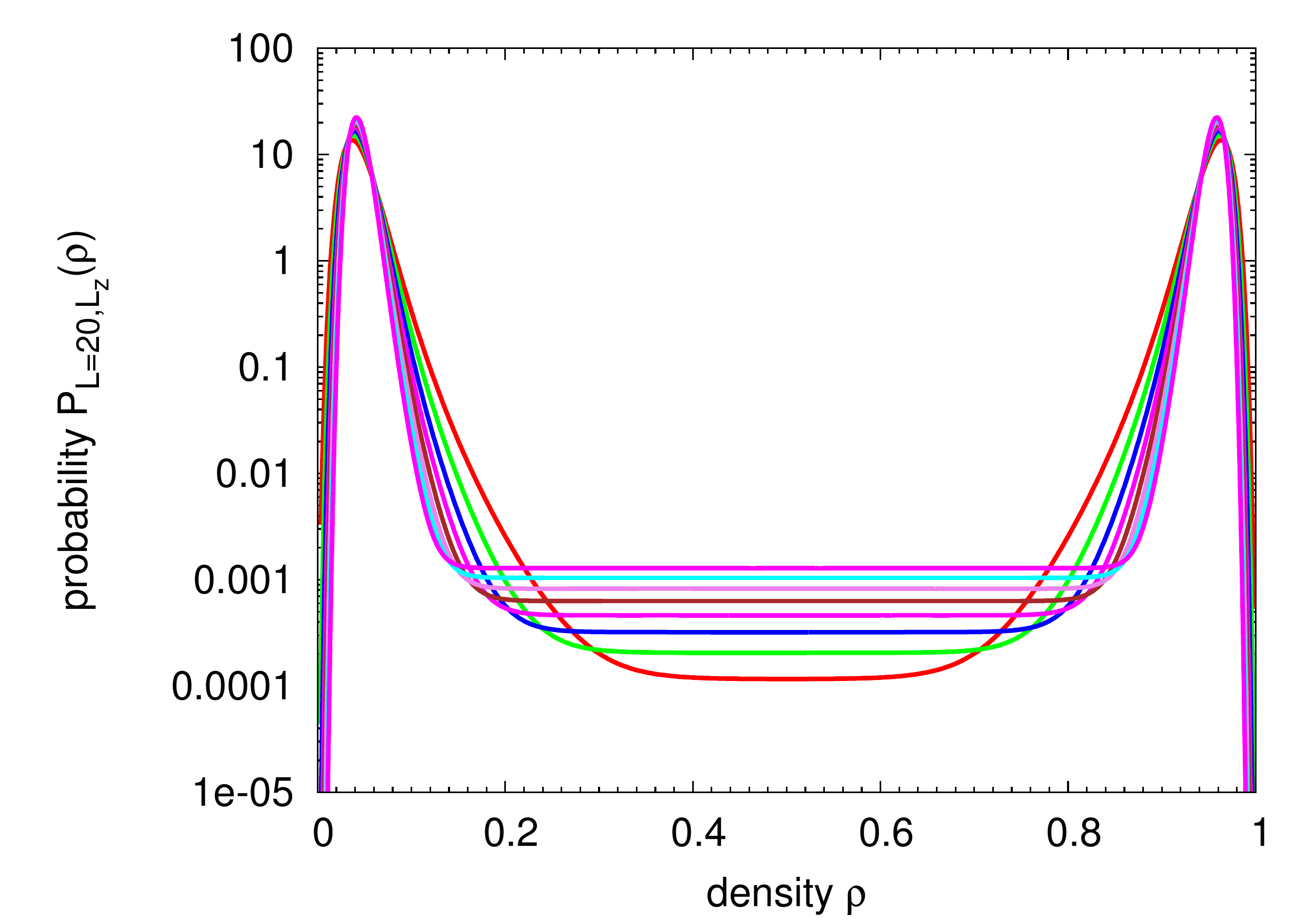}}
	\subfigure[\label{fig: ProbDistributionsFixedLz}]{\includegraphics[clip=true, trim=0mm 0mm 0mm 0mm, angle=-90,width=0.49 \textwidth]{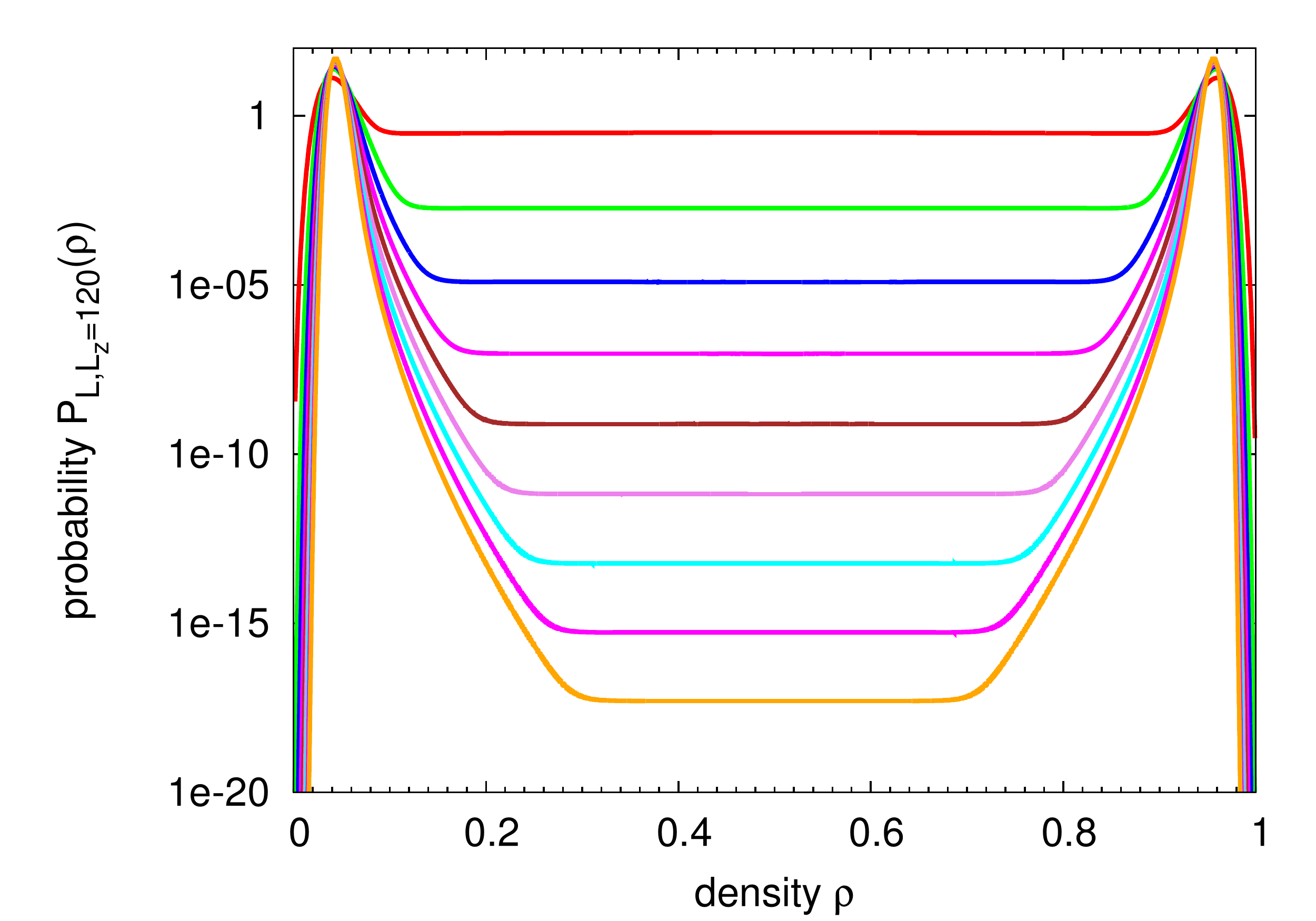}}
	\subfigure[\label{fig: ProbDistributionsLargeLz}]{\includegraphics[clip=true, trim=0mm 0mm 0mm 0mm, angle=-90,width=0.49 \textwidth]{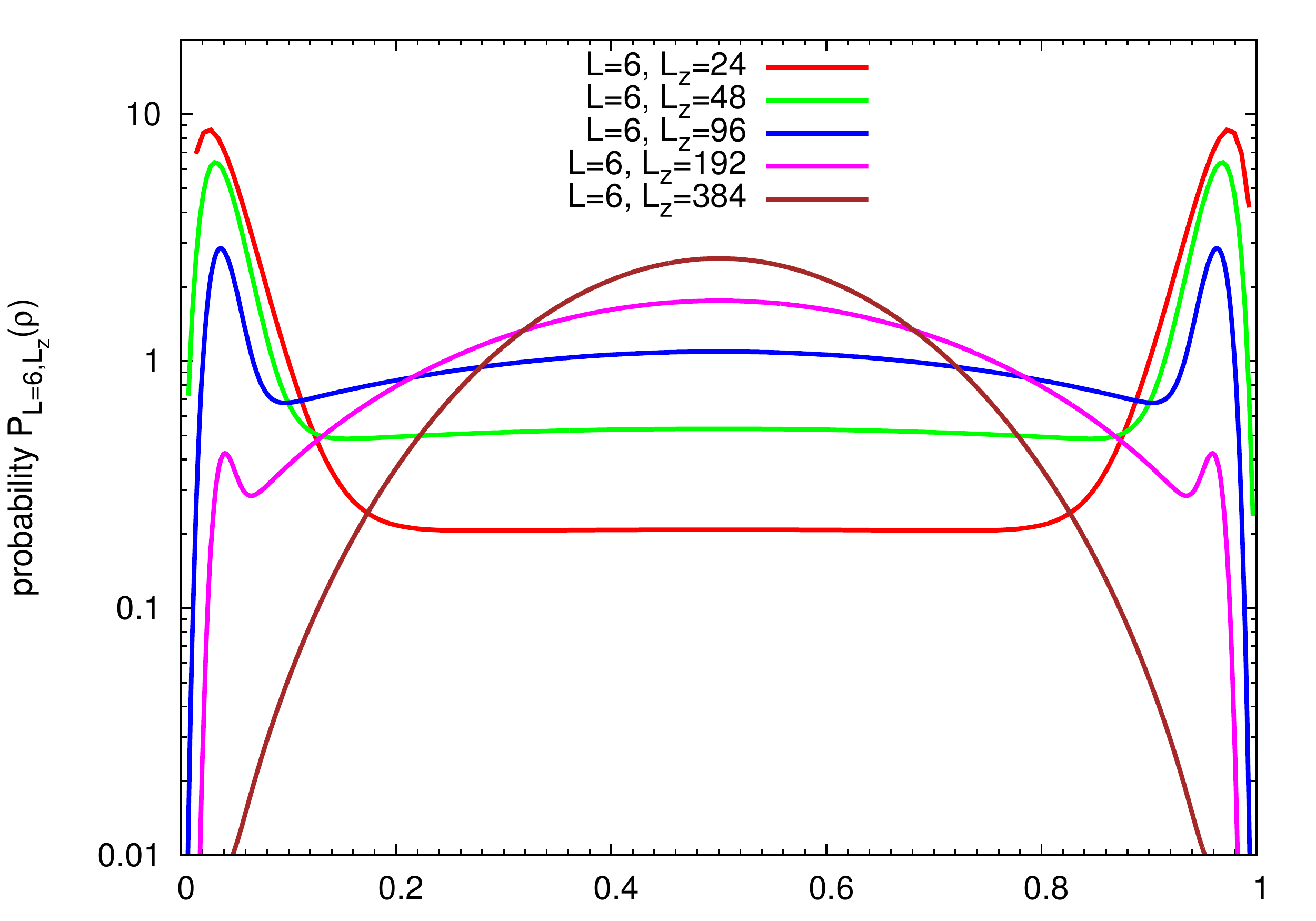}}
\caption{\label{fig: ProbDistributions} Density distribution $P_{L,L_z} (\rho)$ plotted vs. $\rho$ for the two-dimensional Ising (lattice gas) model at a temperature $\kb T/J =2.0$, for the case (a) $L=20$ and varying $L_z$, $L_z=30,40,50,60,70,80,90$ and 100 (from bottom to top at $\rho=0.5$), for the case (b) $L_z$=120, and varying $L$, $L$=10,20,30,40,50,60,70,80 and 90 (from top to bottom at $\rho=0.5$), and for the case (c) $L=6$ and increasing $L_z$ as indicated, illustrating the change from the double-peak distribution for rather small $L_z$ ($L_z=24$) to a three-peak distribution (e.g. $L_z=96,192$) to a distribution with a single central peak ($L_z=384$).}
\end{figure}

\begin{figure}
\centering
\subfigure{\includegraphics[clip=true, trim=2mm 2mm 2mm 2mm, angle=0,width= 0.80 \columnwidth]{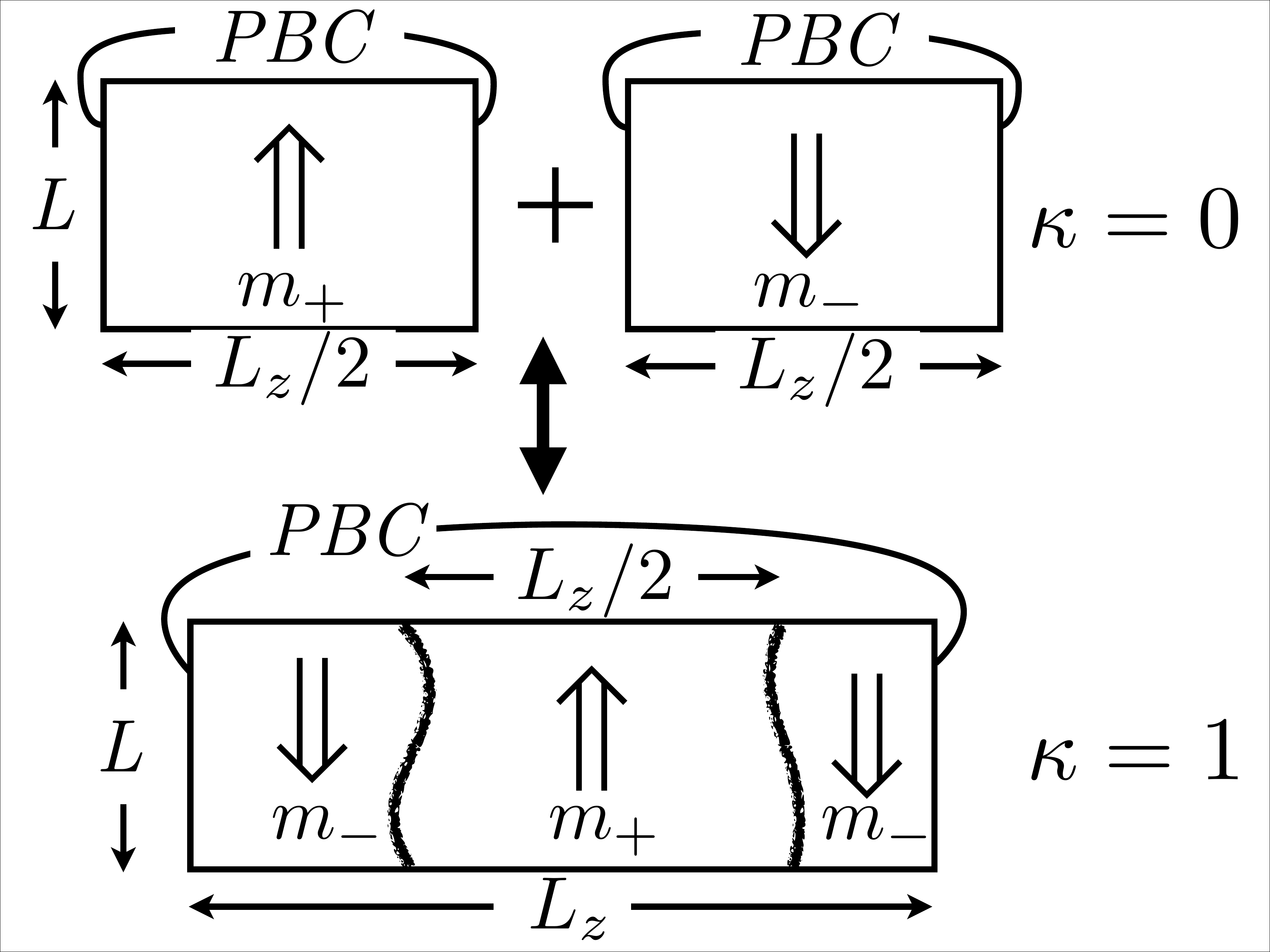}}
\caption{\label{fig: EnsembleSwitchMethodSketch} Schematic explanation of the ``ensemble switch method'' to find the interfacial free energy. A system is constructed as a linear combination of two Hamiltonians, $\mathcal{H}(\kappa)=\kappa \mathcal{H}_1 + (1-\kappa) \mathcal{H}_0$, where $\mathcal{H}_1$ is the desired system of interest (containing two interfaces in the case shown here), and $\mathcal{H}_0$ consists of two separate systems of half the linear dimension $L_z/2$ each, and periodic boundary conditions. The mixing parameter $\kappa$ is in the interval $0 \leq \kappa \leq 1$, and the free energy difference between the states with $\mathcal{H}(\kappa=0)$ and $\mathcal{H}(\kappa=1)$ yields twice the interfacial free energy. If the state $\kappa=1$ is a state with an APBC, one can obtain the free energy associated with a single interface analogously.}
 \end{figure}

All the methods that we discuss here rely on the consideration of the free energy difference between one of the systems discussed above and a system with the same linear dimensions but PBC throughout, so that no interfaces occur. Hence, it is necessary neither to locate where the interface is in the system, nor to characterize its microscopic structure. This free energy difference can either be found from sampling the order parameter distribution function (Fig.~\ref{fig: ProbDistributions}) across the two-phase coexistence region (which is a standard approach used since more than three decades) or from a new variant of the ``ensemble switch'' method (Fig.~\ref{fig: EnsembleSwitchMethodSketch}), described here. In this method, two bulk systems of size $L^{d-1} \times L_z/2$, with PBC containing the two coexisting phases, are connected in phase space via a continuous path to a system of size $L^{d-1} \times L_z$, where now the two phases coexist in one box, being separated by two interfaces.

We stress that these techniques by no means are the only methods from which interfacial tensions can be found: It is also possible to study the $L^{d-1} \times L_z$ system in the grandcanonical ensemble, and analyze the correlation function along the $z$-direction very precisely. Most of the time the system will reside in one of the pure phases, but the rare fluctuation where the system explores slab configurations gives rise to a nontrivial behavior of the correlation function, from which the interfacial tension can be extracted \cite{50,120}. This method is out of our scope here, as well as the possibility to extract the interfacial stiffness from an analysis of the capillary wave spectrum or from the size-dependence of the interfacial broadening. In both these methods the error estimation is a very subtle problem. Alternative algorithms due to Mon \cite{42} and Caselle et~al. \cite{50,60,61} which are particularly valuable to study the interfacial tension near the bulk critical point, have also been out of our consideration.

\begin{figure}
\centering
\subfigure[\label{fig: TranslationalEntropySketch}]{\includegraphics[clip=true, trim=25mm 30mm 30mm 20mm, angle=0,width=0.49 \textwidth]{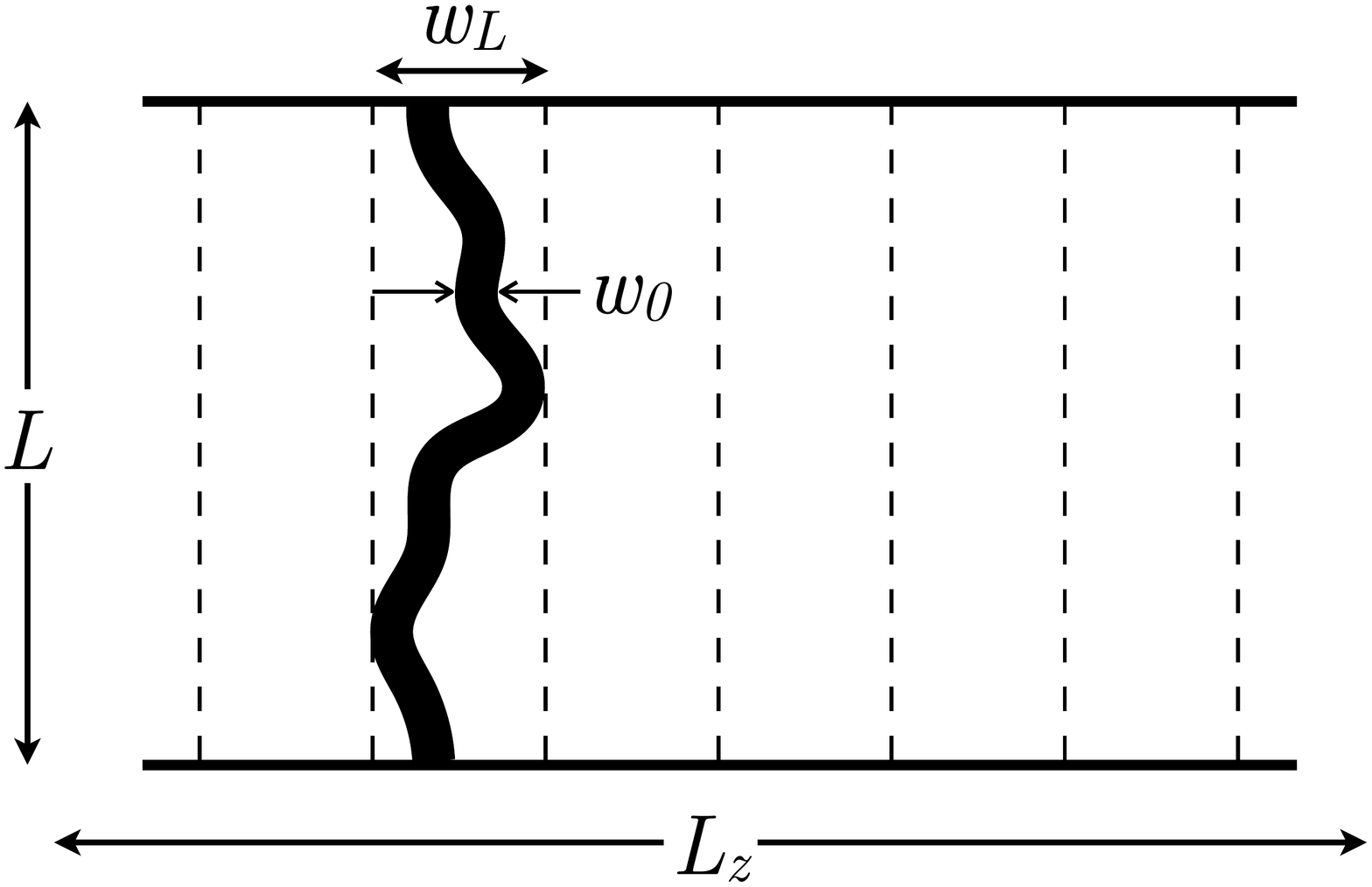}}
\subfigure[\label{fig: TranslationalEntropySnapshot1}]{\includegraphics[clip=true, trim=0mm 0mm 85mm 0mm, angle=90,width=0.49 \textwidth]{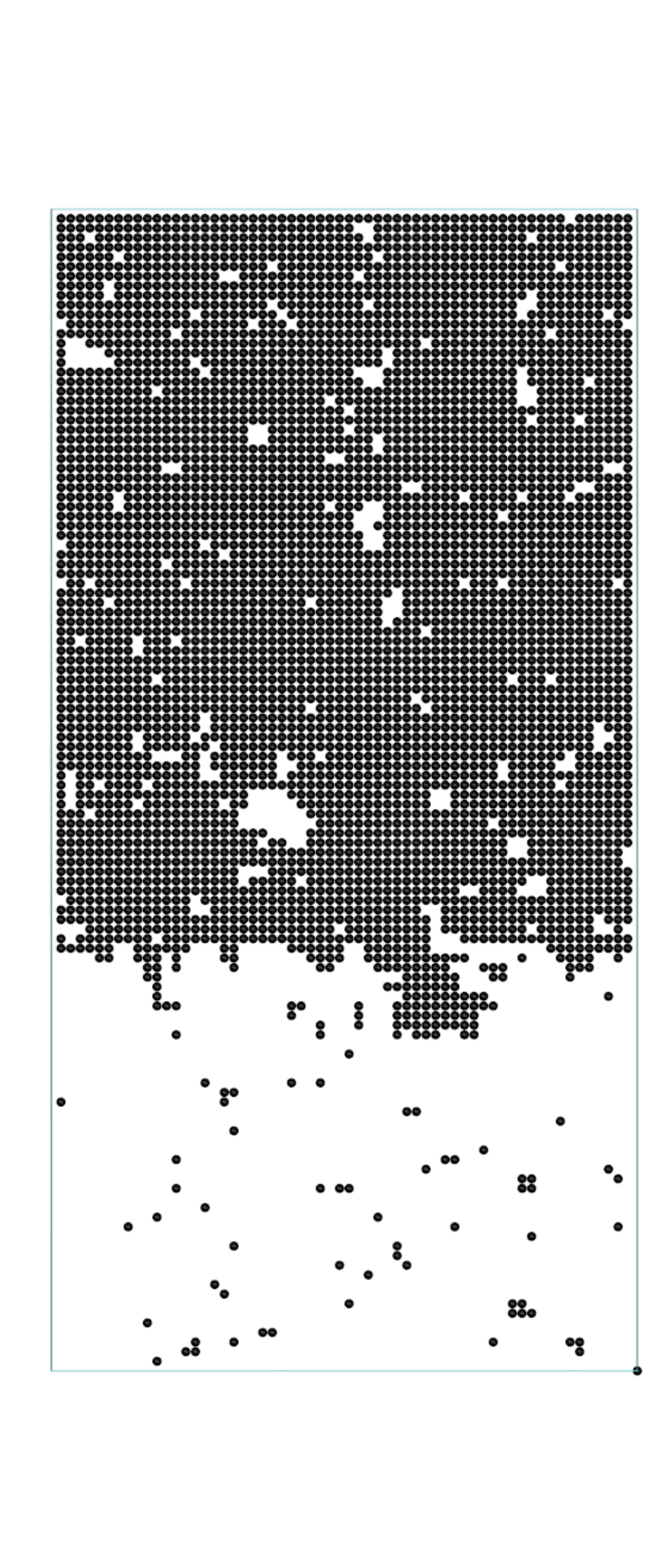}}
\subfigure[\label{fig: TranslationalEntropySnapshot2}]{\includegraphics[clip=true, trim=0mm 0mm 85mm 0mm, angle=90,width=0.49 \textwidth]{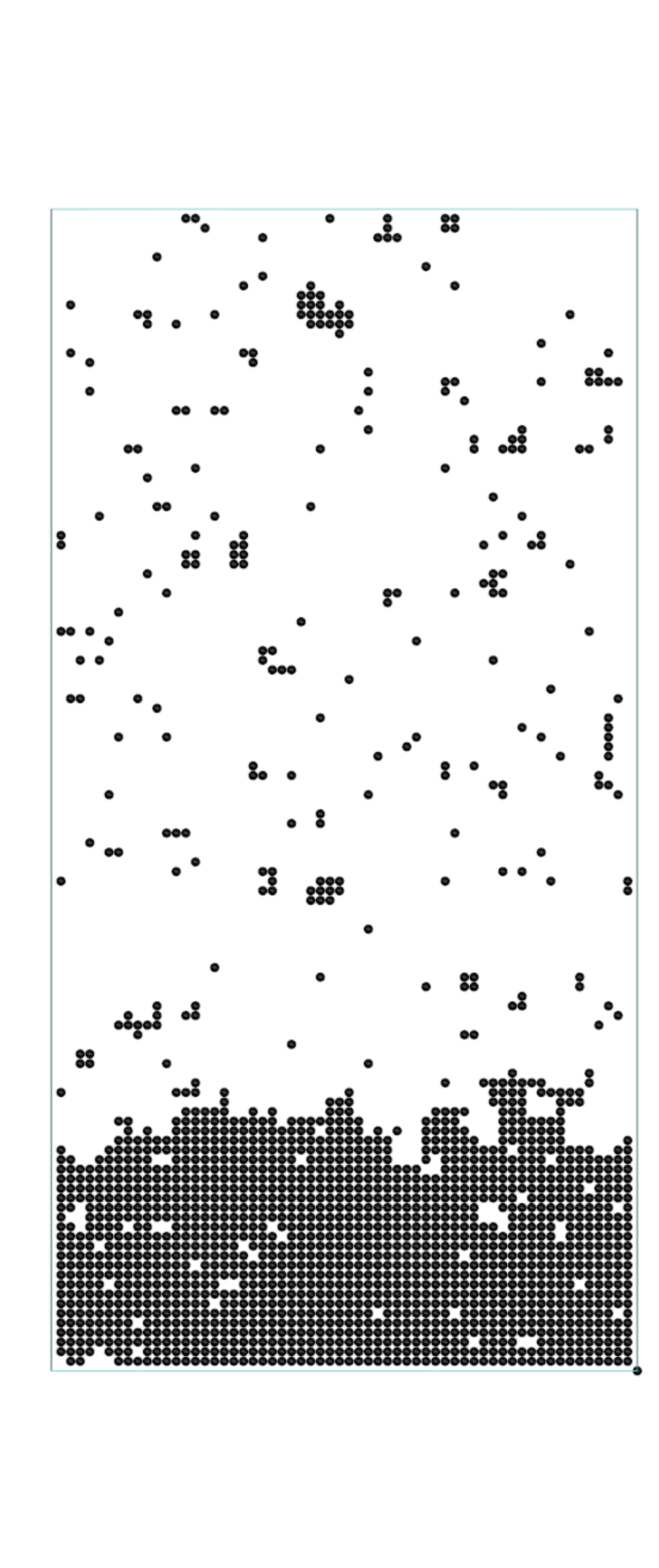}}
\caption{\label{fig: TranslationalEntropy} (a) Coarse-grained view of a single interface on a length scale $L_z$, to illustrate the counting of configurations to estimate its translational entropy. The scale $L_z$ is partitioned into cells of width $w_L$. Coarse-graining in $x$-direction over a length $l_x$ of order $\xi$ (not shown), to eliminate overhangs, bubbles etc. present in the original microscopic configurations (see parts (b), (c)), one obtains a smooth interface with intrinsic width $w_0$ comparable to $l_x$. This coarse-grained interface shows undulations on all length scales from $l_x$ to $L$ (capillary waves) and thereby exhibits a width $w_L > w_0$. (b), (c) show two snapshots resulting from a simulation with $L=60$, $L_z=120$, $\kb T/J=2.0$, APBC in the $z$-direction, and using the grand-canonical ensemble. The interface position ``explores'' the entire length $L_z$ of the system. Up spins are shown as black dots, down spins are not shown.}
\end{figure}

However, also for the methods described here the assessment of errors is rather difficult. Referring to Fig.~\ref{fig: TranslationalEntropy}, it is clear that the translational entropy of the interface is only sampled correctly if the simulation has lasted long enough that the slowly diffusing interface has in fact sampled the full extension $L_z$ of the sample. We have seen in the last section that in particular for off-lattice models of fluids (such as the Lennard-Jones system) this is difficult to achieve. In analytical theories \cite{88,89}, this problem is avoided by putting the system into a potential that localizes the interface. The price to be paid is that a correlation length $\xi_z$ is created, that characterizes the extent of interfacial motions around its average position in $z$-direction \cite{88,89}. While the theory from the outset is based on the concept of an effective interfacial Hamiltonian, it is desirable to avoid this concept in a simulation context. Of course, using the PBC(c) method where a liquid slab occurs, we can ``localize'' the whole slab, e.g.~by using a weak harmonic potential, centered around the center of mass position of the liquid slab. But one needs to carefully check that this potential does not affect other properties, apart from eliminating the translational motion of the slab as a whole.

Such ideas probably are indispensable, when one considers the extension of the method to liquid-solid interfaces, where it is simply too time-consuming to sample the translational motion of the crystalline slab. As a caveat we note, however, that we do not see an obvious recipe to suppress the ``domain breathing'' mechanism. Of course, if one uses a model based on the effective interface Hamiltonian concept, this mechanism has been disregarded from the outset; but the step linking explicitly atomistic Hamiltonians to effective interface Hamiltonians is problematic as well.

An extension that would also be very interesting to consider already for the Ising systems is the consideration of interfaces that are inclined relative to the simple (100) or (001) lattice planes: this extension would allow to study the anisotropy of the interface tension, which is well understood in $d=2$ \cite{123} but not explicitly known in $d=3$, apart from special cases \cite{44,63}. Such inclined interfaces naturally arise in the context of heterogeneous nucleation at walls \cite{118,119}, for instance. Another aspect of interest are finite-size effects on the interface tension of spherical droplets. We plan to report on such extensions in the future.

\acknowledgements{This research was supported by the Deutsche Forschungsgemeinschaft (DFG), Grant No VI 237/4-3. We thank M.~Caselle for useful literature hints. We are grateful to the Centre for Data Processing ZDV of the University of Mainz for a generous grant of computing time on the high-performance computer Mogon.}

\clearpage

\clearpage

\clearpage

\clearpage

\clearpage

\clearpage

\clearpage

\clearpage

\end{document}